\documentclass[useAMS,usenatbib]{mn2e}
\usepackage{graphicx}

\def\mathbi#1{\textbf{\em #1}}

\def\MMSN{\bowtie}
\def\correction{\alpha}

\newcommand{\unit}[1]{\ \mathrm{#1}}
\newcommand{\abs}[1]{\left| #1 \right|}
\def\llt{<\!\!<}
\def\ggt{>\!\!>}
\def\eq{\hspace{-6pt}=\hspace{-6pt}}
\def\+{\hspace{-8pt}+\hspace{-8pt}}

\def\cou{{cou}}
\def\ori{{cou}} 
 
\def\DP{{\mathbf{I}}}
\def\DPSub{{ \textbf{\tiny  {I}}}}
\def\dDP{d\mathbf {I}}
\def\A{{{}_{A}}}
\def\C{{{}_{C}}}
\def\la{{\hspace{-0.5pt}_{L}}} 
\def\sm{{\hspace{-0.5pt}_{S}}}

\def\Wion{\Delta W_{\mathrm {ion}}}

\def\charge{{ch}}
\def\mfp{{\mathrm {mfp}}}
\def\dis{{\mathrm {dis}}}
\def\max{{\mathrm {max}}}
\def\min{{\mathrm {min}}}
\def\crit{{\mathrm {crit}}}
\def\contact{{\mathrm {kiss}}}

\title[Dust-dust Collisional Charging and Lightning]{Dust-Dust Collisional Charging and Lightning in Protoplanetary Discs}
\author[Takayuki Muranushi]{Takayuki Muranushi$^{1}$\thanks{E-mail:
muranushi@tap.scphys.kyoto-u.ac.jp; muranushi@gmail.com}\\
$^{1}$ Department of Physics, Kyoto University, Sakyo-ku, Kyoto, 606-8502, Japan}
\begin{document}

\date{Submitted 2009 Aug 11}

\pagerange{\pageref{firstpage}--\pageref{lastpage}} \pubyear{2009}

\maketitle

\label{firstpage}

\begin{abstract}

  We study the role of dust-dust collisional charging in protoplanetary discs.
  We show that dust-dust collisional charging becomes an important process in
  determining the charge state of dust and gas, if there is dust enhancement
  and/or dust is fluffy, so that dust surface area per disc volume is locally
  increased.

  We solve the charge equilibrium equations for various disc
  environments and dust number density $\eta$, using general purpose
  graphic processors (GPGPU) and {\sc cuda} programming language.  We
  found that as dust number density $\eta$ increases, the charge
  distribution experience four phases. In one of these phases the
  electrostatic field $E$ caused by dust motion increases as $E
  \propto \eta^4$.  As a result, macroscopic electric discharge takes
  place, for example at $\eta = 70$ (in units of minimum-mass solar
  nebula (MMSN) values, considering two groups of fluffy dust with
  radii $10^{-2}\unit{cm}$, $10^{2}\unit{cm}$). We present a model
  that describes the charge exchange processes in the discs as an
  electric circuit. We derive analytical formulae of critical dust
  number density for lightning, as functions of dust parameters.

  We estimate the total energy, intensity and event ratio of such discharges
  (`lightning').  We discuss the possibility of observing lightning and sprite
  discharges in protoplanetary discs by Astronomically Low Frequency ({\em
    ALF}) waves, {\em IR} images, {\em UV} lines, and high energy gamma rays.
  We also discuss the effects of lightning on chondrule heating, planetesimal
  growth and magnetorotational instability of the disc.
 \end{abstract}

 \begin{keywords}
   methods:numerical --- 
   planetary systems:formation --- 
   planetary systems:protoplanetary discs ---
   meteors, meteoroids ---
   plasmas --- 
   turbulence
 \end{keywords}

\section{Introduction} \label{introduction}

Planets are formed in protoplanetary discs from interstellar dust.  The
electric charge state of the dust aggregates in the protoplanetary discs is
one of the key parameters in understanding a number of aspects of
protoplanetary discs and protoplanetary formation.

Planet formation begins with mutual sticking of$\unit{\mu m}$-sized dust, most
probably leading to extremely low density, fluffy structure of the dust
\citep[e.g.][]{1993A&A...280..617O}.  The occurrence of fluffy dust is
suggested by laboratory experiments \citep[e.g.][]{1998Icar..132..125W,
  1998EM&P...80..285B}, by theories
\citep[e.g.][]{2007A&A...461..215O,2008A&A...489..931Z}, by N-body simulations
\citep[e.g.][]{1999Icar..141..388K,2008ApJ...684.1310S,2008ApJ...677.1296W},
and by observations, including optical observations of dust in star forming
region \citep[e.g.][]{2001ApJ...557..193E} and observations of dust linear
polarization in comet comae \citep{2007P&SS...55.1010L}.  For a review of this
field, see e.g. \citet{2004ASPC..309..369B}.

The inner structure of the dust aggregates,
relative velocity, and
electric charge are key parameters that determine the
growth and migration of dust aggregates.
Dust relative velocity \citep{2008A&A...480..859B} includes 
random motion caused by turbulence \citep{2007A&A...466..413O}
and Brownian motion \citep{1996Icar..124..441B},
and bulk motion caused by 
vertical sedimentation \citep{2004A&A...421.1075D}
and radial migration \citep{1977MNRAS.180...57W}. 
The collision velocity governs the
growth rate \citep{2008ApJ...684.1310S}, 
compactification \citep{2009ApJ...696.2036W},
and disruption  \citep{2008ApJ...677.1296W},
of the dust. 

\citet{2009ApJ...698.1122O} considered the charge state of the dust aggregates
in protoplanetary discs.
They assumed that the charge state is determined by 
absorption equilibrium of ions and free electrons.
Since electrons have much larger thermal velocity compared to positive ions,
plasma absorption makes all dust to
charge weakly negative.
The repulsive Coulomb force may suppress dust-dust 
collisional growth for all but the heaviest dust species
who can overcome the Coulomb barrier.

It is also possible that the charge state of the dust is affected by
dust-dust collision.  The effect has been simply ignored in most
research due to the fact that in protoplanetary discs, dust has low
number density, and is surrounded by weakly ionised plasma.  We give
quantitative estimates of the effect of dust-dust collision as a
function of dust size, fractal dimension, and number density, and show
for the first time that the dust-dust collision is actually an
important factor in high dust number density regions of protoplanetary
discs.

One of the possible dust-dust collisional charging mechanisms is known as the
triboelectric process, where two bodies exchange electrons and sometimes
molecular ions when they come into contact
\citep[e.g.][]{2001JGR...106.8343S}.  Another mechanism is possible for
materials with spontaneous surface charge, such as $\mathrm {H_2O}$ ice
crystals \citep[e.g.][]{10.1021/ja077205t}. In this mechanism the surface matter
within typical depth $\sim 1.0 \times {10}^{-4}\unit{cm}$ \citep{JGR105(D16)10185}
is exchanged together with contained charge. 

Surface space charge due to electron spill-out is widely known among
metals and semiconductors \citep{1994.somorjai.book}, the charge
separation being $\sim 10^{-7}\unit{cm}$ deep for metals and $\sim
10^{-5}\unit{cm}$ deep for semiconductors.  $\mathrm {H_2O}$ is unique
in that molecular ions $\mathrm {OH^{-}}$ and $\mathrm {H_3O^{+}}$ holds
the charge, and that proton exchange between the molecules
\citep[Grotthus mechanism; c.f.][]{Agmon1995456} allows charge diffusion much faster
than molecular ion diffusion. Thus surface charge separation develops
as deep as $\sim 2.0 \times {10}^{-4}\unit{cm}$ \citep{JGR106(D17)2039520402}.  For
example, ammonia lacks the mechanism \citep{Goncalves1999140}.  It is
important that charge separation layer is deeper than exchange depth,
because if the entire charge separation layer is exchanged, charge
transport is neutral and collisional charging do not take place.

The dust-dust collisional charging due to the exchange of this
spontaneous surface charge of ice crystals, is an established model in
the context of meteorology
\citep[e.g.][]{TakahashiThunder,10.1002/qj.49711347807,JGR106(D17)2039520402}
that explains lightning on earth.  When two ice dust of different
surface states collide, they exchange their surface charge, producing
charged dust.  When the charged particles within nonconducting gas are
separated by some external force, electric field grows between
them. At the point the electric field is larger than the dielectric
field strength of the gas, rapid ionisation of the gas occurs,
converting the electrostatic energy into kinetic energy of the
electrons and ions.  This is electric discharge.  Lightning in the
earth's atmosphere is one of the most prominent, and well studied
examples of electric discharge phenomena; in thunderclouds, typically
$3.0 \times {10}^{10}\unit{esu}$, or $1.0 \times {10}^{1}\unit{C}$ of electric charge is
repeatedly separated and neutralized with typical length scales
$1.0 \times {10}^{5}\unit{cm}$ \citep{JGR1989.94.1165}.

In  protoplanetary discs, lightning is one of the candidate mechanisms for chondrule
heating, although compared to other models e.g. heating by shock wave
\citep[e.g.][]{2008Icar..197..269M}, some difficulties have been
pointed out \citep{1997LPI....28.1515W}.  For example, electric field cannot
grow large enough to cause electrostatic breakdown in standard discs
\citep{1997Icar..130..517G}.  Moreover, when$\unit{mm}$-sized silicate aggregates
made of$\unit{\mu m}$-sized monomers are subject to electric discharge, they
generally fragment without being thermally processed
\citep{2008Icar..195..504G}.

Lightning in protoplanetary discs is strongly related to turbulence.  The
relative random velocity between the charged dust species that sets the dust
to collide, results from the turbulence.  Also the difference of the bulk
velocities between the charged dust species that leads to macroscopic charge
separation results from the turbulence.

The turbulent state of the accretion discs is often expressed in terms of 
viscous $\alpha$ parameter introduced by \citet{1973AnA....24..337S}.
Since the specific angular momentum increases outward in Keplerian discs,
they satisfy Rayleigh's hydrodynamical stability criterion,
and there are no clear mechanism for hydrodynamic turbulence
in protoplanetary discs \citep{2004ApJ...605..321S}.
On the other hand, the angular velocity decreases outward in Keplerian discs,
they satisfy criterion for magnetorotational instability (MRI).

Therefore, if a protoplanetary disc is ionised enough to sustain
magnetic field, MHD turbulence is excited and  $\alpha$ parameter can be
as large as $1.0 \times {10}^{-3}\sim1.0 \times {10}^{-1}$ \citep{1998ApJ...506L..57S}.
If the ionisation is suppressed, on the other hand, 
 $\alpha \simeq 1.0 \times {10}^{-5}$.
For a typical protoplanetary disc 
it is believed that so-called `dead zones' form 
between $0.1\unit{AU}$ and $10\unit{AU}$
where instabilities are damped and gas flow is almost laminar
\citep[e.g.][]{1996ApJ...457..355G}.
But it is possible that MRI is active in the whole disc, if 
sufficient ionisation degree is maintained, for example 
by turbulent mixing \citep{2007ApJ...659..729T}
or by self-sustained ionisation \citep{2005ApJ...628L.155I}.
Thus ionisation state of the protoplanetary discs is
critical in determining $\alpha$ and understanding the fate of planetesimals
and protoplanetary discs
\citep[e.g.][]{2007ApJ...664L..55K, 2008A&A...480..859B}.

The purpose of this paper is twofold: One is to solve the local charge
exchange equilibrium of gas and dust numerically, for various dust parameters
such as radii, fractal dimensions and dust number density, with dust-dust
collisional charging taken into consideration; Given the results, the other
goal is to determine the critical dust number density $\eta_\crit$ under which
lightning to take place, as analytical functions of other dust parameters such
as radii, fractal dimensions and disc environment parameters such as
temperature and gas number density.

This paper is organized as follows.  We define the terms we use in
Table \ref{table-terminology}, and we list the symbols we frequently
use in Table \ref{table-symbols}.  In \S \ref{section-two-dust-model}
we introduce the dynamic charge exchange equations and its equilibrium
solution in schematic forms. We introduce circuit diagram to depict
them (Fig. \ref{figure-circuit}).  In \S \ref{section-scenario} we
examine the processes in protoplanetary discs that set the parameters
for the charge equilibrium equations.  Crucial parameters are dust
number density, the amount of charge exchange in single dust-dust
collision, and relative velocity.  In \S \ref{section-lightning} we
estimate the electrostatic field strength, and define the critical
number density $\eta_\crit$ for lightning in the protoplanetary discs.
At this point all the equations are specified, and we solve them
numerically.  In \S \ref{section-results} we show the results of the
simulations.  We describe four distinct phases of the charge
distribution and explain the results using circuit diagrams.  We also
give analytical estimates for electric field strength in
protoplanetary discs and critical number density $\eta$ for lightning
to occur.  In \S \ref{section-discussion} we discuss the possibility
of various phenomena caused by the highly charged dust and lightning
in protoplanetary discs, and their observations.

\begin{table*}
\begin{center}
  \small
  \begin{tabular}{cccccc}
    \multicolumn{6}{c}{particle} \\
    \hline
    \multicolumn{3}{c||}{gas} & & \multicolumn{2}{c}{dust} \\
    \cline{1-3}    \cline{5-6}
    & \multicolumn{2}{|c||}{plasma} & & \multicolumn{2}{c}{ } \\
    \cline{2-3} 
    neutral gas & ion & electron & &smaller dust & larger dust \\
    $\mid$ & $\mid$ & $\mid$ & & \multicolumn{2}{|c}{$\mid$\hspace{16pt}$\times$\hspace{16pt}$\mid$}  \\
    neutral & cation & anion & &cationic & anionic \\
  \end{tabular}
\end{center}
\caption{\small
  The terminology we use in this paper. 
  `Particle' is generic term for all components in the protoplanetary discs.
  Solid components are `dust,' and the others are `gas.'
  `Gas' components are further subdivided into `neutral gas,' 
  and charged components, or `plasma.'
  Finally, `plasma' consists of `electron,' the negative charge carrier, 
  and various molecular `ion,' the positive charge carrier.
  On the right side of the table, `dust' is classified by their size
  as `smaller dust' and `larger dust.' 
  Either can be `anionic' or `cationic' dust, depending on the
  material they consist of.
  We also use the one-letter symbols
  `g', `e', `i', `S', and `L' for 
  neutral {\bf g}as, {\bf e}lectron, {\bf i}on, 
  {\bf S}maller and {\bf L}arger dust.
  The symbols for `{\bf C}ationic' and `{\bf A}nionic' dust are
  `C' and `A'. 
  We use variable $\DP$ to represent one of these symbols. 
} \label{table-terminology}
\end{table*}


\def\spc{\vspace{3pt}}
\begin{table*}
\begin{center}
  \small
  \begin{tabular}{cccc}
    {\bf symbol }& {\bf value/dimension} & {\bf meaning} & {\bf definition }\\
    \hline
    \multicolumn{3}{c}{\bf --- constants ---} \\\spc 
    $r$ & $2.7\unit{AU}$ & orbital radius considered & - \\ \spc 
    $\Sigma_g^{\MMSN} $ & $3.8 \times {10}^{2}\unit{g\ cm^{-3}}$ & 
    gas surface density of MMSN & (\ref{mmsn-Sigma}) \\ \spc 
    $h^{\MMSN}$ & $1.6 \times {10}^{-1}\unit{AU}$ &
    scale height of MMSN & (\ref{mmsn-h}) \\ \spc 
    $T^{\MMSN}$ & $1.7 \times {10}^{2}\unit{K}$ 
    & temperature of MMSN & (\ref{mmsn-T}) \\ \spc 
    $\rho_g^{\MMSN}$& $1.6 \times {10}^{-10}\unit{g\ cm^{-3}}$ & 
    gas density of MMSN & (\ref{mmsn-rho}) \\ \spc
    $\rho_\sm^{\MMSN}$ & $1.6 \times {10}^{-12}\unit{g\ cm^{-3}}$  & 
    spatial density of smaller dust  in MMSN & (\ref{rho-small}) \\ \spc
    $\rho_\la^{\MMSN}$ & $1.6 \times {10}^{-13}\unit{g\ cm^{-3}}$  &
    spatial density of larger dust in MMSN & (\ref{rho-large}) \\ \spc
    $\eta_{\charge}$  &
    $0.1$ &
    charge exchange efficiency  & \S \ref{section-surface-charge-exchange-I}\\ \spc
    $\sigma_{\charge}$  &
    $6.2 \times {10}^{9}\unit{e\ cm^{-2}}$ &
    charge surface density& \S \ref{section-surface-charge-exchange-I}\\ \spc
    $u_\la$ & $3.4 \times {10}^{3}\unit{cm\ sec^{-1}}$& bulk velocity of larger dust
    to other species & \S \ref{section-model}\\ \spc
    $v_\DPSub$ & $\sqrt{k_B T/m_\DPSub}$ & random velocity of particles of
    species $\DP$ & \S \ref{section-model} \\ \spc
    $\Delta v_{\la,\sm}$ & $3.4 \times {10}^{3}\unit{cm\ sec^{-1}}$& mean collision velocity between
    a smaller dust and a larger dust & \S \ref{section-model} \\ 
    \multicolumn{3}{c}{\bf --- independent variables ---} \\ 
    $\eta$ & 1 & dust number density of the considered region &  \\ \spc
    & & divided by that of the MMSN model &  -\\ \spc
    $r_\DPSub$ & $\unit{cm}$ & radius of a dust aggregate of species $\DP$&  
    (\ref{fractal-dimension-definition})\\ \spc
    $D_\DPSub$ & 1 & fractal dimension of a dust aggregate of species $\DP$ &  
    (\ref{fractal-dimension-definition})\\ 
    \multicolumn{3}{c}{\bf --- dependent variables ---} \\ \spc
    $m_\DPSub$ & $\unit{cm}$ & mass of a dust aggregate of species $\DP$&  
    (\ref{dust-mass-definition})\\ \spc    
    $\rho_\sm$ & $\unit{g\ cm^{-3}}$  &
    condensed density of smaller dust & $\eta  \, \rho_\sm^{\MMSN}$ \\ \spc
    $\rho_\la$ & $\unit{g\ cm^{-3}}$  &
    condensed density of larger dust & $\eta \, \rho_\la^{\MMSN}$  \\ \spc
    $n_\DPSub$ & $\unit{cm^{-3}}$  &
    number density of dust of species $\DP$ in condensed regions  &
    $\rho_\DPSub / m_\DPSub$  \\ \spc
    $q_\DPSub$ &
    $\mathrm{esu}$ &
    The charge carried by a single particle of species $\DP$ & - \\ \spc
    $Q_\DPSub$ &
    $\mathrm{esu \ cm^{-3}}$ &
    The charge density carried by species $\DP$ & $q_\DPSub n_\DPSub $ \\
    $J_{\DPSub,\DPSub'}$  &
    $\mathrm{esu \ cm^{-3} \ s^{-1}}$ &
    charge transferred from species $\DP$ to  & \\ \spc
    & & species $\DP'$ per unit time per unit volume & 
    (\ref{specific-J-L-S}-\ref{specific-J-i-e})\\ \spc
    $S_\contact$ &
    $\unit{cm^{2}}$ &
    contact surface area within a dust-dust collision &
    (\ref{cross-section-contact}) \\ \spc
    $\Delta q_{\A,\C}$  &
    $\unit{esu}$ &
    amount of charge exchanged within a dust-dust collision&
    $\eta_{\charge} \, \sigma_{\charge} \, S_\contact$ \\ \spc
    $\sigma_\cou$ &
    $\unit{cm^{2}}$ &
    cross section between two charged particles&
    (\ref{sigma_coulomb_focusing+}),(\ref{sigma_coulomb_focusing-}) \\ \spc
    $j_D$ & $\unit{esu\ cm^{-2}\ s^{-1}}$ & current carried by dust particles
    & \S \ref{section-lightning}
    \\ \spc  
    $j_p$ & $\unit{esu\ cm^{-2}\ s^{-1}}$ & current carried by plasma
    particles 
    & \S \ref{section-lightning} \\ \spc
    $E_\dis$ & $\unit{G}$ & critical electric field strength for lightning & (\ref{equation-E_dis})\\ 
    $E_\max$ & $\unit{G}$ & 
    (local maximum of) electric field &\\      \spc
    &&generated in the protoplanetary disc & (\ref{equation-E_max})\\
    $\chi$ & 1 & whether the collision cross section between smaller dust and & \\     \spc
    &&plasma particles are geometric$(\chi \llt 1)$ or Coulomb $(\chi \ggt 1)$& 
    (\ref{definition-of-chi})
    \\      \spc
    $\eta_\crit$ & 1 & the dust number density at which lightning takes place& 
    (\ref{analytic-eta-critical}) and
    (\ref{analytic-eta-critical-c}-\ref{analytic-eta-critical-d})
   \end{tabular} 
  \caption{The list of symbols frequently used in this paper.}
  \label{table-symbols}
\end{center}
\end{table*}

\section{Model description}
 \label{section-two-dust-model}

In this section we describe our models. 
In \ref{two-component-model}
we model the disc and the dust at the unperturbed
state, then introduce the models for dust number density.
In \ref{section-circuit-model}
we model the charge density and charge separation processes.

\subsection{Disc Model} \label{two-component-model}

Unless otherwise mentioned, we focus on a local, uniform box at
certain orbital radius $r$ near the equatorial plane of the
protoplanetary disc.  We model the protoplanetary disc based on the
minimum-mass solar nebula (MMSN) model \citep{1981PThPS..70...35H}.
The gas surface density $\Sigma_g^{\MMSN}(r)$, disc scale height
$h^{\MMSN}(r)$ , and the temperature $T^{\MMSN}(r)$ of the disc are
\begin{eqnarray}
  \Sigma_g^{\MMSN} \left( r \right) &\eq& 1.7 \times {10}^{3} \left( \frac{r}{AU} \right) ^{-\frac{3}{2}}
  \unit{g \ cm^{-2}}, \label{mmsn-Sigma}\\
  h^{\MMSN} \left( r \right) &\eq& 4.7 \times {10}^{-2}  \left( \frac{r}{AU} \right) ^{\frac{5}{4}} \unit{AU},
  \label{mmsn-h} \\
  T^{\MMSN} \left( r \right) &\eq& 2.8 \times {10}^{2}  \left( \frac{r}{AU} \right) ^{-\frac{1}{2}} \unit{K},
  \label{mmsn-T}
\end{eqnarray}
where $r$ is the distance from the central star.
This leads to gas density distribution
\begin{eqnarray}
  \rho_g^{\MMSN} \left( r \right) &\eq& 2.4 \times {10}^{-9} \left( \frac{r}{AU} \right) ^{-\frac{11}{4}} 
  \unit{g \ cm^{-3}}.  \label{mmsn-rho}
\end{eqnarray}
The dust-to-gas ratio in MMSN is approximately $1.0 \times {10}^{-2}$. 

We use the model by \citet{2004ApJ...614..490C}, and introduce two
species of dust, the smaller dust and the larger dust (see Table
\ref{table-terminology}.)  We further assume that surface density of
the larger dust is $10$ per cent of the total dust surface density.
These two species are also either `cationic' and `anionic.'  The
`cationic' species receives the positive electric charge through
dust-dust collision.  See Appendix \ref{cationic-anionic-definition}
for the justification of this two-dust model.  We can also represent
the role of various molecular ions by one abstract ion species
`${\mathrm i}$,' according to \citet{2009ApJ...698.1122O}.

The motivation for this two-dust model is twofold.  First, the two dust model
is the simplest model that can handle the dust-dust collisional
charge separation and the macroscopic relative velocity between the dust
species.  Second, the charge tendency of the dust and their size are strongly
correlated.  In one scenario, older dust are larger and also anionic.  In
another scenario, dust made of ice is larger and also cationic compared to
dust made of silicate.  (see \S \ref{section-charge-separation} for the
details.)  Therefore, we expect that instead of considering four (cationic
smaller dust, cationic larger dust, anionic smaller dust, and anionic larger dust) species of dust, we
can correlate the two size species with the two charge tendency species,
(Table \ref{table-terminology}), although both correspondences (smaller dust is
cationic / larger dust is cationic) are possible.

To summarise, we define the reference 
density of the smaller dust $\rho_\sm^{\MMSN}(r)$ and
the density of the larger dust $\rho_\la^{\MMSN}(r)$ as
\begin{eqnarray}
  \rho_\sm^{\MMSN}\left(r\right) &\eq& 1.0 \times {10}^{-2} \rho_g\left(r\right), \label{rho-small} \\
  \rho_\la^{\MMSN}\left(r\right) &\eq& 1.0 \times {10}^{-3} \rho_g\left(r\right). \label{rho-large}
\end{eqnarray}

We further assume that within a local condensation region, density for
each component of the disc are multiplied. Alternatively, we can think
of protoplanetary discs with different gas or dust density than MMSN.
We denote the ratio of the density of gas, smaller dust, and larger
dust by  $\eta_g, \eta_\sm, \eta_\la$, respectively.  Then the density
of gas, smaller dust and larger dust is given by
\begin{eqnarray}
  \rho_g   &\eq& \eta_g \rho_g^{\MMSN}\left(r\right)      \label{enhancement-gas}, \\
  \rho_\sm &\eq& \eta_\sm \rho_\sm^{\MMSN}\left(r\right) \label{enhancement-smaller dust}, \\
  \rho_\la &\eq& \eta_\la \rho_\la^{\MMSN}\left(r\right) \label{enhancement-larger dust}.
\end{eqnarray}
Mass of the smaller dust and the larger dust are $m_\sm$ and  $m_\la$,
respectively. The number density is density divided by dust mass:
\begin{eqnarray}
  n_\sm &\eq& \frac{\rho_\sm}{m_\sm}, \label{number-density-small} \\
  n_\la &\eq& \frac{\rho_\la}{m_\la}. \label{number-density-large}
\end{eqnarray}
We estimate the mass as a function of the dust radius and the fractal dimension
in \S \ref{section-fluffy-dust-model}.



\subsection{Charge exchange equations}\label{section-circuit-model}

\begin{figure}
  \begin{center}
    \includegraphics[scale=0.5]{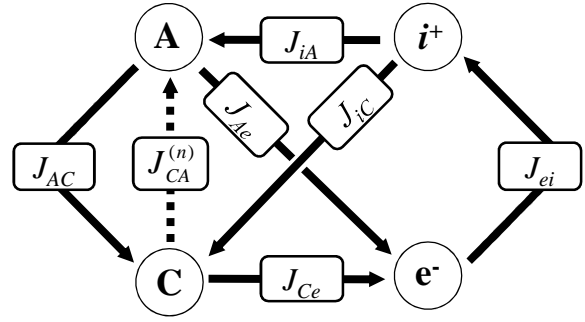}
  \end{center}
  \caption{ The circuit diagram of the charge exchange process in 
    dust plasma. Each arrow represents `current' density $J$, which has the unit
    $ \unit{esu \, cm^{-3} s^{-1}}$, 
    the amount of charge passed from one component to the other per unit disc
    volume per unit time. The arrow points from the component that receives
    negative charge to the component that receives positive charge.
    In this figure, $i$ and $e$ are ions and electrons created from ionising
    neutral disc gas. $C$ and $A$ are cationic and anionic dust defined in
    \S \ref{cationic-anionic-definition}.  }  \label{figure-circuit}
  \vspace{-6pt}
  {\small    
    \ \ $J_{ei}$ represents gas ionisation as `current' from $e$ vertex to $i$
    vertex; $J_{iA}$, $J_{iC}$, $J_{Ae}$, and $J_{Ce}$ are ion and electron
    absorption to dust; $J_{AC}$ is dust-dust collisional charge separation
    and $J^{(n)}_{CA}$ is neutralization current of charged dust-dust 
    absorption.}
\end{figure}

There are four species of charge carrier in our model --- ions,
electrons, cationic, and anionic dust (Table \ref{table-terminology}).
Charge exchange processes between these species are ionisation, plasma
absorption, and dust-dust collision.  The ionisation of the neutral
gas molecules generates the ions and the electrons. Plasma absorption
decreases the number of plasma particles and passes the lost charge to
the dust aggregates. The dust aggregates also get charged by dust-dust
collision.

We label the particle species with letter $\DP$.  The charge density carried
by species $\DP$ is $Q_\DPSub$ (the unit is $\mathrm{esu \ cm^{-3}}$), and the
charge transferred from species $\DP$ to species $\DP'$ is
$J_{\DPSub,\DPSub'}$ (the unit is $\mathrm{esu \ cm^{-3} \ s^{-1}}$).

The charge density $Q_\DPSub$ of a species $\DP$ is the product of
their number density $n_\DPSub$ and their average charge per particle
$q_\DPSub$.  For dust species, we assume that $n_\DPSub$ is known from
number density model while $q_\DPSub$ is unknown; for ion and electrons
we know $q_\DPSub$ but do not know $n_\DPSub$. This constitutes the
four dynamical equations for four unknown variables ${q_\A, q_\C, n_i,
  n_e}$ :
\begin{eqnarray}
  \frac{d q_\A}{dt_{\ }} &\eq& \frac{1_{\ }}{n_\A}\left(-J_{A,C}+J_{i,A}-J_{A,e}+J^{\left(n\right)}_{C,A}\right) \label{basic-eq-dynamic-begin}, \\
  \frac{d q_\C}{dt_{\ }} &\eq& \frac{1_{\ }}{n_\C}\left(J_{A,C}+J_{i,C}-J_{C,e}-J^{\left(n\right)}_{C,A}\right) \\
  \frac{d n_i }{dt_{\ }} &\eq& \frac{1}{q_i}\left(J_{e,i}-J_{i,A}-J_{i,C}\right) \\
  \frac{d n_e }{dt_{\ }} &\eq& \frac{1}{q_e}\left(-J_{e,i}+J_{A,e}+J_{C,e}\right).
  \label{basic-eq-dynamic-end} 
\end{eqnarray}

The current terms $J_{\DPSub,\DPSub'}$ 
are 
\begin{eqnarray}
J_{\A,\C} &\eq& \Delta q_{\A,\C} n_\A n_\C \sigma_{\A,\C} \Delta v_{\A,\C} \label{J-A-C} \\
J_{\C,\A}^{\left(n\right)} &\eq& \abs{q_{\sm}} n_\A n_\C \sigma_{\A,\C}^{\left(n\right)} \Delta v_{\A,\C} \label{J-A-C-n} \\
J_{i,\A}  &\eq& e n_i n_\A \sigma_\ori\left(q_\A,e\right) v_i                \label{J-i-A} \\
J_{i,\C}  &\eq& e n_i n_\C \sigma_\ori\left(q_\C,e\right) v_i                \label{J-i-C} \\
J_{\A,e}  &\eq& e n_e n_\A \sigma_\ori\left(q_\A,-e\right) v_e               \label{J-A-e} \\
J_{\C,e}  &\eq& e n_e n_\C \sigma_\ori\left(q_\C,-e\right) v_e               \label{J-C-e} \\
J_{e,i}   &\eq& \zeta n_g                                        \label{J-e-i}
\end{eqnarray}
where we have included neutral gas ionisation $J_{e,i}$, dust-plasma
absorption $J_{A,i}$, $J_{C,i}$, $J_{A,e}$, $J_{C,e}$, dust-dust
collisional charge-up $J_{A,C} $, and dust-dust collisional
neutralization $J_{C,A}^{(n)}$ terms.  Here, $v_i$ and $v_e$ are the
thermal velocity of the ions and the electrons, $n_g$ is the number
density of the neutral gas, $\zeta$ is the ionisation rate, which is
dominated by cosmic ray ionisation near equatorial, $r=2.7\unit{AU}$ of
MMSN  \citep{2009ApJ...690...69U}.  The exact value for these terms
are given in \S \ref{section-scenario}.  We have neglected, for
example, the gas-phase recombination.

We want to solve the equilibrium equations for the dynamic equations
(\ref{basic-eq-dynamic-begin}-\ref{basic-eq-dynamic-end}): 
\begin{eqnarray}
-J_{A,C}+J_{i,A}-J_{A,e}+J^{\left(n\right)}_{C,A}&\eq&0   \label{basic-eq-begin} \\
  J_{A,C}+J_{i,C}-J_{C,e}-J^{\left(n\right)}_{C,A}&\eq&0 \\
  J_{e,i}-J_{i,A}-J_{i,C}&\eq&0 \\
  -J_{e,i}+J_{A,e}+J_{C,e}&\eq&0,   \label{basic-eq-fourth} 
\end{eqnarray}
together with charge neutrality equation:
\begin{eqnarray}
  Q_A+Q_C+Q_i+Q_e&\eq&0 \label{charge-neutrality}.
  \label{basic-eq-end} 
\end{eqnarray}

We use circuit diagram (Fig. \ref{figure-circuit}) to depict the
dynamical equations
(\ref{basic-eq-dynamic-begin}-\ref{basic-eq-dynamic-end}), and to
interpret the numerical equilibrium solutions
(\ref{basic-eq-begin}-\ref{basic-eq-end}) in \S \ref{section-results}.
The circuit diagram represents charge-exchange processes; each vertex
represents the species of charge reservoir and each arrow represents
the charge exchange process.  The size of the vertex circles
represents the amount of charge $Q_\DPSub$. The thickness of the
arrows represents the amount of charge transfer $J_{\DPSub, \DPSub'}$.
We define the direction of the arrows so that the arrows point to the
positive charge receivers.

In the system of equations depicted by a circuit diagram, charge density of
each vertex $Q_\DPSub$ corresponds to an unknown quantities.  Therefore, the
number of unknown quantities is equal to the number of vertexes $N_V$. 
On the other hand, at the equilibrium, sum of the current flowing into each
vertex is required to be zero (Kirchhoff's Laws); this gives us $N_V$
equations but only $N_V-1$ of them are independent. Charge neutrality gives us
$1$ equation.  Thus we have $N_V$ equations for $N_V$ unknown values.

\section{Charge equilibrium of gas and dust} \label{section-scenario}

In this section we specify the current terms of the dynamic equations 
(\ref{J-A-C}-\ref{J-e-i}),
especially the
dust-dust collisional charging terms $J_{A,C} - J^{(n)}_{C,A}$,
by modelling the dust number density, structure, collisional cross section,
surface charge exchange, and relative velocity.

\subsection{Fluffy dust model} \label{section-fluffy-dust-model}

We use model of dust aggregates by \citet{2008LPI....39.1545W}.
We consider dust aggregates composed of a large number
of spherical monomers with radius $r_m = 0.1 \unit{\mu m}$. 
Each dust species $\DP$ has its mass $m_\DPSub$, the number of
monomers that constitute the dust $N_\DPSub$, and representative
radius $r_\DPSub$. We define the fractal dimension of the fluffy dust $D_\DPSub$ in the
following simple manner:
\begin{eqnarray}
  N_\DPSub = \left(\frac{r_\DPSub}{r_m}\right)^{D_\DPSub} \label{fractal-dimension-definition}.
\end{eqnarray}
The dust mass is expressed in terms of monomer mass $m_m$ as follows:
\begin{eqnarray}
  m_\DPSub = m_m N_\DPSub = m_m\left(\frac{r_\DPSub}{r_m}\right)^{D_\DPSub} \label{dust-mass-definition}.
\end{eqnarray}

\citet{2008LPI....39.1545W} studies the collision of the fluffy dust
of the radii $1.0 \times {10}^{-5} \sim 9.1 \times {10}^{-4}\unit{cm}$.
The effect of offset collisions, collision between dust of much
different sizes, and dust much larger than
$9.1 \times {10}^{-4}\unit{cm}$ are yet to be confirmed. Therefore
we make the following assumptions on smaller dust-larger dust
collision.
\begin{itemize}
  \item If the smaller dust graze at the larger dust, i.e. if the line that passes the
    gravitational centre of the smaller dust and is parallel to the relative velocity
    vector do not intersect with the larger dust, the two dust aggregates 
    do not stick to
    each other. Therefore the grazing cross section is of the order of $r_\sm r_\la$.
    In this case they separate $\Delta q_{\A,\C}$ 
    of charge, which is the product of charge surface density  $\sigma_{ch}$ 
    and contact surface area $S_\contact$. 
    This contributes to the dust-dust charging current, $J_{C,A}$.
  \item If the smaller dust bump into the larger dust, i.e. if the line that passes the
    gravitational centre of the smaller dust and is parallel to the relative velocity
    vector do intersect with the larger dust, the smaller dust do not penetrate the larger dust
    but becomes a part of the larger dust. The cross section is of the order
    ${r_\la}^{2}$. In this case all the charges the smaller dust have are removed from
    the smaller dust charge density and added up to the larger dust charge density.
    This contributes to the dust-dust neutralization current, $J^{(n)}_{C,A}$.
\end{itemize}

\subsection{Collisional cross section of charged spherical object}
\label{charged-cross-section-homoginius}

In this section, we estimate collisional cross sections for 
dust.
The collisional cross sections for 
two electrically charged spherical particle
is given by
\begin{eqnarray}
  \sigma_\cou\left(q\right)=\pi a^2 \exp\left(-\frac{qq'}{ak_BT}\right)  & & \left(qq'>0\right), \ \ 
  \label{sigma_coulomb_focusing+} \\
  \sigma_\cou\left(q\right)=\pi a^2 \left(1-\frac{qq'}{ak_BT}\right) & & \left(qq'<0\right)  \ \ 
  \label{sigma_coulomb_focusing-}  
\end{eqnarray}
where $q$,$q'$ is each particle's charge, $T$ is the temperature
of their relative motion and 
$\pi a^2$ is the geometric cross section
\citep[e.g.][]{1941ApJ....93..369S}.

Equation (\ref{sigma_coulomb_focusing-}) represents the effect of Coulomb
focusing: particles of the opposite charge attract each other and collide more
often than when they are neutral.
On the limit $\abs{qq' a^{-1}} \ggt k_BT$ we can approximate the cross
section as $\sigma_\cou(q) \simeq - \pi a qq'(k_BT)^{-1}$, which is bi-linear
on $q$ and $q'$.
On the other hand, cross section (\ref{sigma_coulomb_focusing+}) represents 
the effect of Coulomb repulsion:
for the collision between particles of the same charge
only a portion of particles that belongs to the long tail of Boltzmann's distribution
for temperature $T$  can overcome the Coulomb barrier and collide.
On the limit $qq' a^{-1} \ggt k_BT$ the cross section vanishes quickly, but
never reaches $0$.

We use Coulomb cross sections (\ref{sigma_coulomb_focusing+}),
(\ref{sigma_coulomb_focusing-}) to estimate the event rate of gas-dust
collision and dust-dust collision.

\subsection{Collisional cross section and contact surface of fluffy
dust} \label{section-contact-surface-density}

\begin{figure}
  \begin{center}
    \includegraphics[scale=0.25]{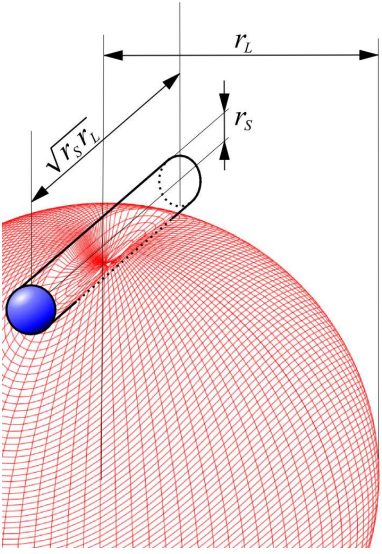}
  \end{center}
  \caption{A grazing collision between a smaller dust (the blue
    solid sphere) and a larger dust (the red wire-frame sphere). The smaller dust
    creates a trench on the larger dust (the black cylinder).} \label{figure-dust-collision}
\end{figure}

The amount of charge exchanged in a collision, $\Delta q_{\A,\C}$,
is product of area of contact $S_\contact$, upper limit of charge exchanged
per unit surface area of contact $\sigma_\charge$,
and the non-dimensional efficiency factor $\eta_\charge$.

We leave the detailed argument to determine $\eta_\charge \sigma_\charge$ to
\S \ref{section-charge-separation}. Here we assume that $\eta_\charge
\sigma_\charge$ is known and describe how to estimate contact surface area
$S_\contact$.  Since it requires another detailed simulation to estimate
$S_\contact$ qualitatively, we resort to an order-of-magnitude estimate for
this part of the work.

We illustrate the collision between a smaller dust and a larger dust in
Fig. \ref{figure-dust-collision}. The smaller dust grazes the larger dust, pushes away the
monomers that belong to the larger dust and creates a trench on the larger dust.  The
trench is a portion of the black cylinder in the figure.  The radius and the
length of the cylinder is $r_\sm$ and $(r_\sm r_\la)^{1/2}$, respectively.
Therefore, the surface area of the trench $S_C$ is of order
\begin{eqnarray}
S_C \simeq {r_\sm}^{3/2} {r_\la}^{1/2} \label{contact-surface-cut},
\end{eqnarray}
and the number of monomers $N_C$ required to fill the surface of the trench is
\begin{eqnarray}
N_C \simeq {r_m}^{-2} {r_\sm}^{3/2} {r_\la}^{1/2} \label{contact-number-compact}.
\end{eqnarray}
Their total surface area is also of the order of $S_C$.

However, $S_\contact \simeq {r_\sm}^{3/2} {r_\la}^{1/2}$  overestimates 
the actual contact surface area if
the large dust is so fluffy that there is not enough
monomers in the trenched volume to fill the trench surface.

From the definition of the fractal dimension (\ref{fractal-dimension-definition}),
the number density of monomers within the larger dust material is 
\begin{eqnarray}
  n_\la^{\left(M\right)} = N {r_\la}^{-3} = {r_m}^{-D_\la} {r_\la}^{D_\la - 3} . 
\end{eqnarray}
On the other hand the volume of the trench is
\begin{eqnarray}
  V_F \simeq {r_\sm}^{5/2} {r_\la}^{1/2} .
\end{eqnarray}
Therfore, the number of particle contained in the trench is
\begin{eqnarray}
  N_F = n_\la^{\left(M\right)} V_F \simeq {r_m}^{-D_\la} {r_\sm}^{5/2} {r_\la}^{D_\la - 5/2}  \label{contact-number-fluffy},
\end{eqnarray}
and their total surface area is
\begin{eqnarray}
  S_F \simeq {r_m}^2 N_F \simeq {r_m}^{2-D_\la} {r_\sm}^{5/2} {r_\la}^{D_\la - 5/2} \label{contact-surface-fluffy} .
\end{eqnarray}

If $N_F < N_C$, the surface of the trench is only partially covered by the
monomers, and we estimate $S_\contact \simeq S_F$. 
On the other hand, if $N_F > N_C$, $N_F$ monomers are crushed onto the trenched
surface, and since they overlap, about $N_C$ monomers will take part in the
charge exchange. In this case we estimate $S_\contact \simeq S_C$. 
To summarize, we assume that $S_\contact$ is the smaller of
(\ref{contact-surface-cut}) or (\ref{contact-surface-fluffy}): 
\begin{eqnarray}
  S_\contact = 
\min\left({r_\sm}^{3/2} {r_\la}^{1/2}, 
 {r_m}^{2-D_\la} {r_\sm}^{5/2} {r_\la}^{D_\la - 5/2}\right) \label{cross-section-contact}.
\end{eqnarray}

\subsection{Charge separation processes} \label{section-charge-separation}
There are generally two classes of possible charge separation processes 
in protoplanetary discs.

One is surface charge exchange, where each dust has some kind of spontaneous
charge separation \citep{10.1021/ja077205t}, so at the initial condition each
dust charge is zero as a whole (globally neutral), but there are charge
separation within the dust particles (locally charged).  For example, water
ice crystals tend to gather negative charge at its surface and positive charge
inside. When two dust aggregates with different charge collide and melt
partially, they exchange molten material and the charge included in the molten
material. As a result each dust gets globally charged.

The other charge separation mechanism may be triboelectric processes
\citep[e.g.][]{2000Icar..143...87D}. In this case, at the initial condition
each dust is both globally and locally neutral. When two dust aggregates made
of materials with different electron affinity collide, the surface electrons
move from one material to the other. As a result each dust gets globally
charged.

\subsubsection{Surface charge exchange I --- larger dust is anionic} \label{section-surface-charge-exchange-I}

The mechanism we consider the most plausible for the dust-dust collisional
charge separation is surface charge exchange between ice dust.  For the dust
aggregate of ice mantled silicate, \citet{2004ApJ...614..490C} proposed a
condensation scenario, that at the snow line ice larger dust drifting inward 
dissociate and many smaller dust form.

There are established models on charge separation caused by ice-ice dust
collision in the context of thundercloud meteorology (for review, see
e.g. \citet{JGR106(D17)2039520402}).  We will carefully import them as a
charge separation model in protoplanetary discs.  The essential steps to cause
lightning on earth are (1) spontaneous charge separations on ice crystal
surfaces, (2) existence of different dust species with different spontaneous
charge separation per surface area, (3) collisions between the different dust
that leads to global charging of each dust and (4) relative motion between the
globally charged dust to create electrostatic field.  

For (1), we argue that the charge separation per surface area is
quantitatively the same as the values measured in laboratory
experiments.  For (2), dominating dust species in charge separation
process in protoplanetary discs is uncertain, and we discuss two
possibilities (c.f.  \S \ref{section-surface-charge-exchange-I}, \S
\ref{section-surface-charge-exchange-II} ) in this work. For (3) and
(4), we make simple estimations for the collision rate and relative
velocity in protoplanetary discs.

Ice crystal surface is intrinsically charge-separated.  Ice is negatively
charged near the surface, and the inside is positive.  The typical charge
surface density for stable ice surface is $\sigma_\charge \simeq
3.0\unit{esu \ cm^{-2}}$ or $\sigma_\charge \simeq 6.2 \times {10}^{9}\unit{e \ cm^{-2}}$ and the typical skin depth of the charged layer is
$d_\charge \simeq 2.0 \times {10}^{-4} \unit{cm}$, though charge surface density for
fast-growing ice surfaces are larger and shallower
\citep{JGR106(D17)2039520402}.  This charge separation has a general
explanation as a result of interaction between hydroxide($OH^{-}$) and
hydronium ($H_3O^{+}$) ions and a hydrophobic surface
\citep{10.1021/ja077205t}, and the above value of typical charge surface
density is observed at liquid water-air surfaces as well as at ice crystal-air
surfaces \citep{TakahashiBubble}. Therefore we use the value for ice-vacuum
surfaces as well.

In the thundercloud, there are varieties of ice crystals with different
surface charge densities, depending on the surface history of the ice
crystals.  Newly formed surfaces have larger charge surface density than old
surfaces, because they have higher fractal dimension and deeper amorphous
layers.

We now consider how surface charge exchange works in the model of
\citet{2004ApJ...614..490C}.  Larger dust that migrate towards the snow line has
old surface and has less negative charge surface density, while smaller dust formed
at the snow line have new surface and larger negative charge surface density,
as in meteorological case. Note that before collision each dust is globally
neutral.

At the collision, the surface of the dust aggregates melts and the surface
charge density is exchanged, and averaged. The larger dust, having less surface
charge density than the smaller dust, receives more negative charge than it gives.
Therefore the larger dust becomes anionic, smaller dust becomes cationic.

Laboratory experiments \citep{TakahashiThunder}, in-situ observations
and meteorological estimates \citep{QJRMetSoc1978.104.447,
  QJRMetSoc1980.106.159} suggest that for mm-size ice crystals, at
least $10$ per cent of the total surface charge within contact surface
is exchanged in a single collision; experiments by
\citet{JGR105(D16)10185,JGR106(D17)2039520402} suggests almost $\eta_\charge =1.0$.
As a conservative estimate, we use $\eta_\charge =0.1$ unless mentioned otherwise.

\subsubsection{Surface charge exchange II --- larger dust is cationic} \label{section-surface-charge-exchange-II}
It may be possible that charge separation processes occurring in
protoplanetary discs are different from those occurring in the terrestrial
thunderclouds.  The collision time-scale in the protoplanetary discs is much
longer than that in a thundercloud, so long that sintering may take place
\citep{1999A&A...347..720S}.  As a result, The surface state of old ice
larger dust and young ice smaller dust might resemble each other. If they are identical,
some random charge exchange by collision is still possible, but they do not
exchange charge on average.

However, compared to thundercloud, protoplanetary discs are more dirty and
fine-grained; they contain much dust made of materials other than ice such as
silicates, and the monomer size is $0.1 \unit{\mu m}$ rather than $1 \unit{mm}$.  Since the
monomer size is smaller than typical skin depth of the charge separation
$d_\charge \simeq 2.0 \times {10}^{-4} \unit{cm}$ mentioned above, it is possible that ice
smaller dust and silicate smaller dust with thin ice mantles formed at the snow line is
inefficient in separating charge. There may be silicate aggregates with no
surface charge separation. Meanwhile old larger dust that have travelled from the
far end of the protoplanetary disc have undergone sintering and have developed
thick mantles with full surface charge separation.

In such scenario, the larger dust has more surface charge separation than the
smaller dust.  Therefore, collision between a larger dust and a smaller dust still leads to
charge separation but the larger dust becomes cationic, and the smaller dust is anionic in
this case. We assume that $\eta_\charge = 0.1$ and $\sigma_\charge \simeq
-3.0\unit{esu \ cm^{-2}}$ in this case (The charge exchange
rate has the same magnitude but the opposite sign compared to that of \S
\ref{section-surface-charge-exchange-I}.)

Both scenarios, the larger dust is anionic and the larger dust is cationic are
plausible. They may even take place in the different parts of the same disc
simultaneously. Therefore, we have decided to take both scenarios into
consideration. To that end, we treat the concept of cationic and anionic dust
separately from the size of the dust.

\subsubsection{Triboelectric charge separation} \label{section-triboelectric}

\citet{2000Icar..143...87D} have proposed that collision between large
silicate grains and fine iron metal grains leads to triboelectric charge
separation. For instance, silicate dust of radius $3.0 \times {10}^{-2}\unit{cm}$ will gain 
$5.4 \times {10}^{3}\unit{e}$ charges per dust.  The process can be built into our
model in the same manner as we treat surface charge exchange processes.

\begin{table*}
\begin{center}
  \small
  \begin{tabular}{cccccc}
    $r_\la$& $D_\la$ & $m_\la$ & St &$u_\la^{mig}$ & $u_\la^{turb}$ \\
    \hline
    $1.0 \times {10}^{2}\unit{cm}$ & $3.0$ & $3.9 \times {10}^{6}\unit{g}$ &
    $1.6 \times {10}^{1}$&
    $7.8 \times {10}^{2}\unit{cm\ s^{-1}}$ & 
    $1.1 \times {10}^{5}\unit{cm\ s^{-1}}$
    \\
    $1.0\unit{cm}$ & $3.0$ & $3.9\unit{g}$ &
    $1.6 \times {10}^{-3}$&
    $2.1 \times {10}^{1}\unit{cm\ s^{-1}}$ & 
    $5.9 \times {10}^{3}\unit{cm\ s^{-1}}$ 
    \\
    $1.0 \times {10}^{2}\unit{cm}$ & $2.4$ & $1.5 \times {10}^{2}\unit{g}$ &
    $6.2 \times {10}^{-4}$&
    $7.9\unit{cm\ s^{-1}}$ & 
    $3.6 \times {10}^{3}\unit{cm\ s^{-1}}$ 
    \\
  \end{tabular}
  \caption{ \small The estimated Stokes number, the bulk velocity due to the
    inward migration $u_\la^{mig}$ \citep{2008A&A...480..859B}, and the
    turbulent speed $u_\la^{turb}$ \citep{2007A&A...466..413O} for some
    typical large dust parameters, for MMSN equatorial at $r=2.7\unit{AU}$.
  } \label{table-uL-estimates}
\end{center}
\end{table*}

\subsection{Relative velocity}\label{section-relative-velocity}

When a cloud of positively and negatively charged dust is separated much
larger than plasma Debye length 
\begin{eqnarray}
\lambda_D &\eq& \sqrt{\frac{T}{4 \pi n_i e^2}} \nonumber \\
&\eq& 4.0 \times {10}^{2}\unit{cm}
\left(\frac{T}{170K}\right) ^ {\frac 1 2}\left(\frac{n_i}{5.0 \times {10}^{-2} cm^{-3}}\right) ^ {- \frac 1 2}
\end{eqnarray}
the electrostatic field between them become observable.  In order to
cause such macroscopic charge separation, there must be a significant
relative bulk motion between anionic and cationic dust. Inward
migration of large dust is a source of this bulk motion. The
sedimentation may act in the same way.  Also
\citet{2000Icar..143...87D} have proposed that largest eddies in
turbulence of protoplanetary discs cause bulk motion between smaller
dust and larger dust.  Such effects on the relative velocity between
dust species in MMSN has been studied (see
\citet[][]{2008A&A...480..859B} and references therein).

Here, we simply assume that the largest contribution to the smaller
dust-larger dust relative velocity is the bulk motion of the larger
dust, and the velocity is $\Delta v_{\la,\sm} \equiv u_\la \equiv
3.4 \times {10}^{3}\unit{cm \ s^{-1}}$, the catastrophic collision velocity of the
ice dust aggregates of $9.1 \times {10}^{-4}\unit{cm}$ size dust
\citep{2008LPI....39.1545W}. Note that the non-sticking velocity
threshold decrease as the monomer size increase
\citep{2000Icar..143..138B}. We also check our analytic formulae with
smaller values of $\Delta v_{\la,\sm}$ and $u_\la$ assumed.

Dust migration speed are comparable to this value at some stages of
the dust growth. On the other hand, turbulent motion is faster than
the value for most of our parameter range (c.f. Table
\ref{table-uL-estimates}). Turbulent mode that is larger than the
scale of interest can be treated as bulk motion, and can be used to
explain the charge separations of the scale. The scale can be as large
as of order of disc scaleheight \citep{1991ApJ...376..214B}.

\subsection{The charge equilibrium equations}\label{section-model}

By substituting the results of analyses up to here into 
(\ref{basic-eq-dynamic-begin}-\ref{basic-eq-dynamic-end})
we have the following dynamic 
equation for charge transport:

\begin{eqnarray}
  \frac{dQ_\la}{dt} &=&  \ \ \ \ - J_{\la, \sm} - J_{\la, i} - J_{\la, e} \label{specific-eq-dynamic-begin} \\
  \frac{dQ_\sm}{dt} &=& J_{\la, \sm}\ \ \ \ \ - J_{\sm, i} - J_{\sm, e} \\
  \frac{dQ_i}{dt} &=& J_{\la,i} + J_{\sm,i}\ \ \ \ \ \ - J_{i, e} \\
  \frac{dQ_e}{dt} &=& J_{\la,e} + J_{\sm,e} + J_{i,e}\ \ \ \ \label{specific-eq-dynamic-end},
\end{eqnarray}
where the current density terms (\ref{J-A-C}-\ref{J-e-i}) become:
\begin{eqnarray}
  J_{\la,\sm} &=& \left(\frac{2 r_\sm}{r_\la} \Delta q_{\A,\C} -
  \frac{Q_\sm}{n_\sm} \right) n_\sm n_\la \Delta v_{\la, \sm}  \nonumber \\ 
  & & \hspace{-10pt}  \sigma_\cou\left(\frac{Q_\la}{n_\la},
  \frac{Q_\sm}{n_\sm}, r_\la,  \frac{1}{2}m_\sm{\Delta v_{\la, \sm}}^{2}\right)  \label{specific-J-L-S} \\
  J_{\la, i} &=& -Q_i n_\la \sigma_\cou\left(\frac{Q_\la}{n_\la}, e, r_\la, k_B T\right) v_i \\
  J_{\la, e} &=& -Q_e n_\la \sigma_\cou\left(\frac{Q_\la}{n_\la},-e, r_\la, k_B T\right) v_e \\
  J_{\sm, i} &=& -Q_i n_\sm \sigma_\cou\left(\frac{Q_\sm}{n_\sm}, e, r_\sm, k_B T\right) v_i \\
  J_{\sm, e} &=& -Q_e n_\sm \sigma_\cou\left(\frac{Q_\sm}{n_\sm},-e, r_\sm, k_B T\right) v_e \\
  J_{i, e} &=& -e \zeta n_g \label{specific-J-i-e}.
\end{eqnarray}

In (\ref{specific-J-L-S}), the amount of current exchange $\Delta
q_{\A,\C}$ is product of contact surface area $S_\contact$ and surface
charge density $\sigma_\charge$, each described in \S
\ref{section-contact-surface-density} and \S
\ref{section-charge-separation}.  The contact surface area
$S_\contact$ is the function of dust radii and dust fractal
dimensions; see equation (\ref{cross-section-contact}).  The surface
charge density $\sigma_\charge$ depends on the dust material.  The
relative velocity of the larger dust and the smaller dust is $ \Delta
v_{\la, \sm} = 3.4 \times {10}^{3} \unit{cm s^{-1}}$, as we have discussed in \S
\ref{section-relative-velocity}.  The cross section term $\sigma_\cou$
is the Coulomb cross section introduced in \S
\ref{charged-cross-section-homoginius}.  We assume $v_i$ and $v_e$ to
be thermal velocities of ions and electrons.  For ionisation in MMSN
at $r=2.7\unit{AU}$, cosmic ray ionisation is the main contributor and
$\zeta \simeq 10^{-18}$ \citep{2009ApJ...690...69U}.  We introduce the
nondimensional dust number density $\eta$ (dust number density in unit
of MMSN values), so that in equations
(\ref{enhancement-gas}-\ref{enhancement-larger dust}), $\eta_g=1$, and
$\eta_\sm=\eta_\la=\eta$.  From those density term, the number density
terms $n_g, n_\sm, n_\la$ are given as $\rho_g / m_g, \rho_\sm /
m_\sm, \rho_\la / m_\la$.  The masses of dust aggregates $m_\sm,
m_\la$ are function of their radii and fractal dimensions; see
equation (\ref{dust-mass-definition}).

All the variables that appear in the current density terms
(\ref{specific-J-L-S}-\ref{specific-J-i-e}) are controlled by five
parameters; radii of the dust aggregates ($r_\sm, r_\la$), their
fractal dimension ($D_\sm, D_\la$), and the nondimensional dust number
density $\eta$.

The equilibrium equations (\ref{basic-eq-begin}-\ref{basic-eq-end})
become:
\begin{eqnarray}
  -J_{\la,\sm} - J_{\la, i} - J_{\la, e} &\eq&0 \label{specific-kirch-L} \label{specific-equilibrium-begin}\\
  J_{\la, \sm} - J_{\sm, i} - J_{\sm, e} &\eq& 0 \label{specific-kirch-S} \\
  J_{\la,i} + J_{\sm,i} - J_{i, e} &\eq& 0    \label{specific-kirch-i} \\
  J_{\la,e} + J_{\sm,e} + J_{i, e} &\eq& 0    \label{specific-kirch-e} \\
  Q_\la+Q_\sm+Q_i+Q_e &\eq&0 \label{specific-charge-neutrality}   \label{specific-equilibrium-end}.
\end{eqnarray}
Again note that, out of four Kirchhoff's Laws
(\ref{specific-kirch-L}-\ref{specific-kirch-e}) only three of them are
independent, and the charge neutrality condition (\ref{specific-charge-neutrality}) is necessary.

\section{Critical dust number density for lightning}
\label{section-lightning}

In this section we derive the strength of electric field generated by
the relative motion of the large and small dust, 
and set conditions for macroscopic electric discharge events, or lightning.

Lightning occurs when the maximum electric field in the plasma $E_\max$
exceeds the critical value $E_\dis$.  The critical electric field $E_\dis$ is
determined by the condition that an electron accelerated by the field has
kinetic energy large enough to ionise a neutral gas molecule.  Let $l_\mfp$ be
the mean free path for electron.  Then an electron accelerated in electric
field of strength $E$ receive the energy of order $e \, E \, l_\mfp$.  The
ionisation potentials $\Wion$ for $\mathrm{H}$, $\mathrm{H_2}$, and
$\mathrm{He}$ molecules are $13.6\unit{eV}$, $15.4\unit{eV}$, and
$24.6\unit{eV}$ respectively \citep{1984inch.book.....D}. We use $\Wion =
15.4\unit{eV}$ in this work. Therefore the critical value $E_\dis$ of electric
field for the lightning satisfies:
\begin{eqnarray}
  e \,  E_\dis \, l_\mfp = \Wion, \\
  E_\dis = \frac{\Wion}{e \, l_\mfp}  \label{equation-E_dis} .
\end{eqnarray}

Next we derive the value of $E_\max$. When the differential motion between the
oppositely charged dust species continues much longer than the plasma Debye
length, it can be interpreted as current carried by the dust $j_D$ generating
electrostatic field, and the plasma counter-current $j_p$ is induced in the
neutralizing direction . We consider that $j_p$ is carried by electrons, and
neglect current carried by positive ions because it is at most the same order
as that by electrons.  Moreover, even if positive ions are accelerated to
$\Wion$ and ionise other molecules, they increase the electron number density
only linearly, not exponentially.

The dust current $j_D$ is estimated simply, by the product of dust charge
density $Q_\la$ and macroscopic motion $u_\la$, as:
\begin{eqnarray}
  j_D = Q_\la u_\la \label{equation-macrocurrent-jD}.
\end{eqnarray}

On the other hand the particle current $j_p$ is determined by the Ohm's law:
\begin{eqnarray}
  j_p = \nu E_\max \label{equation-macrocurrent-jp},
\end{eqnarray}
where $\nu$ is the electric conductivity,
\begin{eqnarray}
  \nu = \frac{n_e \, l_\mfp \, e^2}{m_e \, v_e} \label{equation-conductivity}.
\end{eqnarray}
$E_\max$ is determined at the equilibrium of these two
currents  $j_D$ and $j_p$:
\begin{eqnarray}
  j_D + j_p = 0  \label{equation-macrocurrent-equilibrium}.
\end{eqnarray}
By substituting 
(\ref{equation-macrocurrent-jD}),
(\ref{equation-macrocurrent-jp}), and
(\ref{equation-conductivity}) into
(\ref{equation-macrocurrent-equilibrium}), we obtain
\begin{eqnarray}
  E_\max = - \frac{m_e \, v_e \, Q_\la \, u_\la }{n_e \, l_\mfp \, e^2} .
  \label{equation-E_max}
\end{eqnarray}

Now that we know both $E_\max$ and $E_\dis$, the condition for
electric discharge is
\begin{eqnarray}
\abs{E_\max} \ge E_\dis \label{equation-if-lightning}.
\end{eqnarray}
By substituting (\ref{equation-E_dis}) and (\ref{equation-E_max})
into (\ref{equation-if-lightning}) , we have the following form of 
the condition for electric discharge:
\begin{eqnarray}
\abs{\frac{Q_\la}{Q_e}} \ge \frac{\Wion}{m_e v_e  \abs{u_\la}} \label{equation-macroscopic-discharge}.
\end{eqnarray}

Within our parameter range of interest, the behaviour of the left hand side of
(\ref{equation-macroscopic-discharge}) as we increase $\eta$ is that it first
keeps values much smaller than the right hand side and then it monotonically
increases (c.f. Figure. \ref{figure-four-phase},
\ref{figure-four-phase-IceSi}).  Thus there is a unique value of $\eta$ at
which the equality for (\ref{equation-macroscopic-discharge}) holds.  We
define this value to be $\eta_\crit$, the critical dust number density at which
lightning takes place.  Note that the condition doesn't depend on the detail
of the electron stopping processes because we can eliminate $l_\mfp$ from the
condition.

\section{Results} \label{section-results}

We have performed two sets of numerical experiments.  In the first set of
experiments, we fixed the set of parameters, $r_\sm$, $r_\la$, $D_\sm$, and
$D_\la$ to some typical values.  We varied the dust number density $\eta$, and
calculated charge density for each species of particles at the equilibrium.

In the second set of numerical experiments, we varied the set of input parameters,
$r_\sm$, $r_\la$, $D_\sm$, and $D_\la$, and for each set of input parameters
we calculated the dust number density required to cause electric discharge
$\eta_\crit$.

For all these simulations we assumed the environment at the equatorial
plane and the snowline of the MMSN model; $r = 2.7\unit{AU}$,
$T^{\MMSN} = 1.7 \times {10}^{2}\unit{K}$, $\rho_g^{\MMSN} =
1.6 \times {10}^{-10}\unit{g\ cm^{-3}}$, $\rho_\sm^{\MMSN} =
1.6 \times {10}^{-12}\unit{g\ cm^{-3}}$, $\rho_\la^{\MMSN} =
1.6 \times {10}^{-13}\unit{g\ cm^{-3}}$.

The results of the first set of experiments are in \S
\ref{section-result-equilibrium}. We found that the dust-plasma system
experience four phases as we increase $\eta$. We interpret this result
in \S \ref{section-result-phases}. The results of the second set of
experiments are in \S \ref{section-result-lightning}.
We derive the analytic formula for $\eta_\crit$ in
 \S \ref{section-result-lightning-analytic}.

\subsection{Equilibrium charge density of particles as a function of dust number density}
\label{section-result-equilibrium}

\begin{figure*}
  \vspace{30pt}
  \begin{center}
    \begin{minipage}[t]{.69\textwidth}
      \includegraphics[scale=0.450,angle=270]{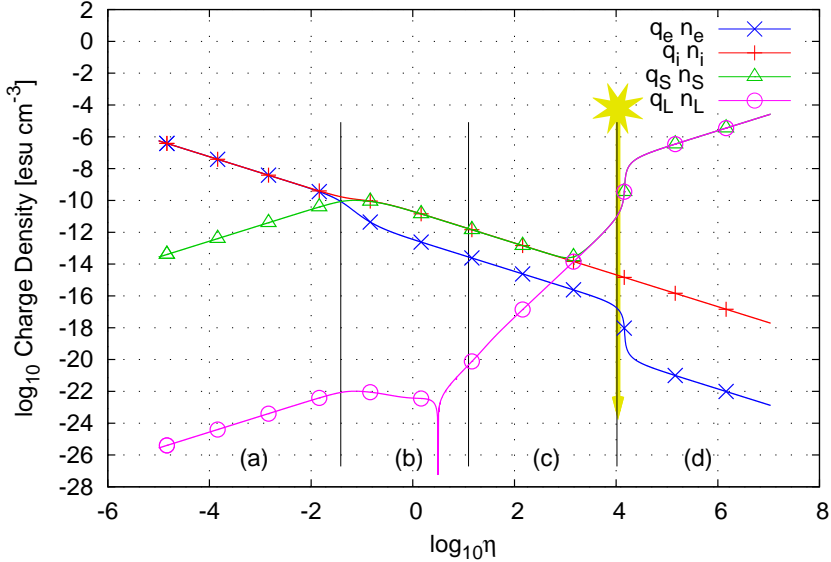}
    \end{minipage}
    \begin{minipage}[t]{.29\textwidth}
      \vspace{20pt} 
      \begin{tabular}{cc}
        symbol & value \\
        \hline
        $r$ & $2.7\unit{AU}$ \\
        $\Sigma_g^{\MMSN} $ & $3.8 \times {10}^{2}\unit{g\ cm^{-3}}$ \\
        $h^{\MMSN}$ & $1.6 \times {10}^{-1}\unit{AU}$ \\
        $T^{\MMSN}$ & $1.7 \times {10}^{2}\unit{K}$ \\
        $\rho_g^{\MMSN}$& $1.6 \times {10}^{-10}\unit{g\ cm^{-3}}$ \\
        $\rho_\sm^{\MMSN}$ & $1.6 \times {10}^{-12}\unit{g\ cm^{-3}}$  \\
        $\rho_\la^{\MMSN}$ & $1.6 \times {10}^{-13}\unit{g\ cm^{-3}}$  \\
        $\rho_m$ & $9.3 \times {10}^{-1}\unit{g\ cm^{-3}}$ \\
        $r_m$ & $1.0 \times {10}^{-5}\unit{cm}$  \\
        $r_\sm$ & $1.0 \times {10}^{-4}\unit{cm}$ \\
        $r_\la$ & $1.0 \times {10}^{2}\unit{cm}$  \\
        $D_\sm$ & $3.0$ \\
        $D_\la$ & $3.0$ \\
        $\zeta$ & $1.0 \times {10}^{-18}$ \\
        $\Delta v_{\la,\sm} $ & $ 3.4 \times {10}^{3}\unit{cm \  s^{-1}}$ \\
        $ u_\la $ & $ 3.4 \times {10}^{3}\unit{cm \  s^{-1}}$ \\
        $\sigma_\charge$ & $6.2 \times {10}^{9}\unit{e\ cm^{-2}}$\\
        $\eta_\charge$ & $1.0 \times {10}^{-1}$ \\
      \end{tabular}
    \end{minipage}
  \end{center}
  \caption{\small Amount of charge stored in each species, $e n_e$, $e
    n_i$, $\abs{q_\sm} n_\sm$, and $\abs{q_\la} n_\la$, as functions
    of $\eta$. This figure is for ice dust-ice dust case, so larger
    dust is anionic and smaller dust is cationic. The polarity matches
    that of Fig. \ref{figure-circuit-phases}.  The radius of smaller
    dust, radius of larger dust, fractal dimension of smaller dust,
    fractal dimension of larger dust are $1.0 \times {10}^{-4}\unit{cm}$,
    $1.0 \times {10}^{2}\unit{cm}$, $3.0$, and $3.0$ respectively.  (a), (b),
    (c), and (d) corresponds to the four phases described in \S
    \ref{section-result-phases}.  The yellow arrow denotes the
    critical number density $\eta$ where the macroscopic electric
    discharge condition (\ref{equation-macroscopic-discharge}) is met.
    The settings of the simulation that produces this figure is in the
    right table.  }
  \label{figure-four-phase}
\end{figure*}

\begin{figure*}
  \begin{center}
    \begin{minipage}[t]{.69\textwidth}
      \includegraphics[scale=0.45,angle=270]{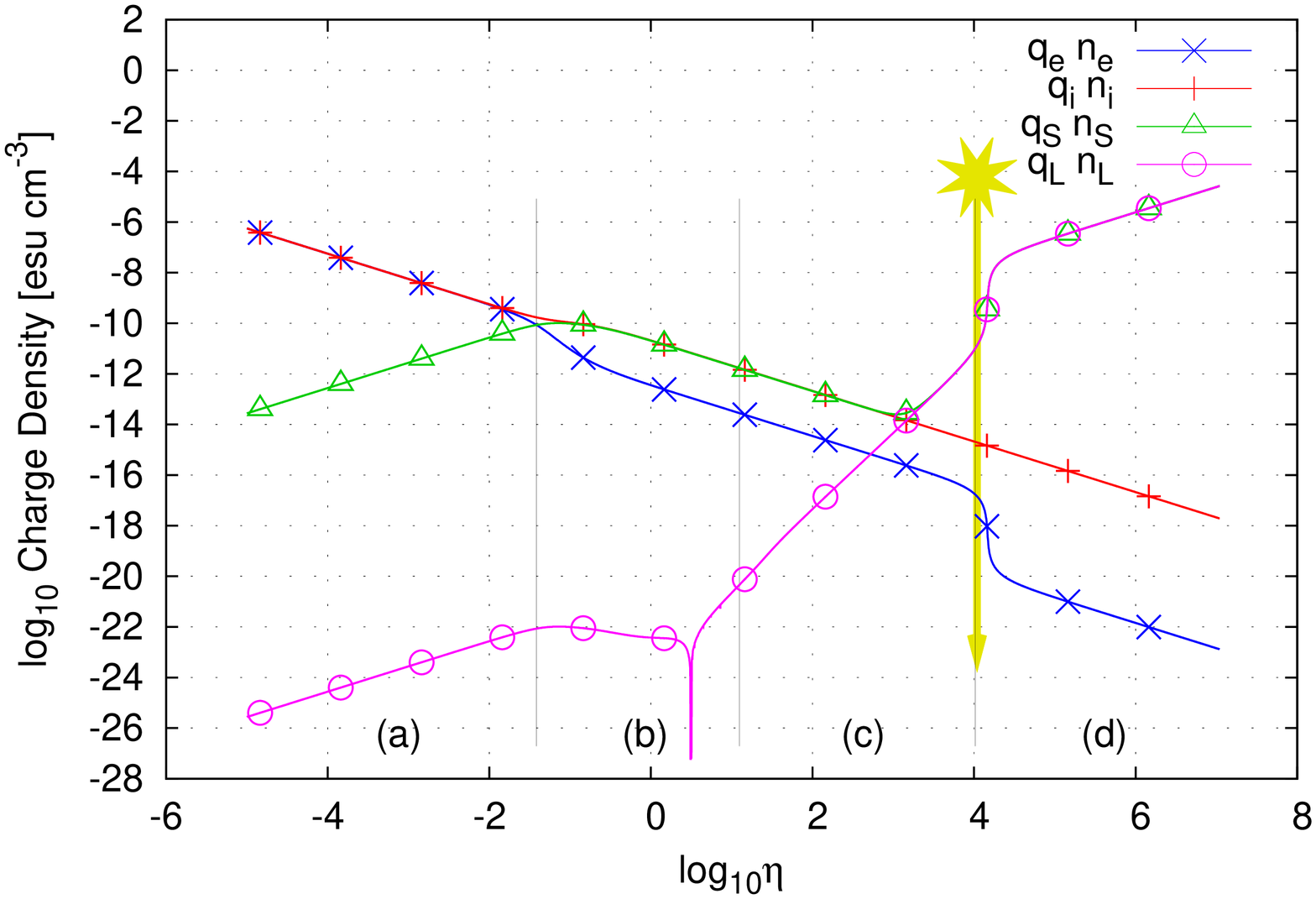}
    \end{minipage}
    \begin{minipage}[t]{.29\textwidth}
      \vspace{20pt} 
      \begin{tabular}{cc}
        symbol & value \\
        \hline
        $r$ & $2.7\unit{AU}$ \\
        $\Sigma_g^{\MMSN} $ & $3.8 \times {10}^{2}\unit{g\ cm^{-3}}$ \\
        $h^{\MMSN}$ & $1.6 \times {10}^{-1}\unit{AU}$ \\
        $T^{\MMSN}$ & $1.7 \times {10}^{2}\unit{K}$ \\
        $\rho_g^{\MMSN}$& $1.6 \times {10}^{-10}\unit{g\ cm^{-3}}$ \\
        $\rho_\sm^{\MMSN}$ & $1.6 \times {10}^{-12}\unit{g\ cm^{-3}}$  \\
        $\rho_\la^{\MMSN}$ & $1.6 \times {10}^{-13}\unit{g\ cm^{-3}}$  \\
        $\rho_m$ & $9.3 \times {10}^{-1}\unit{g\ cm^{-3}}$ \\
        $r_m$ & $1.0 \times {10}^{-5}\unit{cm}$  \\
        $r_\sm$ & $1.0 \times {10}^{-4}\unit{cm}$ \\
        $r_\la$ & $1.0 \times {10}^{2}\unit{cm}$  \\
        $D_\sm$ & $3.0$ \\
        $D_\la$ & $3.0$ \\
        $\zeta$ & $1.0 \times {10}^{-18}$ \\
        $\Delta v_{\la,\sm} $ & $ 3.4 \times {10}^{3}\unit{cm \  s^{-1}}$ \\
        $ u_\la $ & $ 3.4 \times {10}^{3}\unit{cm \  s^{-1}}$ \\
        $\sigma_\charge$ & $-6.2 \times {10}^{9}\unit{e\ cm^{-2}}$ * \\
        $\eta_\charge$ & $1.0 \times {10}^{-1}$ \\
      \end{tabular}
    \end{minipage}
  \end{center}
  \vspace{-10pt}
  \caption{\small Amount of charge stored in each species,
$e n_e$, $e n_i$, $\abs{q_\sm} n_\sm$, and $\abs{q_\la} n_\la$,
as functions of $\eta$. This figure is for ice dust-silicate dust case, so
the larger dust is cationic.
Radii and fractal dimensions of dust, and other parameters are all same
as in Fig. \ref{figure-four-phase}, except that the amount of charge
exchanged in a collision has the opposite sign,
so larger dust is cationic and smaller dust is anionic.
}
  \label{figure-four-phase-IceSi}

  \vspace{30pt}
\end{figure*}

We found that
as we increase $\eta$ while keeping other dust parameters constant, 
the equilibrium charge densities $Q_\DPSub = q_\DPSub n_\DPSub$ 
experience four phases (Table \ref{table-four-phase}).
Fig. \ref{figure-four-phase} and
Fig. \ref{figure-four-phase-IceSi} shows the typical four phases behaviour.

In this and the next sections, we explain the origin of the four phases, using the circuit
diagrams (Fig. \ref{figure-circuit-phases}) as a great help. 
The four-phase behaviour we describe here is 
independent of most of the details of charge exchange processes.
In fact Fig. \ref{figure-four-phase} model and
Fig. \ref{figure-four-phase-IceSi} model have the opposite sign for 
dust-dust collisional charge exchange, but the evolutions are almost similar.
The rest of the discussion
in following sections is based on the former case, which we consider is most
plausible (see \S \ref{section-surface-charge-exchange-I}). 
The discussion is easily generalized to the other case.

To analyse the result,
we first identify the dominant processes by comparing the competitive current
in circuit diagram, then write down all the unknown values in simple
polynomials of $\eta$.
Fig.  \ref{figure-circuit-phases} illustrates the transition of dominant
process in the circuit as dust number density $\eta$ increases.
The two particles with the largest charge density is marked by larger circle.
There are always two of them, one carrying most of the system's positive charge 
and the other negative, thus charge neutrality holds.
The arrows and their line width represents 
direction and amount of currents.
Labels for dominant currents are marked with thick rectangle,
sub-dominant currents with thin rectangle,
negligible currents with dashed rectangle.
The names and conditions for each phase is listed in Table
\ref{table-four-phase}. 

There are two major consequences of the size difference.  Larger dust
is much fewer in number density.  So in the fewer dust limit ($\eta
\llt 1$) the larger dust carries much less charge density than smaller
dust do.  Since larger dust is the fewer, one larger dust collides
with smaller dust much more often than one smaller dust does with
larger dust.  Therefore larger dust are the species that experience
the quick charge density raise in (c)charge-up phase.  The main
role of the smaller dust is to absorb plasma and keep the charge
neutrality.

\begin{table}
    \begin{tabular}{|c|l|c|}
      \hline
    (a) & ion-electron   & $\abs{Q_e} \simeq Q_i$                    \\
        & plasma phase   & (in this paper $2\abs{Q_e} > Q_i$) \\
      \hline
    (b) & ion-dust       & $2\abs{Q_e} < Q_i$ \ \  ,               \\
        & plasma phase   & $ \abs{J_{\A,\C}} < \abs{J_{\A,e}}$              \\
      \hline
    (c) & charge-up      & $\abs{J_{\A,e}} < \abs{J_{\A,\C}} < \abs{J_{i,\C}}$  \\
        & phase          &                                        \\
      \hline
    (d) & dust phase     & $\abs{J_{i,\C}} < \abs{J_{\A,\C}}$ \\
      \hline
    \end{tabular}
    \caption{The names and conditions for four phases of charge separation.
They are basically named after dominant charge
carrier of each phase. }
    \label{table-four-phase}
\end{table}

\subsection{Four phases of charge separation as a function of dust number density}
\label{section-result-phases}

\subsubsection{Ion-electron plasma phase}
In ion-electron plasma phase (Fig. \ref{figure-circuit-phases} (a)), 
the dominant path of charge transfer is 
\begin{eqnarray}
e^{-} \to  i^{+}  \to C \to e^{-} \label{path-IE-1},
\end{eqnarray}
the next-dominant path is 
\begin{eqnarray}
i^{+}  \to A \to e^{-} \label{path-IE-2}.
\end{eqnarray}

Therefore, we have following current hierarchy:
\begin{eqnarray}
\begin{array}{cll}
                   & J_{e,i} \simeq J_{i,\C} \simeq J_{\C,e} & \propto  \eta^{0} \label{J-IE-1} \\ 
\ggt \hspace{-6pt} & J_{i,\A} \simeq J_{\A,e} &   \label{J-IE-2} \\ 
\ggt \hspace{-6pt} & J_{\A,\C} .& 
\end{array}
\end{eqnarray}
The amount of current for path (\ref{path-IE-1}) is constrained by
edge $e^{-} \to  i^{+}$; since we have assumed that $\zeta$ and $n_g$ is
independent of $\eta$, so is $J_{e,i}$.

From charge neutrality (\ref{charge-neutrality}), $Q_e = Q_i$ and therefore $n_e = n_i$.
So equation $J_{i,\C} \simeq J_{\C,e}$ is satisfied
by setting, in equations (\ref{J-i-C}) and (\ref{J-C-e}),
\begin{eqnarray}
  \sigma_\ori\left(q_\C,e\right) v_i = \sigma_\ori\left(q_\C,-e\right) v_e , \\
  \frac{\sigma_\ori\left(q_\C,e\right)}{\sigma_\ori\left(q_\C,-e\right)} \simeq  \frac{v_e}{v_i}
  \propto \eta^0 \label{q-IC-1}.
\end{eqnarray}
Equation (\ref{q-IC-1}) tells us that $\sigma_\cou(q_\C,e) /
\sigma_\cou(q_\C,-e) $ is constant of $\eta$. This means $q_\C \propto \eta^0$
because the only $\eta$-dependent term in $\sigma_\cou$
is $q_\C$. By definition of dust number density factor $\eta$, 
$n_\C \propto \eta^1$,  so $Q_\C \propto \eta^1$.

By similar argument we can deduce $Q_\A \propto \eta^1$ from $J_{i,\A} \simeq J_{\A,e}$.

In other hand, to satisfy $J_{i,\C} \propto \eta^{0}$ and 
$\simeq J_{\C,e} \propto \eta^{0}$ 
we need $n_i , n_e \propto \eta^{-1}$. And since $q_i ,q_e \propto \eta^0$, 
we have $Q_i, Q_e \propto \eta^{-1}$.

In this phase, ions and electrons are the major carriers of 
positive and negative charge.
Equation (\ref{q-IC-1}) also tells us that $\sigma_\cou(q_\C,e) /
\sigma_\cou(q_\C,-e) \simeq v_e/v_i \ggt 1$. This is interpreted as follows:
Since thermal velocity of electron is much faster than that of molecular
ions, electron is more rapidly absorbed to neutral dust than ions.
Therefore dust continues to acquire negative charge, until its negative charge is
enough to repulse most of the electrons inflow to attain a current equilibrium.
Both cationic and anionic dust are forced to charge negative to hold back the
overwhelming electron absorption.

To summarise,
\begin{eqnarray}
Q_i \propto \eta^{-1}, \\
Q_e \propto \eta^{-1}, \\
Q_\A \propto \eta^{+1}, \\
Q_\C \propto \eta^{+1}. 
\end{eqnarray}

\subsubsection{Ion-dust plasma phase}
The system enters ion-dust plasma phase when 
the negative charge in dust $Q_\C$ become comparable to that in plasma $Q_e$. 
Charge neutrality (\ref{charge-neutrality}) requires free electrons to decrease.
So the Coulomb barrier of dust species become weaker until Coulomb cross section
approximates geometric cross section 
$\sigma_\cou(q_\C,e) \simeq \sigma_\cou(q_\C,e)\simeq \pi {a_\C}^2 \propto \eta^{0}$
where electrons and ions are equally absorbed to the dust.

In ion-dust plasma phase (Fig. \ref{figure-circuit-phases} (b)), the dominant path is still
\begin{eqnarray}
e^{-} \to  i^{+}  \to C \to e^{-} \label{path-GD-1},
\end{eqnarray}
and the next-dominant path is still
\begin{eqnarray}
i^{+}  \to A \to e^{-} \label{path-GD-2},
\end{eqnarray}
and the same current hierarchy holds:
\begin{eqnarray}
\begin{array}{cll}
                   & J_{e,i} \simeq J_{i,\C} \simeq J_{\C,e} & \propto \eta^{0} \label{J-GD-1} \\ 
\ggt \hspace{-6pt} & J_{i,\A} \simeq J_{\A,e} &  \label{J-GD-2} \\ 
\ggt \hspace{-6pt} & J_{\A,\C}. & 
\end{array}
\end{eqnarray}

However, now that $\sigma_\cou(q_\C,e) \simeq \sigma_\cou(q_\C,e)$,
equation $J_{i,\C} \simeq J_{\C,e}$ is satisfied
by setting, in equations (\ref{J-i-C}) and (\ref{J-C-e}),
\begin{eqnarray}
  n_i v_i = n_e v_e, \\
  \frac{n_i}{n_e} \simeq  \frac{v_e}{v_i}
  \propto \eta^0 \label{q-IC-2}.
\end{eqnarray}

So the ratio ${n_i}/{n_e}$  is kept constant to ${v_e}/{v_i} = 6.1 \times {10}^{1}$.
Still, in order to have $J_{i,\C} \propto \eta^{0}$ and 
$\simeq J_{\C,e} \propto \eta^{0}$ 
we need $n_i, n_e \propto \eta^{-1}$. Since $q_i ,q_e \propto \eta^0$, 
we have $Q_i, Q_e \propto \eta^{-1}$.

In this phase 
the cationic dust carry most of the negative charge while
ions carry most of the positive charge of the system.
Therefore, the charge neutrality equation (\ref{charge-neutrality}) 
is dominated by these two components, and 
$Q_\C \propto Q_i \propto \eta^{-1}$.

In this phase anionic dust also feels the same environment as cationic dust,
so $Q_\A \propto \eta^{-1}$. 
However as $\eta$ approaches to (c)charge-up
phase, dust-dust collisional charge separation $J_{\A,\C}$ gradually comes 
into play and $Q_\A$ increases.
Therefore in Fig. \ref{figure-four-phase} we can see the power law 
$Q_\A \propto \eta^{-1}$  
only at the beginning of (b)ion-dust plasma phase.

To summarise,
\begin{eqnarray}
Q_i \propto \eta^{-1}, \\
Q_e \propto \eta^{-1}, \\
Q_\A \propto \eta^{-1}, \\
Q_\C \propto \eta^{-1}. 
\end{eqnarray}

In ion-electron plasma phase and ion-dust plasma phase the dust-dust
collisional charging is ineffective. 
So we can understand these two phase
without dust-dust collisional charging
(see \citet{2009ApJ...698.1122O} and references therein.)

\subsubsection{Charge-up phase} \label{section-dust-chargeup-phase}
The system enters (c)charge-up phase when $J_{\A,\C}$ becomes larger
than $J_{\A,e}$. Now anionic dust has their own negative charge supply from
dust-dust collision, their negative charge grow quickly, and
$\sigma_\cou(q_\A,-e) $ become rapidly small. At this point, the circuit
switches one of its current path.


In charge-up phase (Fig. \ref{figure-circuit-phases} (c)), the dominant path is still
\begin{eqnarray}
e^{-} \to  i^{+}  \to C \to e^{-} \label{path-DSD-1},
\end{eqnarray}
but the next-dominant path is 
\begin{eqnarray}
i^{+}  \to A \to C \label{path-DSD-2}.
\end{eqnarray}
The amount of current for path (\ref{path-DSD-1}) is constrained by
edge $e^{-} \to  i^{+}$; since we have assumed that $\zeta$ and $n_g$ is
independent of $\eta$, so is $J_{e,i}$.
The amount of current for path (\ref{path-DSD-2}) is constrained by
edge $A \to C$ (\ref{J-A-C}); since we have assumed that $\Delta q_{\A,\C}$ is
independent of $\eta$,  $J_{\A,\C} \propto \eta^2$.

Therefore, we have following  hierarchy:
\begin{eqnarray}
\begin{array}{cll}
                   & J_{e,i} \simeq J_{i,\C} \simeq J_{\C,e} & \propto  \eta^{0} \label{J-DSD-1} \\ 
\ggt \hspace{-6pt} & J_{i,\A} \simeq J_{\A,\C} & \propto  \eta^{2}  \label{J-DSD-2} \\ 
\ggt \hspace{-6pt} & J_{\A,e} .& 
\end{array}
\end{eqnarray}

The path (\ref{path-DSD-1}) is as same in
ion-dust plasma phase, leading to
$Q_i , Q_e \propto \eta^{-1}$, and charge neutrality requires $Q_\C \propto \eta^{-1}$.

In dust charge-up phase, however, anionic dust has so much charge that
electrostatic potential for electron and ion at the surface of larger dust
is larger than their thermal energy;
this is $qq' a^{-1} \ggt k_BT$ limit of
the Coulomb cross section
(\ref{sigma_coulomb_focusing+}), (\ref{sigma_coulomb_focusing+}).
Thus $\sigma_\cou(q_\A,-e) \to 0$ and
 $\sigma_\cou(q_\A,e) \propto q_\A$ in (\ref{J-i-A}).
Substituting $n_i \propto \eta^{-1}$ and $n_\A \propto \eta^1$ into
$J_{i,\A} \propto  \eta^{2}$, we have
$q_\A \propto \eta^2$ and $Q_\A \propto \eta^3$.

To summarise,
\begin{eqnarray}
Q_i  &\hspace{-6pt} \propto\hspace{-6pt} & \eta^{-1}, \\
Q_e  &\hspace{-6pt} \propto\hspace{-6pt} & \eta^{-1} \label{phaseC-electron-charge-density}, \\
Q_\A &\hspace{-6pt} \propto\hspace{-6pt} & \eta^{+3} \label{phaseC-anionic-charge-density}, \\
Q_\C &\hspace{-6pt} \propto\hspace{-6pt} & \eta^{-1}. 
\end{eqnarray}

At this phase, by substituting equations
(\ref{phaseC-electron-charge-density})
(\ref{phaseC-anionic-charge-density}) into equation (\ref{equation-E_max}) we have
\begin{eqnarray}
E_\max \propto \eta^4 \label{equation-fourth-power}.
\end{eqnarray}
The $E_\max$ has the dependency of $\eta^4$ in this phase, instead of
$E \propto \eta^2$ dependence used, for example, in
\citet{1997Icar..130..517G}.  Moreover, at the end of dust charge-up
phase there is a steep increase in $Q_\la$ and steep decrease in
$Q_e$.  These means that the electric discharge condition
(\ref{equation-macroscopic-discharge}) meets at smaller value of
$\eta$.

\subsubsection{Dust phase} \label{section-dust-phase}
The system enters (d)dust phase when $J_{\A,\C}$ becomes larger
than $J_{i,\C}$. Now the charge states of both anionic and cationic dust is
governed by dust-dust collision, and the plasma component is sub-dominant to the dust.

In dust phase (Fig. \ref{figure-circuit-phases} (d)), the dominant path is 
\begin{eqnarray}
A \to  C \to A \label{path-DDD-1},
\end{eqnarray}
the dust-dust collision is now short-circuiting. The next-dominant path is 
\begin{eqnarray}
C \to e^{-}  \to i^{+}  \to A  \label{path-DDD-2}.
\end{eqnarray}
The amount of current for path (\ref{path-DDD-1}) is constrained by
edge $A \to C$ (\ref{J-A-C}); since we have assumed that $\Delta q_{\A,\C}$ is
independent of $\eta$,  $J_{\A,\C} \propto \eta^2$.

The amount of current for path (\ref{path-DDD-2}) is constrained by
edge $e^{-} \to  i^{+}$; since we have assumed that $\zeta$ and $n_g$ is
independent of $\eta$, so is $J_{e,i}$.

Therefore, we have following  hierarchy:
\begin{eqnarray}
\begin{array}{cll}
                   & J_{\A,\C} \simeq J_{\C,\A}^{\left(n\right)} & \propto  \eta^{2}  \label{J-DDD-1} \\ 
\ggt \hspace{-6pt} & J_{\C,e} \simeq J_{e,i} \simeq J_{i,\A} & \propto \eta^{0} \label{J-DDD-2} \\ 
\ggt \hspace{-6pt} & J_{\A,e}. & 
\end{array}
\end{eqnarray}

Equation $J_{\A,\C} \simeq J_{\C,\A}^{(n)}$ (\ref{J-DDD-1}) requires
$\Delta q_{\A,\C} \sigma_{\A,\C} = q_\C
\sigma_{\A,\C}^{(n)}$. Therefore only $\eta$ dependent term $q_\C$
must satisfy $q_\C \propto \eta^0$, leading to $Q_\C \propto
\eta^1$. Charge neutrality leads to $Q_\A \propto \eta^1$.

The path (\ref{path-DDD-2}) gives us $Q_i , Q_e \propto \eta^{-1}$,
same as in ion-dust plasma phase and in dust charge-up phase.

At the boundary of (c)charge-up phase and (d)dust phase there is a
jump of dust charge. This is because when $\eta$ cross the boundary
dust charge grows until dust-dust collisional neutralization can
compensate dust-dust charge separation.

To summarise,
\begin{eqnarray}
Q_i  &\hspace{-6pt} \propto\hspace{-6pt} & \eta^{-1}, \\
Q_e  &\hspace{-6pt} \propto\hspace{-6pt} & \eta^{-1}, \\
Q_\A &\hspace{-6pt} \propto\hspace{-6pt} & \eta^{+1}, \\
Q_\C &\hspace{-6pt} \propto\hspace{-6pt} & \eta^{+1}. 
\end{eqnarray}


\def\phaseDiagramScale{0.48}

\begin{figure}
  \begin{center}
    \includegraphics[scale=\phaseDiagramScale]{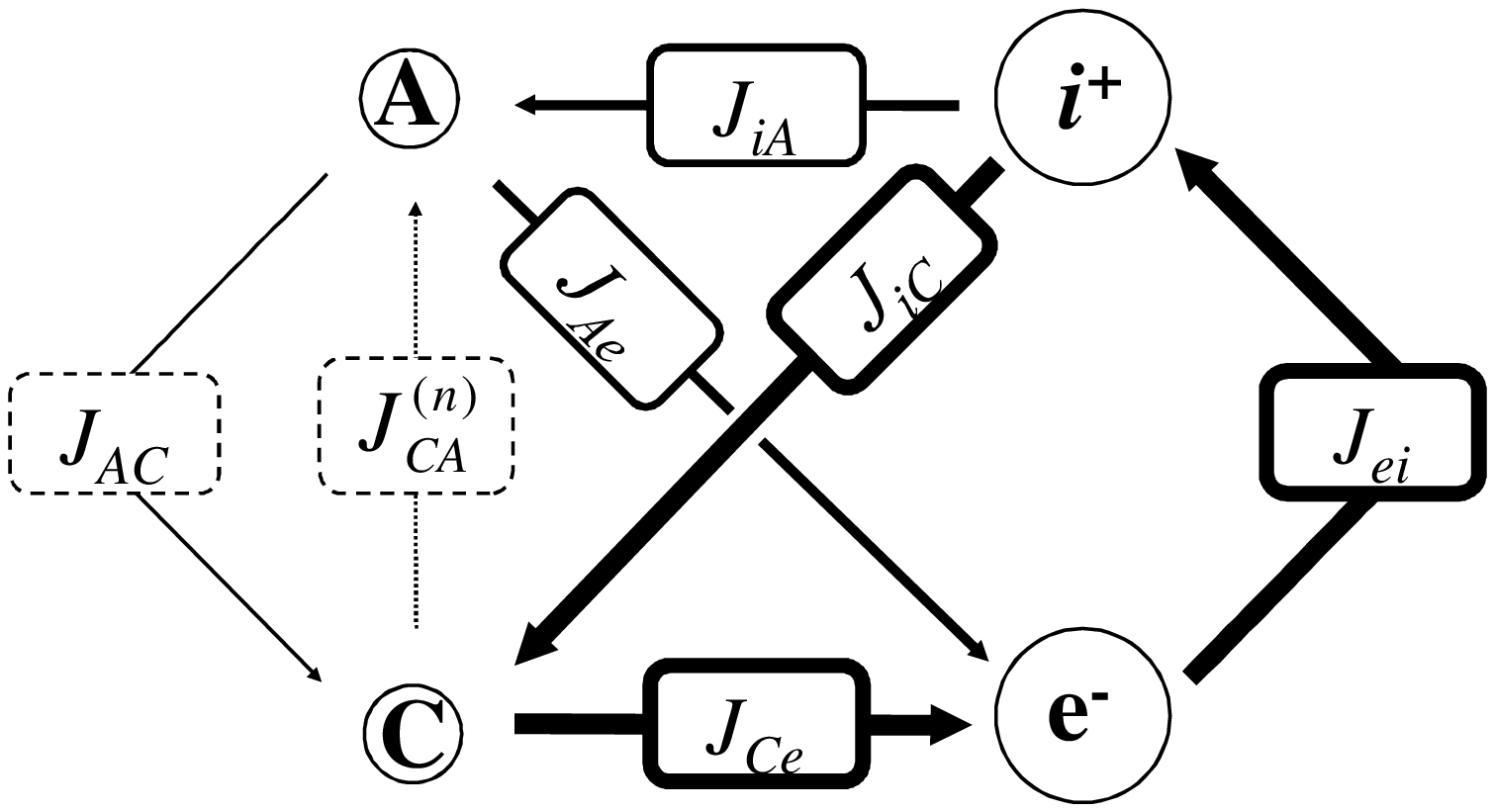}

    (a) ion-electron plasma phase 
  \end{center}
    
  \begin{center}
    \includegraphics[scale=\phaseDiagramScale]{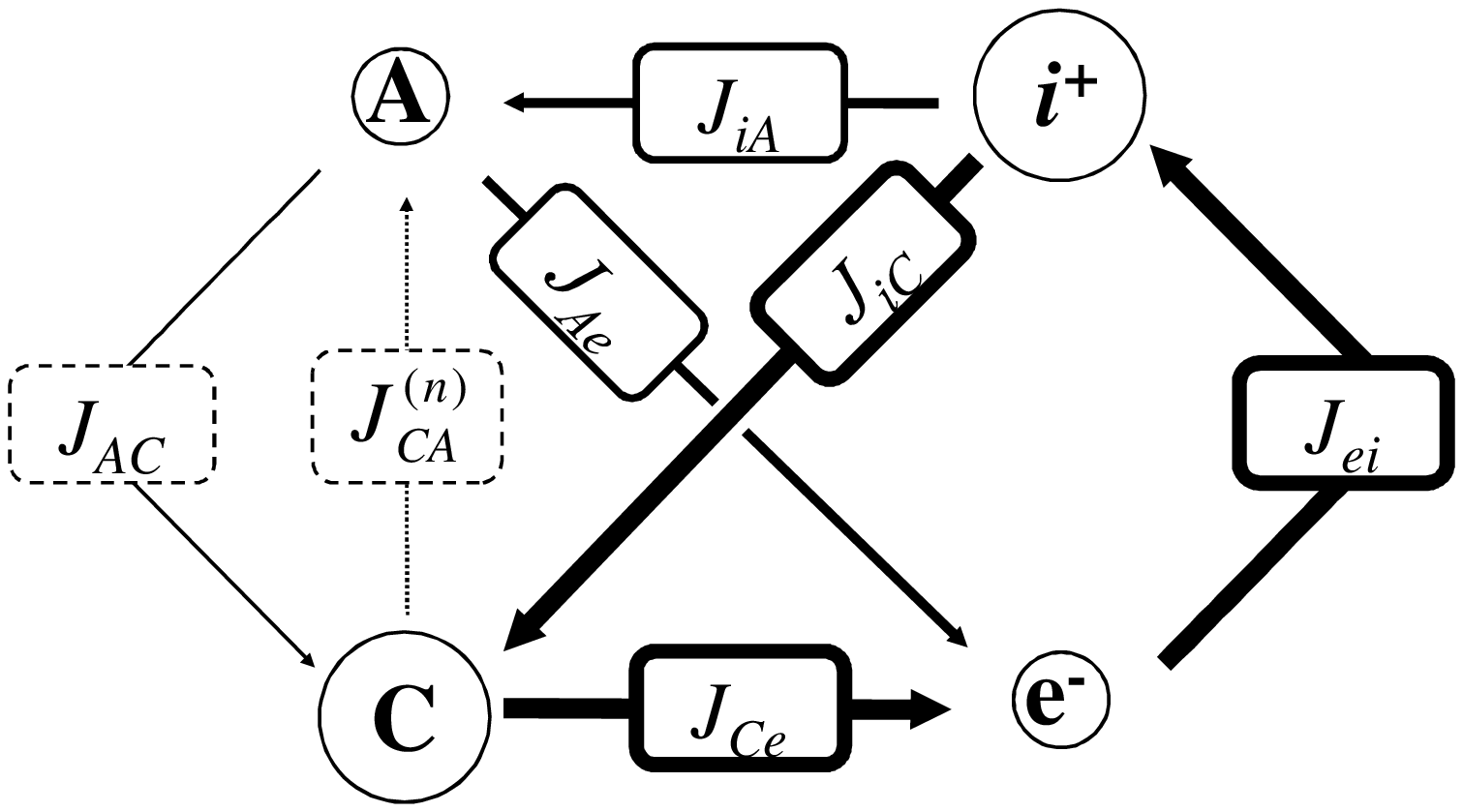}

    (b) ion-dust plasma phase 
  \end{center}

  \begin{center}
    \includegraphics[scale=\phaseDiagramScale]{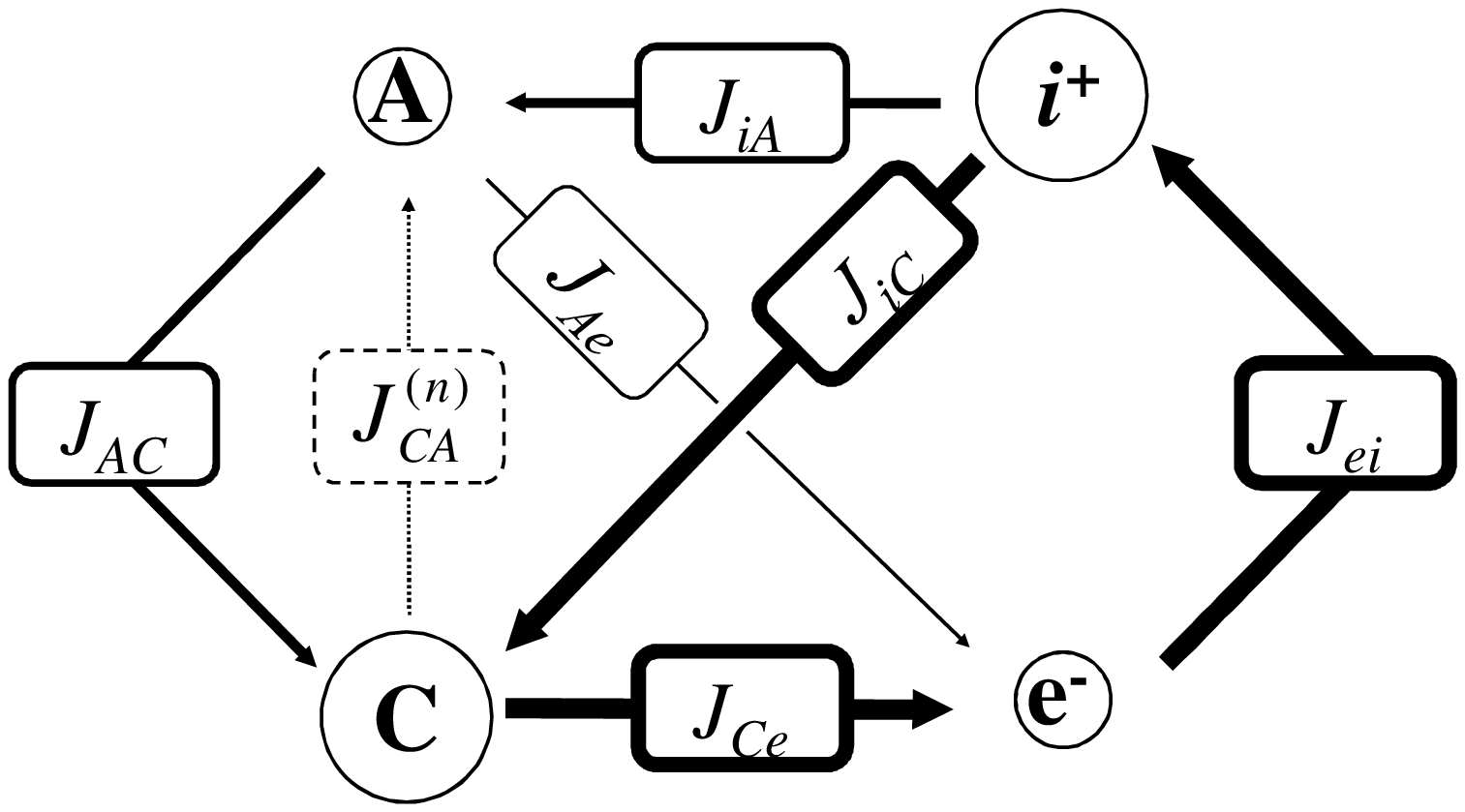}

    (c) charge-up phase
  \end{center}

  \begin{center}
    \includegraphics[scale=\phaseDiagramScale]{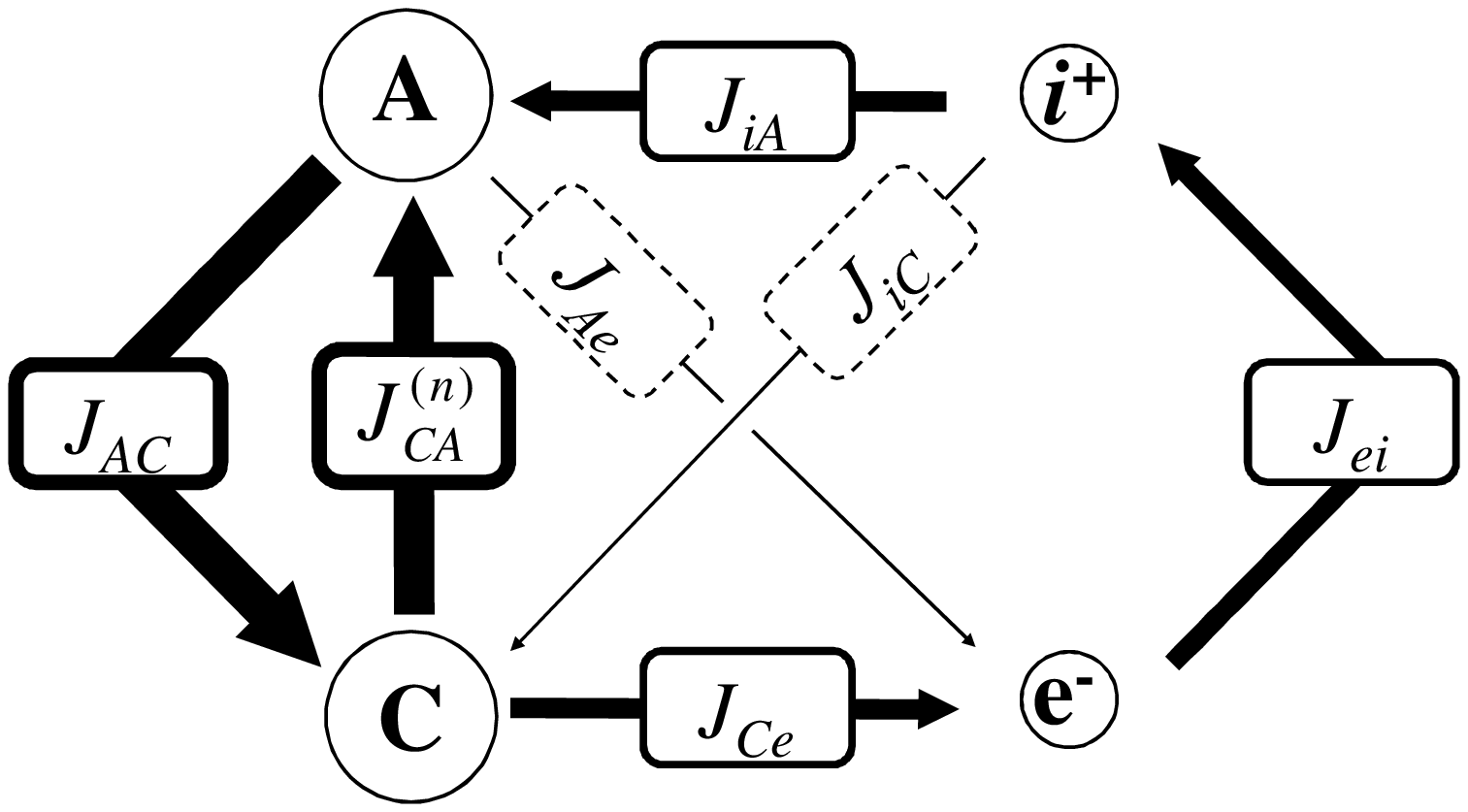}

    (d) dust phase
  \end{center}
  \caption{\small 
    The evolution of the charge density and current density as dust condense.
    As dust number density $\eta$ increase,
    $J_{e,i} \propto \eta^0$ is constant
    while $J_{\A,\C} \propto \eta^2$ grows,
    and the particle experience four phases in order $(a) \to (b) \to (c) \to (d)$.
    (a) At ion-electron plasma phase, most of the charge is carried by plasma species 
    and the charge state of the dust is governed by plasma absorption. 
    (b) At ion-dust plasma phase, the current balances are same as it was 
    in ion-electron plasma phase, but now the negative charge carrier is
    cationic smaller dust.
    (c) At charge-up phase, anionic larger dust has sufficient charge to
    cut off $J_{\A,e}$. 
    (d) At dust phase, most of the charge is carried by dust species
    and the collisional charging current $J_{\A,\C}$ 
    balance with neutralization current $J_{\C,\A}^{(n)}$.
  }
  \label{figure-circuit-phases}
\end{figure}

\subsection{Critical dust number density as function of dust parameters}
\label{section-result-lightning}

We now explain the details of the second numerical experiments, where
we varied the set of input parameters, $r_\sm$, $r_\la$, $D_\sm$, and
$D_\la$, and for each set of input parameters we calculated the
critical dust number density $\eta_\crit$ at which the lightning
strikes.  The numerical results strongly suggest that the parameter
space $(r_\sm, r_\la, D_\sm, D_\la)$ is subdivided into several
regions, at each of which $\eta_\crit$ is a simple analytic function
of parameters $(r_\sm, r_\la, D_\sm, D_\la)$.

The parameter ranges are
\begin{eqnarray}
  1.0 \times {10}^{-4}\unit{cm} <& \hspace{-6pt} r_\sm \hspace{-6pt} &< 1.0 \times {10}^{2}\unit{cm}, 
  \label{constraint-begin} \\
  1.0\unit{cm}  <& \hspace{-6pt} r_\la \hspace{-6pt} &< 1.0 \times {10}^{3}\unit{cm}, \\
  2.0 <& \hspace{-6pt} D_\sm \hspace{-6pt} &< 3.0 ,\\
  2.0 <& \hspace{-6pt} D_\la \hspace{-6pt} &< 3.0 ,
\end{eqnarray}
with additional constraints
\begin{eqnarray}
  r_\sm & \hspace{-6pt} < \hspace{-6pt} & r_\la \label{constraint-radius}, \\
  D_\sm & \hspace{-6pt} < \hspace{-6pt} & D_\la \label{constraint-fractal}, \\
  1.0 \ <& \hspace{-12pt} \eta \hspace{-12pt} & < \ 1.0 \times {10}^{6} 
  \label{constraint-eta}   \label{constraint-end}.
\end{eqnarray}
Constraint (\ref{constraint-radius}) requires that the smaller dust is smaller
than the larger dust.  Constraint (\ref{constraint-fractal}) comes from
empirical fact that larger dust aggregates have experienced more
compactification, and have higher fractal dimension
\citep{2008ApJ...684.1310S,2008ApJ...677.1296W}.  Constraint
(\ref{constraint-eta}) is cutoff value of our computation.

We visualize the four-dimensional field $\eta_\crit (r_\sm, r_\la,
D_\sm, D_\la)$ in the figures at the last of this paper, by choosing
some representative points and presenting several 2-dimensional
sections that passes the point.  As the mass, the radius, and the
fractal dimension of a dust is related by equation
(\ref{dust-mass-definition}), we have some freedom of choosing the
direction of 2-dimensional section.  We keep $m_\DPSub$ constant when
we vary $D_\DPSub$ (the dust puff up with constant mass); we keep
$D_\DPSub$ constant when we vary $r_\DPSub$ (the dust mass increase
with constant fractal dimension).

First, Fig. \ref{figure-result-tanaka} shows the `fluffy dust'
cross sections, where the representative dust are $r_\sm =
1.0 \times {10}^{-2}\unit{cm}$, $r_\la=1.0 \times {10}^{2}\unit{cm}$, $m_\sm =
3.9 \times {10}^{-9}\unit{g}$, $m_\la = 1.5 \times {10}^{2}\unit{g}$, $D_\sm = 2.0$, and
$D_\la = 2.4$.  The critical number density is $\eta_\crit =
7.37 \times {10}^{1}$ for this representative parameter.

The second set of Fig. \ref{figure-result-hayashi} uses the `hard dust'
cross sections, where $r_\sm = 1.0 \times {10}^{-4}\unit{cm}$,
$r_\la=1.0\unit{cm}$, $m_\sm = 1.9 \times {10}^{-12}\unit{g}$, $m_\la =
3.9\unit{g}$, $D_\sm = 2.7$, and $D_\la = 3.0$.  The
critical number density is $\eta_\crit = 3.01 \times {10}^{2}$
for this representative parameter.

The third set of Fig.
\ref{figure-result-averaged}, is the 
$\eta_\crit$ averaged over the parameters that do not appear in the axes,
to show the tendency of overall dependence on the parameters, and to
demonstrate the precision of the analytic formulae.

The fourth set of Fig. \ref{figure-result-hypertanaka} uses the same
representative dust as in Fig. \ref{figure-result-tanaka}, but is the
result of another simulations, where we are now extremely pessimistic
and assume that the charge exchange is four orders of magnitude
inefficient ($\eta_\charge = 1.0 \times {10}^{-5}$ instead of $\eta_\charge =
1.0 \times {10}^{-1}$). Even though, the number density $\eta_\crit$ required for
lightning has raised only by two order of one magnitude.  The critical
number density is $\eta_\crit = 6.59 \times {10}^{3}$ for the
representative parameter.

The fifth set of Fig. \ref{figure-result-averaged-pessimistic} shows
the averaged $\eta_\crit$ for the pessimistic case $\eta_\charge =
1.0 \times {10}^{-5}$.  We later examine the accuracy of our formulae by
fitting Fig. \ref{figure-result-averaged-pessimistic} with the
formulae using correction factors determined by
Fig. \ref{figure-result-averaged} data.

\subsection{Analytic formulae for lightning conditions}
\label{section-result-lightning-analytic}

In this section we derive the analytic form, of $\eta_\crit$ and
lightning conditions.  Numerical results obtained in \S
\ref{section-result-lightning} are of great help in deriving these
analytic formulae.  We show at the end of \S
\ref{section-analytic-combined} that by our analytic formulae we can
fit $364325$ numerically-obtained points distributed among six decades
with 21 per cent precision.  Moreover, the
formulae `predicts' results of another simulation with
59 per cent precision, where charge
exchange is $10^4$ times inefficient.  These agreements are good
evidences for correctness of both numerical and analytical results.

We made plots like Fig. \ref{figure-four-phase} and
Fig. \ref{figure-four-phase-IceSi}
for many points within our parameter space,
and found that $\eta=\eta_\crit$ is met at the boundary of
(c)charge-up phase and (d)dust phase in most cases, and sometimes
in (d)dust phase or (c)charge-up phase.
Therefore we derive analytic form of $\eta$ corresponding to these three cases
in \S \ref{section-analytic-cd} to \S \ref{section-analytic-c},
and combine them in \S \ref{section-analytic-combined}

\subsubsection{Analytic formulae for charge-up phase / dust phase boundary}\label{section-analytic-cd}
We first calculate $\eta^{(cd)}$, the value of $\eta$ corresponding to the (c)charge-up phase / (d)dust phase boundary.

The boundary satisfies $J_{\la,\sm} = J_{i,\sm}$ (Table \ref{table-four-phase}).
We use the approximations in (c)charge-up phase to find the break point
of the phase. Then  
\begin{eqnarray}
  J_{\la,\sm} &\simeq& 2 \pi r_\sm r_\la \Delta q_{\A,\C} n_\sm n_\la 
  \Delta v_{\la, \sm}  \label{analytic2-J-LS} \\
  J_{\sm, i}  &\simeq& -Q_i n_\sm  \pi {r_\sm}^2 \sigma_{\sm, i} v_i  \label{analytic2-J-Si} 
\end{eqnarray}
because we can ignore the neutralization current $J_{\la,\sm}^{(n)}$, 
approximate $\sigma_{\la, \sm} = \pi {r_\la}^2$,
$\sigma_{\sm, i} = \pi {r_\sm}^2$ (geometric
cross sections) and
$\sigma_{\la, i} = \pi {r_\la}^2 \abs{eQ_\la} (r_\la k_B T n_\la)^{-1}$ 
($qq' a^{-1} \ggt k_BT$ limit of Coulomb cross section
(\ref{sigma_coulomb_focusing-})).

In dust charge-up phase, both ions and electrons are mainly absorbed
by smaller dust, so from
(\ref{specific-kirch-i}) and (\ref{specific-kirch-e}) we have
\begin{eqnarray}
  Q_i &\eq& \frac{e \zeta n_g}
  {n_\sm \sigma_{\sm, i} v_i } \label{analytic2-Q-i}  \\
  Q_e &\eq& \frac{-e \zeta n_g}
  {n_\sm \sigma_{\sm, e} v_e } \label{analytic2-Q-e}  
\end{eqnarray}
and the absorption cross sections are geometric : 
$\sigma_{\sm, i} = \sigma_{\sm, e} = \pi {r_\sm}^2$.

By substituting (\ref{analytic2-Q-i})  into (\ref{analytic2-J-Si}) 
\begin{eqnarray}
  J_{\sm, i} = e \zeta n_g = J_{e, i} \label{analytic2-J-Si-2}
\end{eqnarray}
cf. Fig. \ref{figure-four-phase}(c) and in equation (\ref{J-DSD-1}).

We have come to a simple result, that $\eta_\crit$ satisfies
\begin{eqnarray}
  J_{\la, \sm} = J_{e, i} \label{analytical-boundary}.
\end{eqnarray}
Substituting equations (\ref{analytic2-J-LS}) ,
(\ref{analytic2-J-Si-2}), together with $n_\sm = \eta^{(cd)}
n_\sm^{\MMSN}$ and $n_\la = \eta^{(cd)} n_\la^{\MMSN}$ into
equation (\ref{analytical-boundary}) and solving for dust number density
$\eta$, we have
\begin{eqnarray}
  \eta^{\left(cd\right)} = \left(\frac{\correction^{\left(cd\right)}}{2 \pi}
  \frac{e}{\Delta q_{\A,\C}}
  \frac{n_g}
  {n_\sm^{\MMSN} n_\la^{\MMSN} r_\sm r_\la}
  \frac{\zeta}
  {\Delta v_{\la, \sm}}
  \right)^{\frac 1 2} 
 \label{symbolic-eta-critical-cd}
\end{eqnarray}
We have introduced a nondimensional correction factor $\correction^{(cd)}$,
 a constant that does not depend on $r_\sm, r_\la, D_\sm, D_\la$.
We need this to compensate the error arising from using the formulae
in (c)charge-up phase to find the break point of itself.
The actual value for $\correction^{(cd)}$ is in \S \ref{section-analytic-combined}.

\subsubsection{Analytic formula for $\eta_\crit$ in dust phase}\label{section-analytic-d}

Next, we derive the analytic formula of the critical density
$\eta_\crit^{(d)}$, where the condition for electric discharge
(\ref{equation-macroscopic-discharge}) is met in (d)dust phase
(c.f. \S \ref{section-dust-phase}, Figure \ref{figure-circuit-phases}(d)).

Imposing $J_{\la,\sm} = 0$ in (\ref{specific-J-L-S}), and by
approximating the charge neutrality (\ref{specific-charge-neutrality})
with $Q_\sm + Q_\la = 0$, we have
\begin{eqnarray}
Q_\sm = \abs{Q_\la} = \Delta q_{\A,\C} n_\sm \frac{2 r_\sm}{r_\la}
\end{eqnarray}

By approximating equation (\ref{specific-kirch-e}) with $J_{\sm,e} +
J_{i,e} = 0$, we have
\begin{eqnarray}
Q_e &\eq& - \frac{e n_g \zeta}{\pi n_\sm {r_s}^2 \left(1+\chi\right) v_e} \\
where \ \ 1+\chi &\eq& 1 + \frac{Q_\sm e}{n_\sm r_\sm k_B T}; \label{definition-of-chi}
\end{eqnarray}
the factor $(1+\chi)$ comes from the Coulomb cross section (\ref{sigma_coulomb_focusing-}).

Substituting $Q_\la$ and $ Q_e$, the equality for the lightning
condition (\ref{equation-macroscopic-discharge}) becomes
\begin{eqnarray}
\frac{2 \pi \Delta q_{\A,\C}} {e} 
\frac{{n_s}^2 {r_\sm}^3 \left(1+\chi\right)} {n_g r_\la}
\frac{v_e}{\zeta} = \frac{\Wion} {m_e v_e u_\la}
\end{eqnarray}

By substituting $n_\sm = \eta_\crit^{(d)} n_\sm^{\MMSN}$ and by
solving for $\eta_\crit^{(d)}$, we have the following analytic formula 
for  $\eta_\crit^{(d)}$:
\begin{eqnarray}
  \eta_\crit^{\left(d\right)} = \left(\frac{\correction^{\left(d\right)}}{2 \pi \left(1+\chi\right)}
  \frac{e}{\Delta q_{\A,\C}}
  \frac{n_g r_\la}
  { {n_\sm^{\MMSN}}^2 {r_\sm}^3 }
  \frac{\zeta}
  {u_\la}
  \frac{\Wion}
  {k_B T}
  \right)^{\frac 1 2} .
 \label{symbolic-eta-critical-d}
\end{eqnarray}
We have introduced another nondimensional correction constant
$\correction^{(d)}$ as we did in \S \ref{section-analytic-cd}.

\subsubsection{Analytic formula for $\eta_\crit$ in charge-up phase}
\label{section-analytic-c}

Finally, we derive the analytic formula of the critical density
$\eta_\crit^{(c)}$, where the condition for electric discharge
(\ref{equation-macroscopic-discharge}) is met in (c)charge-up
phase (c.f. \S \ref{section-dust-chargeup-phase}, Figure
\ref{figure-circuit-phases}(c)).

By approximating equations (\ref{specific-kirch-i}) and
(\ref{specific-charge-neutrality}) with $J_{\sm,i} = J_{i,e}$ and
$Q_\sm + Q_i = 0$, we have

\begin{eqnarray}
Q_i = -Q_\sm = \frac{e n_g \zeta} {\pi n_\sm {r_\sm}^2 v_i}
\end{eqnarray}

In equation (\ref{specific-kirch-L}), we can ignore $J_{\la, e}$ and further
ignoring the second term in equation (\ref{specific-J-L-S}), we have

\begin{eqnarray}
  \frac{2 r_\sm}{r_\la} \Delta q_{\A,\C} n_\sm n_\la \Delta v_{\la,\sm} \pi {r_\la}^2
  - Q_i n_\la \sigma_{\la, i} v_i = 0\\
where \ \ \sigma_{\la, i} = \frac{Q_\la e}{n_\la r_\la l_B T} \pi {r_\la}^2
\end{eqnarray}
here we used the $qq' a^{-1} \ggt k_BT$ limit of Coulomb cross section
(\ref{sigma_coulomb_focusing-}).

Solving this for $Q_\la$, we have
\begin{eqnarray}
\abs{Q_\la} = \frac{2 \Delta q_{\A,\C}} {e^2} 
\frac{{n_\sm}^2 n_\la {r_\sm}^3}{n_g}
\frac{\Delta v_{\la, \sm} k_B T}{\zeta}
\end{eqnarray}

And from equation (\ref{specific-kirch-e}) we have
\begin{eqnarray}
Q_e = - \frac{e n_g \zeta} {\pi n_\sm {r_\sm}^2 v_e}
\end{eqnarray}

By substituting these $Q_\la$ and $ Q_e$ to the equality for the
lightning condition (\ref{equation-macroscopic-discharge}), replacing
$n_\sm = \eta_\crit^{(c)} n_\sm^{\MMSN}$ and $n_\la = \eta_\crit^{(c)}
n_\la^{\MMSN}$, and by solving for $\eta_\crit^{(c)}$, we obtain the
following analytic formula for $\eta_\crit^{(c)}$:

\begin{eqnarray}
  \eta_\crit^{\left(c\right)} = \left(\frac{\correction^{\left(c\right)}}{2 \pi}
  \frac{e^3}{\Delta q_{\A,\C}}
  \frac{{n_g }^2}
  { {n_\sm^{\MMSN}}^3 {n_\la^{\MMSN}} {r_\sm}^5 }
  \frac{\zeta^2}
  {\Delta v_{\la, \sm}  u_\la}
  \frac{\Wion}
  {\left(k_B T\right)^2}
  \right)^{\frac 1 4} 
 \label{symbolic-eta-critical-c}
\end{eqnarray}
We have introduced a third nondimensional correction constant $\correction^{(c)}$
as we did in previous sections.

\subsubsection{The combined analytic formula for $\eta_\crit$}
\label{section-analytic-combined}

The critical number density $\eta_\crit$ is either of
$\eta_\crit^{(c)}, \eta^{(cd)}, \eta_\crit^{(d)}$. To choose the
correct one, we have to consider the phase boundary conditions
(c.f. Table \ref{table-four-phase}). Instead, we propose the following
convenient scheme to choose the correct one:
\begin{eqnarray}
  \eta_\crit 
  &\eq& \eta_\crit^{\left(d\right)} \ \ \ \ if \ \ \eta_\crit^{\left(d\right)} > \eta^{\left(cd\right)},
  \nonumber \\
  &\eq& \eta^{\left(cd\right)} \ \ \ \ if \ \ \eta_\crit^{\left(c\right)} > \eta^{\left(cd\right)} > \eta_\crit^{\left(d\right)},
  \nonumber \\
  &\eq& \eta_\crit^{\left(c\right)} \ \ \ \ otherwise
  \label{analytic-eta-critical}.
\end{eqnarray}
This scheme is based on the intuition that the (cd)phase boundary is
included in both (c)charge-up phase and (d)dust phase. We can argue
that if $\eta_\crit^{(d)} > \eta^{(cd)}$, the number density
$\eta^{(cd)}$ is not large enough to cause lightning, and that if
$\eta_\crit^{(c)} < \eta^{(cd)}$, the number density
$\eta_\crit^{(c)}$ is already large enough to cause lightning.

Now, without the correction, e.g. with $\correction^{(c)} = \correction^{(cd)} =
\correction^{(d)} = 1$, the analytic values for $\eta_\crit$ differs from
the numerical values $\eta_\crit^{(num)}$ calculated in \S
\ref{section-result-lightning}, because of approximations used.  For
example, substituting the reference parameter of Fig.
\ref{figure-result-tanaka}: $r_\sm = 1.0 \times {10}^{-2}\unit{cm}$,
$r_\la=1.0 \times {10}^{2}\unit{cm}$, $D_\sm = 2.0$, and $D_\la = 2.368$,
equation (\ref{analytic-eta-critical}) evaluates to $\eta_\crit =
1.47 \times {10}^{2}$.  For the reference
parameter of Fig. \ref{figure-result-hayashi}: $r_\sm =
1.0 \times {10}^{-4}\unit{cm}$, $r_\la=1.0\unit{cm}$, $D_\sm = 2.665$, and
$D_\la = 3.0$, equation (\ref{analytic-eta-critical}) evaluates to
$\eta_\crit = 9.99 \times {10}^{2}$.  The
results of the simulations for those two parameter are
$\eta_\crit^{(num)} = 7.37 \times {10}^{1}$ and
$\eta_\crit^{(num)} = 3.01 \times {10}^{2}$, respectively.  The
analytic and simulational values agree upto a factor of three.

We set the values for $\correction^{(c)},  \correction^{(cd)}, \correction^{(d)}$ by the condition 
that the following squared-error integral over the entire parameter ranges 
(\ref{constraint-begin}-\ref{constraint-end}) is minimum:
\begin{eqnarray}
\int \!\!\! \int \!\!\! \int \!\!\! \int 
dr_\sm \, dr_\la \, dD_\sm \, dD_\la 
\left(\log_{10} \eta_\crit - \log_{10} \eta_\crit^{\left(num\right)}\right)^2
\end{eqnarray}
This gives $\correction^{(c)} = 9.4 \times {10}^{-1}, \correction^{(cd)} =
3.3 \times {10}^{-1}, \correction^{(d)} = 8.5 \times {10}^{-1} $.
Taking these corrections into account, the values for 
 $\eta^{(c)},  \eta^{(cd)}, \eta^{(d)}$ are as follows:

\begin{eqnarray}
  \eta^{\left(c\right)}_\crit &\eq& 1.1 \times {10}^{3} 
  \left(\frac{\Delta q_{\A,\C}}{ 6.2 \times {10}^{2} \unit{e} }\right)^{-\frac 1 4} 
  \left(\frac{n_g}{4.7 \times {10}^{13} \unit{cm^{-3}}} \right)^{\frac 1 2} \nonumber \\ &{}&
  \left(\frac{n_\sm^{\MMSN}}{8.8 \times {10}^{-1} \unit{cm^{-3}}} \right)^{-\frac 3 4}  
  \left(\frac{n_\la^{\MMSN}}{4.0 \times {10}^{-14} \unit{cm^{-3}}} \right)^{-\frac 1 4}  \nonumber \\ &{}&
  \left(\frac{r_\sm}{1.0 \times {10}^{-4} \unit{cm}}\right)^{- \frac 5 4} 
  \nonumber \\ &{}&
  \left(\frac{\zeta}{1.0 \times {10}^{-18}\unit{sec^{-1}}}\right)^{\frac 1 2}
  \nonumber \\ &{}&
  \left(\frac{\Delta v_{\la, \sm}}{3.4 \times {10}^{3}\unit{cm \ sec^{-1}}}\right)^{-\frac 1 4}
  \left(\frac{u_\la}{3.4 \times {10}^{3}\unit{cm \ sec^{-1}}}\right)^{- \frac 1 4}
  \nonumber \\ &{}&
  \left(\frac{\Wion}{15.4\unit{eV}}\right)^{\frac 1 2}
  \left(\frac{T}{1.7 \times {10}^{2}\unit{K}}\right)^{- \frac 1 2} 
 \label{analytic-eta-critical-c},
  \\
  \eta^{\left(cd\right)} &\eq& 3.3 \times {10}^{2} 
  \left(\frac{\Delta q_{\A,\C}}{ 6.2 \times {10}^{2} \unit{e} }\right)^{-\frac 1 2} 
  \left(\frac{n_g}{4.7 \times {10}^{13} \unit{cm^{-3}}}\right)^{\frac 1 2} 
  \nonumber \\ &{}&
  \left(\frac{n_\sm^{\MMSN}}{8.8 \times {10}^{-1} \unit{cm^{-3}}} \right)^{- \frac 1 2}
  \left(\frac{n_\la^{\MMSN}}{4.0 \times {10}^{-14} \unit{cm^{-3}}} \right)^{- \frac 1 2}
  \nonumber \\ &{}&
  \left(\frac{r_\sm}{1.0 \times {10}^{-4} \unit{cm}}\right)^{- \frac 1 2} 
  \left(\frac{r_\la}{1.0 \unit{cm}}\right)^{- \frac 1 2} 
  \nonumber \\ &{}&
  \left(\frac{\zeta}{1.0 \times {10}^{-18}\unit{sec^{-1}}}\right)^{\frac 1 2}
  \nonumber \\ &{}&
  \left(\frac{\Delta v_{\la, \sm}}{3.4 \times {10}^{3}\unit{cm \ sec^{-1}}}\right)^{-\frac 1 2}
 \label{analytic-eta-critical-cd},
 \\
 \eta^{\left(d\right)}_\crit &\eq& 5.9 \times {10}^{1}
  \left(\frac{\Delta q_{\A,\C}}{ 6.2 \times {10}^{2} \unit{e} }\right)^{-\frac 1 2} 
  \left(\frac{n_g}{4.7 \times {10}^{13} \unit{cm^{-3}}} \right)^{\frac 1 2} 
  \nonumber \\ &{}&
  \left(\frac{n_\sm^{\MMSN}}{8.8 \times {10}^{-1} \unit{cm^{-3}}} \right)^{-1}  
  \nonumber \\ &{}&
  \left(\frac{r_\sm}{1.0 \times {10}^{-4} \unit{cm}}\right)^{- \frac 3 2} 
  \left(\frac{r_\la}{1.0 \unit{cm}}\right)^{ \frac 1 2} 
  \nonumber \\ &{}&
  \left(\frac{\zeta}{1.0 \times {10}^{-18}\unit{sec^{-1}}}\right)^{\frac 1 2}
  \left(\frac{u_\la}{3.4 \times {10}^{3}\unit{cm \ sec^{-1}}}\right)^{- \frac 1 2}
  \nonumber \\ &{}&
  \left(\frac{\Wion}{15.4\unit{eV}}\right)^{\frac 1 2}
  \left(\frac{T}{1.7 \times {10}^{2}\unit{K}}\right)^{- \frac 1 2} 
  \label{analytic-eta-critical-d}. 
\end{eqnarray}

Note that $\Delta q_{\A,\C}$, $n_\sm^{\MMSN}$, and $n_\la^{\MMSN}$
also depends on dust parameters: $r_\sm$, $r_\la$, $D_\sm$, and
$D_\la$.  Using equation (\ref{cross-section-contact}) and the
$\eta_\charge, \sigma_\charge$ introduced in \S
\ref{section-surface-charge-exchange-I},
\begin{eqnarray}
  \Delta q_{\A,\C} &\eq& \eta_\charge \sigma_\charge S_\contact \\
  &\eq& 
  6.2 \times {10}^{2}  \unit{e} \cdot
  \frac{\eta_\charge}{0.1} \frac{\sigma_\charge}{6.2 \times {10}^{9}\unit{e\ cm^{-2}}}
  \nonumber \\
  &&\frac{\min \left( {r_\sm}^{3/2} {r_\la}^{1/2}, 
 {r_m}^{2-D_\la} {r_\sm}^{5/2} {r_\la}^{D_\la - 5/2}\right)}{1.0 \times {10}^{-6}\unit{cm^2}} .
\end{eqnarray}
Using equations (\ref{rho-small}-\ref{number-density-large})
and equation (\ref{dust-mass-definition}),
\begin{eqnarray} 
n_\sm^{\MMSN} &\eq& 4.0 \times {10}^{2} \left(\frac{r_\sm}{r_m}\right)^{-D_\sm} \unit{cm^{-3}} 
\nonumber \\
&& \left(\frac{r}{2.7 \unit{AU}}\right)^{-11/4} \left(\frac{m_m}{3.9 \times {10}^{-15}\unit{g}}\right)^{-1} \\
n_\la^{\MMSN} &\eq& 4.0 \times {10}^{1} \left(\frac{r_\la}{r_m}\right)^{-D_\la} \unit{cm^{-3}}
\nonumber \\
&& \left(\frac{r}{2.7 \unit{AU}}\right)^{-11/4}  \left(\frac{m_m}{3.9 \times {10}^{-15}\unit{g}}\right)^{-1},
\end{eqnarray}
and the monomer radius
\begin{eqnarray} 
r_m = 1.0 \times {10}^{-5} \unit{cm}.
\end{eqnarray}

We have plotted these analytic solutions
(\ref{analytic-eta-critical-c}-\ref{analytic-eta-critical-d}) combined
with the condition (\ref{analytic-eta-critical}) in solid-line
contours from Fig. \ref{figure-result-tanaka} to
Fig. \ref{figure-result-averaged}.  The red thin contour represents
the parameter ranges where $\eta^{(cd)}$ contributes. The blue thick
contours represents the parameter ranges where $\eta^{(d)}$
contributes, where blue solid contour means $\chi < 1$ and blue dashed
contour $\chi > 1$. The thick yellow-sleeved red contours represents
the parameter ranges where $\eta^{(c)}$ contributes.  The numerical
solutions, on the other hand, are plotted in colour maps and the black
dashed contours.

The averaged plots, Fig. \ref{figure-result-averaged} shows the
agreement of the numerical and analytic value over the entire
parameter range.  Quantitatively, the root-mean-square error is
\begin{eqnarray}
\sqrt{\frac{
\int \!\!\! \int \!\!\! \int \!\!\! \int 
dr_\sm \, dr_\la \, dD_\sm \, dD_\la 
\left(\log_{10} \eta_\crit - \log_{10} \eta_\crit^{\left(num\right)}\right)^2
}{
\int \!\!\! \int \!\!\! \int \!\!\! \int 
dr_\sm \, dr_\la \, dD_\sm \, dD_\la 
}} \nonumber \\
= 9.2 \times {10}^{-2}.
\end{eqnarray}
Moreover, using the values of $\correction^{(c)}, \correction^{(cd)},
\correction^{(d)}$ obtained only from the `normal' run
(Fig. \ref{figure-result-tanaka}, \ref{figure-result-hayashi} and
\ref{figure-result-averaged}), we can fit the results of the
`pessimistic' simulations (Fig. \ref{figure-result-hypertanaka}) by a
root-mean-square error of $2.6 \times {10}^{-1}$.  We
also perform the simulations with smaller values of relative velocity
and fit the results. The root-mean-square errors were $5.6 \times {10}^{-2}, 6.4 \times {10}^{-2}, 1.1 \times {10}^{-1}$, for
$\Delta v_{\la,\sm} = u_\la = 3.4 \times {10}^{2}\unit{cm\ s^{-1}},
3.4 \times {10}^{1}\unit{cm\ s^{-1}}, 3.4\unit{cm\ s^{-1}}$, respectively.
These fits prove the predictability of our analytic formulae
(\ref{analytic-eta-critical}) and
(\ref{analytic-eta-critical-c}-\ref{analytic-eta-critical-d}).

\section{Conclusions and discussions} \label{section-discussion}

We have shown that as dust number density $\eta$ increase, the charge
density distribution experience four phases: (a)ion-electron plasma
phase, (b)ion-dust plasma phase, (c)charge-up phase and (d)dust phase.
The former two phases are studied in detail by
\citet{2009ApJ...698.1122O}, while the latter two phases are unique
results of taking dust-dust collision into consideration.  We have
calculated the dust number density $\eta_\crit$ at which lightning
strikes, as function of dust radius $r_\sm$, $r_\la$ and fractal
dimension $D_\sm$, $D_\la$ numerically. Using the numerical results we
have derived the analytical formulae for $\eta_\crit$: equations
(\ref{analytic-eta-critical}),
(\ref{analytic-eta-critical-c}-\ref{analytic-eta-critical-d}).
Because the generated electrostatic field $E_\max(\eta)$ grows more
rapidly than estimate by \citet{1997Icar..130..517G} in (c)charge-up
phase and (d)dust phase, lightning in protoplanetary discs are
possible with smaller dust number densities. We discuss the
consequences in this section.

\subsection{Energetics and direct observations}

We estimate the total energy of a lightning event in a protoplanetary
disc at $r = 2.7\unit{AU}$. For MMSN, the number density of the
gas is $4.7 \times {10}^{13} \unit{cm^{-3}}$ in the region.  The typical electron
mean free path at this site is $l_{mfp} \simeq 1.2 \times {10}^{2} \unit{cm}$.  By
equation (\ref{equation-E_dis}) we know the critical electric field
$E_\dis \simeq 4.3 \times {10}^{-4} \unit{G}$.  The sphere with radius of
the disc scale-height $h \simeq 2.4 \times {10}^{12} \unit{cm}$ contains the electric
energy $W \equiv {E_\dis}^2/8 \pi \times 4 \pi h^3/3 \simeq
4.3 \times {10}^{29} \unit{erg}$.  When the lightning strikes, the energy is
concentrated into lightning bolt of radius $w$ and length $h$, where
$w$ is related to $l_{mfp}$ by $w \simeq 5000 \, l_{mfp} \simeq
6.0 \times {10}^{5} \unit{cm}$ \citep{1992ApJ...387..364P}.  If all the
energy is used to heat the gas within the lightning bolt, the gas can be
heated to $1.6 \times {10}^{7} \unit{K}$.

The ultimate energy source for this electric discharge event is the gravitational
energy of the accreting matter. 
In our model the mass accretion ratio of uncondensed larger dust is
$\dot M = 2 \pi r  \Sigma_\la^{\MMSN} u_\la \simeq 3.3 \times {10}^{17} \unit{g \
  sec^{-1}}$.
The gravitational energy released within condensation region $h$ is
$L \equiv G  M_\odot  \dot M  h r^{-2} \simeq 6.6 \times {10}^{28}
\unit{erg \ sec^{-1}}$.
 For the largest energy event $W=4.3 \times {10}^{29} \unit{erg}$,
The upper limit of the event rate is $1.5 \times {10}^{-1}  \unit{sec^{-1}}$.

\subsubsection {Astronomically Low Frequency (ALF) Waves}

The change density evolution, electromagnetic pulse, and electromagnetic waves
accompanying lightning in terrestrial thunderclouds are observed
\citep[e.g.][]{JGR1989.94.1165,JGR1979.84.6307}. 
The typical wavelength of the electromagnetic waves are similar to the scale 
height of the thundercloud. These are called extremely low frequency waves.
The electromagnetic waves from lightning can be basically modelled as
solutions of Maxwell equations, including lightning current as a source
term \citep[e.g.][]{1998IEEE.40.403}.
When we apply these models to the protoplanetary discs, 
the electromagnetic wave spectrum is extend between the event duration
and light crossing time of the system, or
$ 9.6 \times {10}^{-5} \sim 1.2 \times {10}^{-2} \unit{Hz}$.
This frequency range is at least two orders of magnitude lower than 
any frequencies with established observational methods.
It is difficult to make a fair choice for the successor to the frequency list
`very low frequency ({\em VLF}),'  `ultra low frequency({\em ULF}),'
`super low frequency ({\em SLF}),' and  
`extremely low frequency ({\em ELF}).'
We opt for Astronomically Low Frequency (ALF) waves 
and hope that the reader will forgive us!
Anyway the frequency is so low 
that we will need an astronomical budget 
to build an astronomically large detector to receive it, 
considering its wavelength of order of an astronomical unit.

\subsubsection {Infrared (IR) Observations}
The energy of the lightning contributes to the local heating of the
protoplanetary discs, which might be resolved by advanced telescopes
such as Atacama Large Millimetre Array ({\em ALMA}).  The most
possible observational evidence is excess of heating near the
snowline.  To distinguish the cause of the heating with other heating
model candidates, the variability or correlation function of the
heating might be useful. This is because lightning propagates at the
speed of ionised electrons, which is much faster than the speed of
sound.

\subsubsection {Ultraviolet (UV) Observations}
The ionisation electrons of the lightning excite various electron levels
in gas molecules and dust.
There is possibility of observing fluorescence photons from such excited
molecules.
Although the disc gas is generally expected to be thick for ultraviolet
photons, there are categories of lightning that extends toward thin regions
of the gas, known as sprites and elves \citep[e.g.][]{PhysicsToday00319228}. The sprites and elves are phenomena
similar to lightning observed in the mesosphere of the earth, possibly caused
by electric fields induced by the thunderclouds.
Fluorescence lines from such regions can be observed by future ultraviolet
missions like {\it THEIA} \citep{2009AAS...21345804S}. 
Also, some observational results on protostellar and protoplanetary systems
today have difficulties in explaining either lack or excess of {\em UV}
\citep[e.g.][]{2005A&A...438..923N,
  2008A&A...488..565C,2008A&A...486..533P,2008ApJ...681..594H}. 
If excess of {\em UV} photons is observed compared to the model, 
it might be from the sprite discharges and elves 
from the surface of the protoplanetary discs;
on the other hand if the chemical composition model require more {\em UV} photons
than is observed, lightning hidden in the disc mid-plane might be providing
them.

\subsubsection {High Energy Gamma Rays}

Detection of burst-like gamma-ray is reported from terrestrial
thunder clouds.
The burst precedes a cloud-to-ground lightning,
lasts for $\sim 40$ seconds,
extends to $10\unit{MeV}$.
The spectrum can be interpreted as
consisting of bremsstrahlung photons from relativistic electrons
\citep{2007PhRvL..99p5002T,2008ICRC....1..745E}.
These relativistic electrons are 
secondary electrons generated by cosmic rays, 
and accelerated by the electric fields 
through process known as avalanche amplification \citep{1996JGR...101.2297R}.
If a charged particle is accelerated by the protoplanetary thundercloud field,
through similar process, its kinetic energy reaches 
$e E h \simeq 3.1 \times {10}^{11} \unit{eV}$.

\subsection{Chondrule heating by lightning}

Chondrule heating by lightning scenario is now considered unlikely
 \citep{1997LPI....28.1515W, 1997Icar..130..517G, 2008Icar..195..504G}.
The reasons that prohibit the scenario can be summarized as following three
problems. 

\subsubsection{Energetics problem}
The ultimate energy source (gravitational potential of the protoplanetary
disc) is sufficient to melt the chondrules; but most of the energy earned by
ingoing larger dust go to the
outgoing gas by angular momentum exchange \citep{1997LPI....28.1515W};  
little contribute to the random motion, the energy source for the lightning.

\subsubsection{Neutralization problem}
Unlike the earth atmosphere, the protoplanetary discs are filled with weakly
ionised plasma which rapidly responds to electric field. Neutralization effect
can be further subdivided to microscopic neutralization of individual dust and
macroscopic neutralization of large-scale electric field necessary to cause
lightning.  If a dust get charged by dust-dust collision, the dust absorbs
plasma of opposite polarity in $\sim 10\unit{sec}$ and returns to equilibrium
charge state.  Moreover, even if there is charged dust and bulk motion between
the oppositely charged dust, the electric field caused by the dust induces
Ohmic current in the plasma.  The current will quickly neutralize the electric
field.

\subsubsection{Destruction problem}
After all, there is an experimental evidence by \citet{2008Icar..195..504G} 
that lightning destroys the dust aggregates rather than melting them.

\subsubsection{Solution to the problems}
This work can provide answer for the first and second problem.
energetics problem, the larger dust and the gas (containing smaller
dust that are coupled to the gas) is now `harnessed' by electric
field. Outgoing gas is not free in carrying the gravitational energy
away; instead the gas converts its gravitational energy into electric
field energy, fully contributing to lightning.  For the neutralization
problem, we have shown in this work that with reasonably high dust
number density $\eta$, the dust-dust charge separation can dominate
over the plasma neutralization, and the electrostatic field can grow
up to critical value.

For the third problem, we point out that 
in \citet{2008Icar..195..504G}'s experiment, either 
the electron mean free path is many orders of magnitude
shorter, or the electron kinetic energy is much larger
compared to the protoplanetary-disc environment.
They used air at pressures between 
$10$ and $10^5 \unit{Pa}$.
Air consists of 
$78$ per cent nitrogen, 
$21$ per cent oxygen, and
$1$ per cent argon.
Their molecular van der Waals radii are
$1.6 \times {10}^{-8} \unit{cm}$,
$1.5 \times {10}^{-8} \unit{cm}$, and
$1.9 \times {10}^{-8} \unit{cm}$, respectively
\citep{JPhysChem.1964.68.441}.

Therefore, the electron mean free path 
and the electron kinetic energy, $W_e = e \, E \, l_\mfp$, was
$l_\mfp \sim 4.8 \times {10}^{-1} \unit{cm}$, 
$W_e = 1.6 \times {10}^{4} \unit{eV}$ 
 for $10\unit{Pa}$ case,
and $l_\mfp \sim 4.8 \times {10}^{-5} \unit{cm}$ ,
$W_e = 1.6 \unit{eV}$
for $10^5 \unit{Pa}$ case, respectively.
On the other hand in protoplanetary discs, 
typical mean free path and electron kinetic energy are $l_\mfp = 1.2 \times {10}^{2} \unit{cm}$,
$W_e = 15.4 \unit{eV}$.

It might be possible that protoplanetary-disc lightning is effective in melting 
dust aggregates, although experimental lightning
is ineffective in heating and led to disruption of the dust,
due to shorter mean free path or higher energy electron.
The minimum size of the structures that electron can form is of order of its mean free path.
If the electron mean free path is much shorter than the dust aggregates, 
as in $10^5 \unit{Pa}$ case,
the electron current may concentrate on the most conductive part of the dust aggregate,
leading to partial heating and explosion of the dust.
On the other hand if the electron is 
much more energetic, as in $10 \unit{Pa}$ case, 
it may react differently on dust monomers.

To reproduce the mean free path and electron energy simultaneously,
one must reproduce the electric field strength $E = 4.3 \times {10}^{-4} \unit{G}$
of protoplanetary discs; while the electric field used in the experiment
$E = 1.1 \times {10}^{2} \unit{G}$ was much stronger.
This much stronger electric field itself, might be the cause of dust aggregate dissociation,
due to much stronger electric force exerted on electron-absorbed dust monomers.
Also the discharge time-scale in the experiment was much smaller
than that in the protoplanetary discs, which might have led to 
the catastrophic results.

We think that the effect of lighting on dust aggregate in protoplanetary-disc environment
is yet to be confirmed in future experiments and simulations.

\subsection{Effects on magnetorotational instability (MRI) and disc
  environment} 

The dust-dust collisional charging and lightning is not a side-effect of some other processes, 
but is one of the  key processes in protoplanetary discs that affects each other.
The lightning is powered by gravitational energy
of the migrating larger dust.
The migration of the larger dust as well as the long term evolution of the gas
disc is governed by the disc viscosity. The best candidate for providing
the disc viscosity is MRI. And MRI is controlled by gas ionisation degree,
which in turn is controlled by the dust charge state and 
lightning.

Even the longest estimate for time-scale of the lightning $1.0 \times {10}^{4}
\unit{sec}$ is much smaller than the time-scale of MRI, which is at least of
the order of Kepler timescales. Lightning occur in low-ionisation regions
where MRI is prohibited (dead zones), and even if the lightning instantly
raise the ionisation rate, the free electrons and ions will quickly be
absorbed by the dust. Therefore we expect that MRI and lightning cannot
co-exist.  However lot of profound phenomena are possible. Just for an example
let us think of a two-layer dead-active zone model but with dust-dust
collisional charging.  The dead-zone is filled with lightning, inducing sprite
discharges towards active zones, which sustains the ionisation rate and
MRI. The MRI in turn shovels the dust into the dead-zone.

Such global models are beyond the reach of this paper. Nevertheless we
conclude this paper by stating that the dust-dust collisional charging is a necessary
component for understanding the planetesimal formation 
and global behaviour of the protoplanetary discs.

\section*{Acknowledgments}

The authors thank 
Tatsuya Tomiyasu for his useful advice
on protoplanetary discs
and collaboration with him
on study of ice surface charge.
The authors also thank 
Hidekazu Tanaka, Tetsuo Yamamoto and their colleagues at 
Institute of Low Temperature Science, Hokkaido University
for their kind invitation and discussion.
The authors thank Tsuyoshi Hamada for his advice on
GPGPU calculations.
The authors thank Takayuki Muto for his careful reading 
of the first draft of this paper.
The authors also thank 
Shu-ichiro Inutsuka, Hitoshi Miura, Satoshi Okuzumi
and other people for useful comments.
We also thank the anonymous referee for a number of suggestions 
that improved this paper.

The numerical simulations were carried out on 
Tenmon GPGPU cluster ({\it Tengu}) in Kyoto University.
Construction of {\it Tengu} is supported 
by Theoretical Astrophysics Group in Kyoto University,
by Grants-in-Aid (16077202, 18540238) from MEXT of Japan,
and by Global COE Startup Project `Breaking new grounds in numerical
astrophysics with General Purpose Graphic Processors.'
T. M. is supported by grants-in-aid for JSPS Fellows (21-1926)
from MEXT of Japan.
This work was supported by the Grant-in-Aid for the Global COE Programme
`The Next Generation of Physics, Spun from Universality and Emergence'
from the MEXT of Japan.

\def\gridscale{0.4}
\def\gridpush{\hspace{-10pt}}
\def\gridpull{\vspace{-30pt}}
\def\gridcut{\hspace{-100pt}}
\begin{figure*}
  \begin{center}
  \begin{tabular}{cc} \ \ 
    \gridpush \gridpull
    \includegraphics[scale=\gridscale,angle=270]{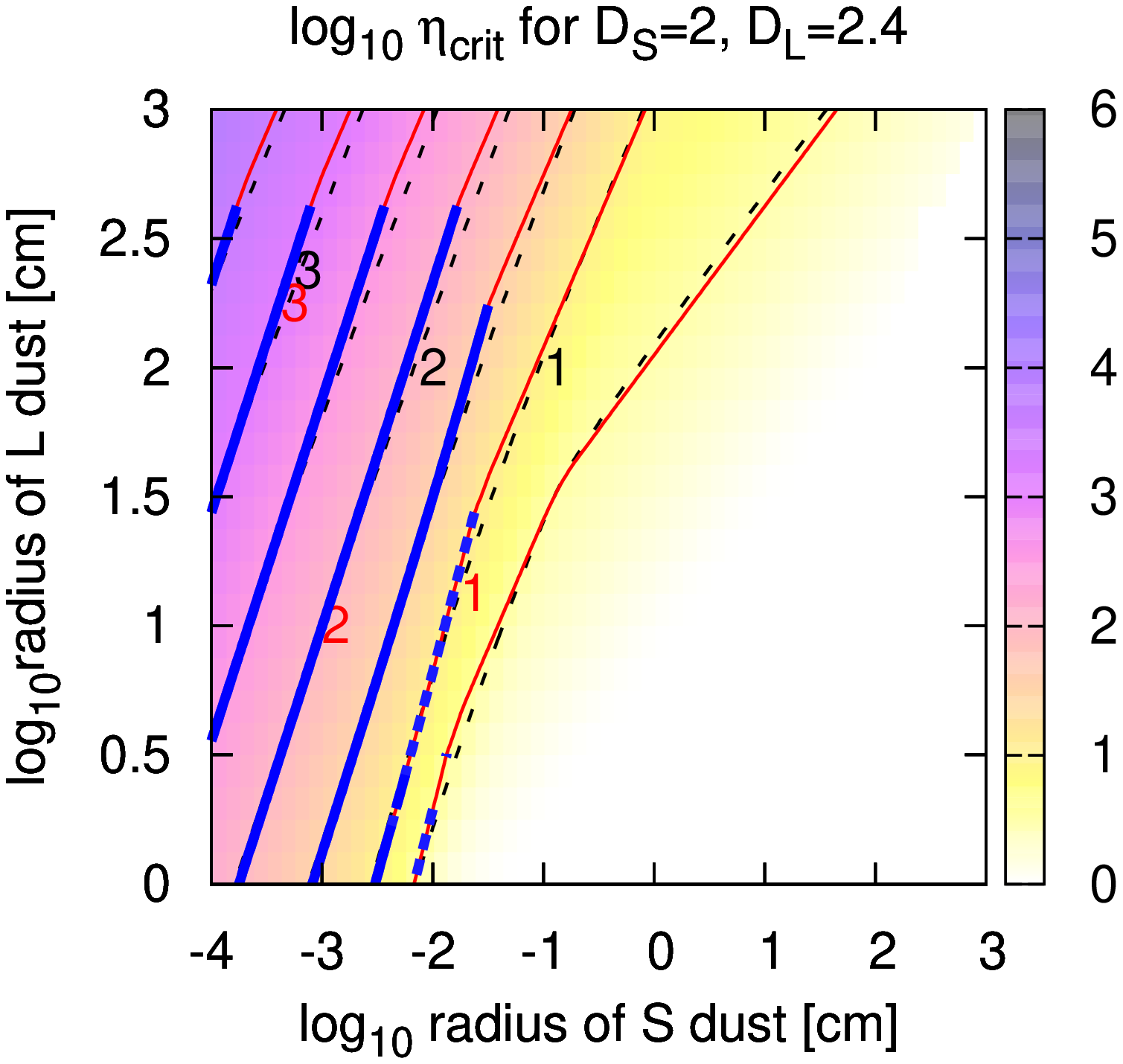} 
    \gridcut & \gridcut
    \includegraphics[scale=\gridscale,angle=270]{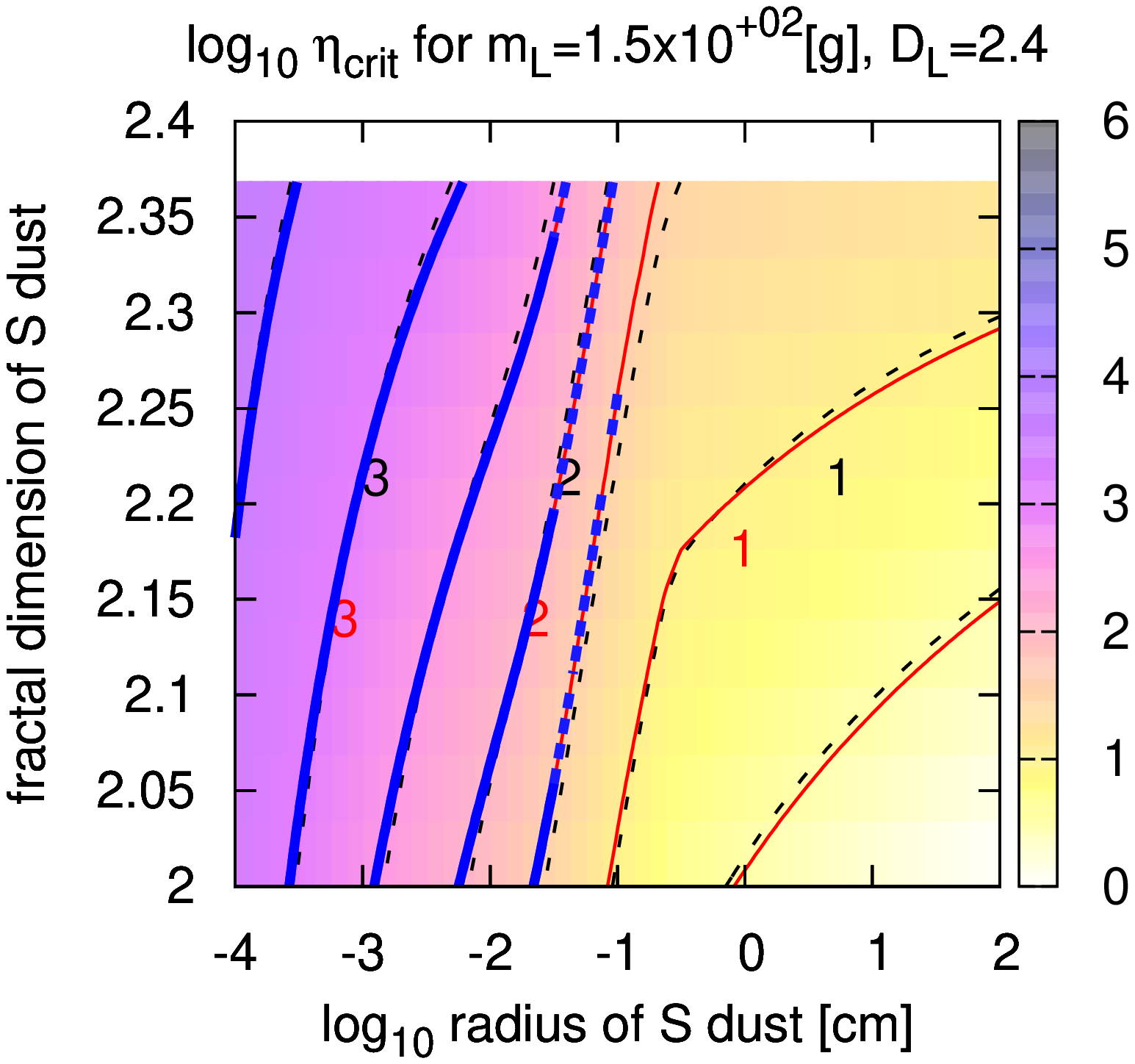} 
    \gridcut \\ \gridpush \gridpull
    \includegraphics[scale=\gridscale,angle=270]{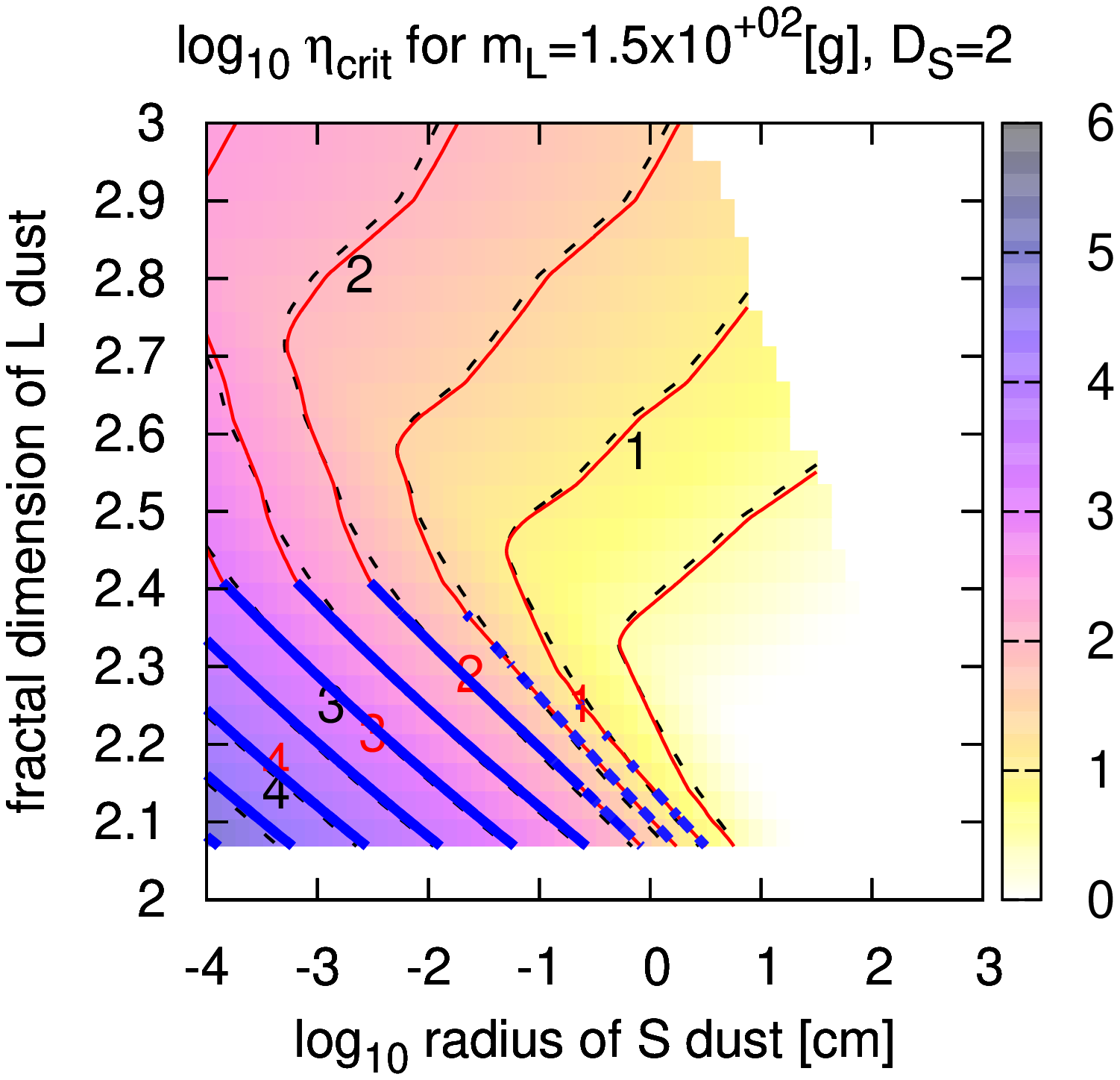} 
    \gridcut & \gridcut
    \includegraphics[scale=\gridscale,angle=270]{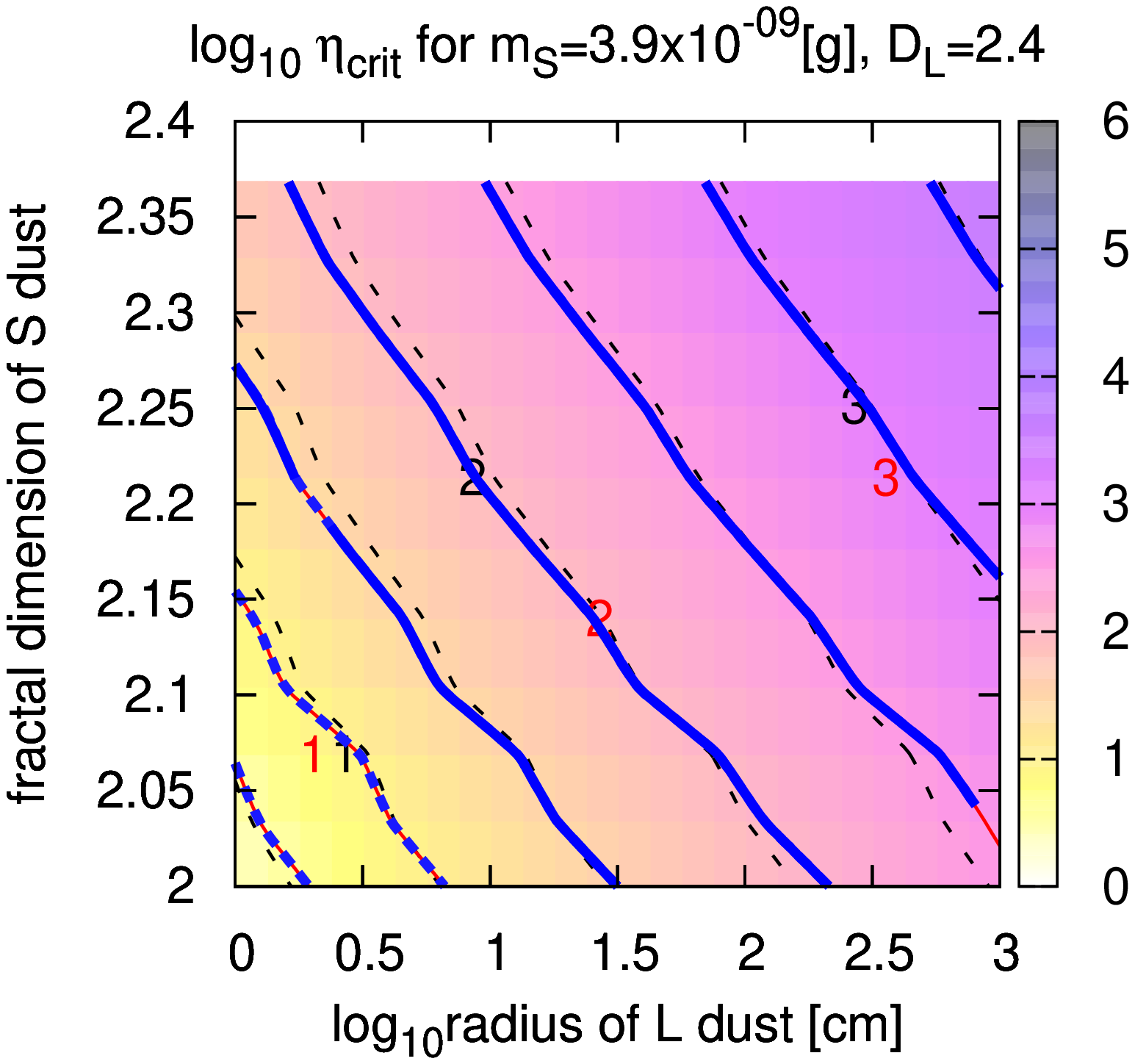} 
    \gridcut \\ \gridpush 
    \includegraphics[scale=\gridscale,angle=270]{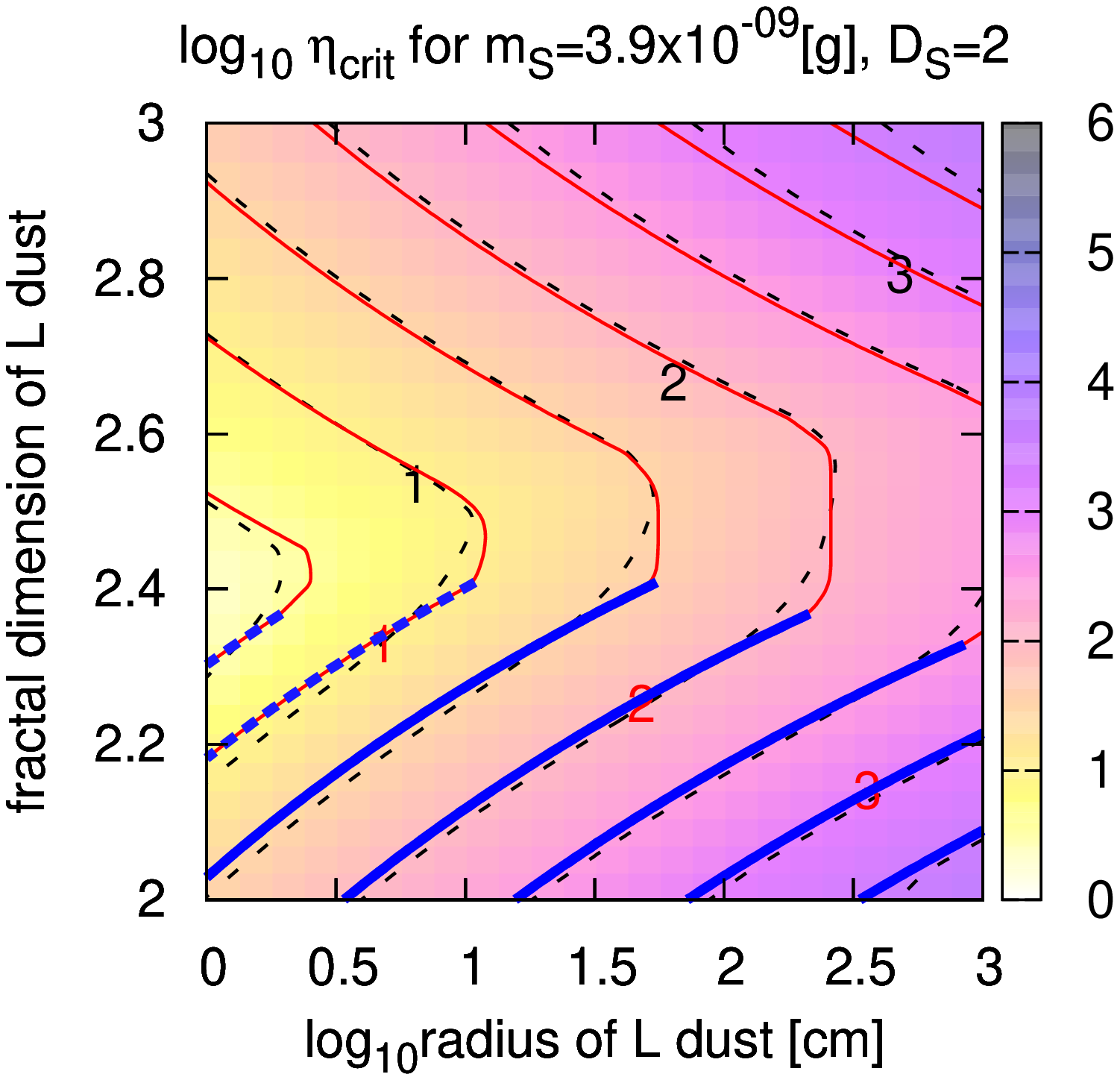} 
    \gridcut & \gridcut
    \includegraphics[scale=\gridscale,angle=270]{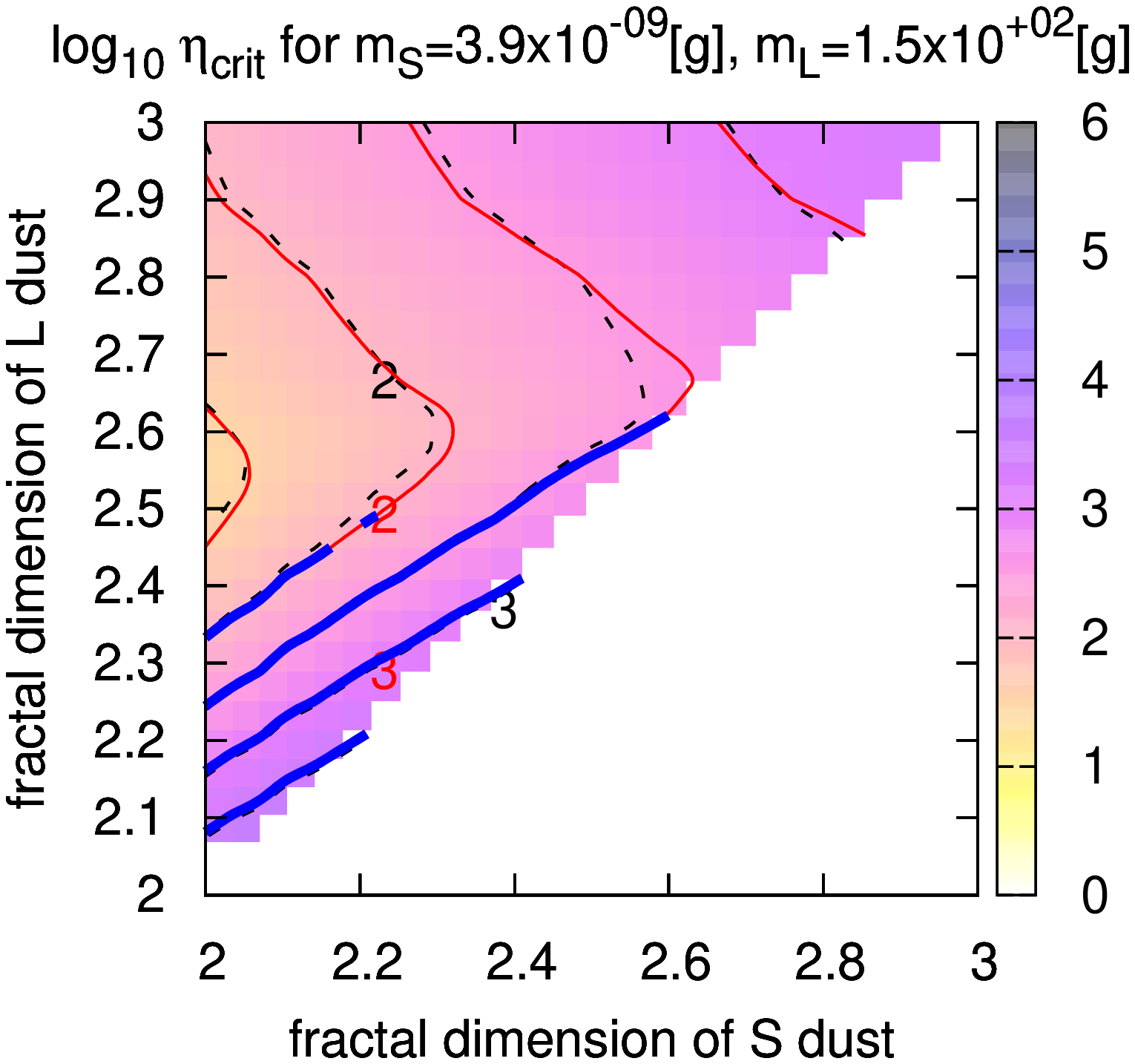} 
    \gridcut \\ 
  \end{tabular}
  \end{center}
  \caption {\small Value of $\eta_\crit$ as function of dust radius
    $r_\sm$, $r_\la$ and fractal dimension $D_\sm$, $D_\la$.  The base
    values are $r_\sm = 1.0 \times {10}^{-2}\unit{cm}$, $r_\la=1.0 \times {10}^{2}\unit{cm}$,
    $m_\sm = 3.9 \times {10}^{-9}\unit{g}$, $m_\la = 1.5 \times {10}^{2}\unit{g}$, $D_\sm =
    2.0$, and $D_\la = 2.368$.  We keep $m_\DPSub$
    constant when we vary $D_\DPSub$; we keep $D_\DPSub$ constant when
    we vary $r_\DPSub$.  Numerical results are in colour maps and black
    dashed contours; analytical values in coloured solid contours
    (c.f. \S \ref{section-analytic-combined} for the details of the
    plots.)  } \label{figure-result-tanaka}
\end{figure*}

\def\gridscale{0.4}
\def\gridpush{\hspace{-10pt}}
\def\gridpull{\vspace{-30pt}}
\def\gridcut{\hspace{-100pt}}
\begin{figure*}
  \begin{center}
  \begin{tabular}{cc} \ \ 
    \gridpush \gridpull
    \includegraphics[scale=\gridscale,angle=270]{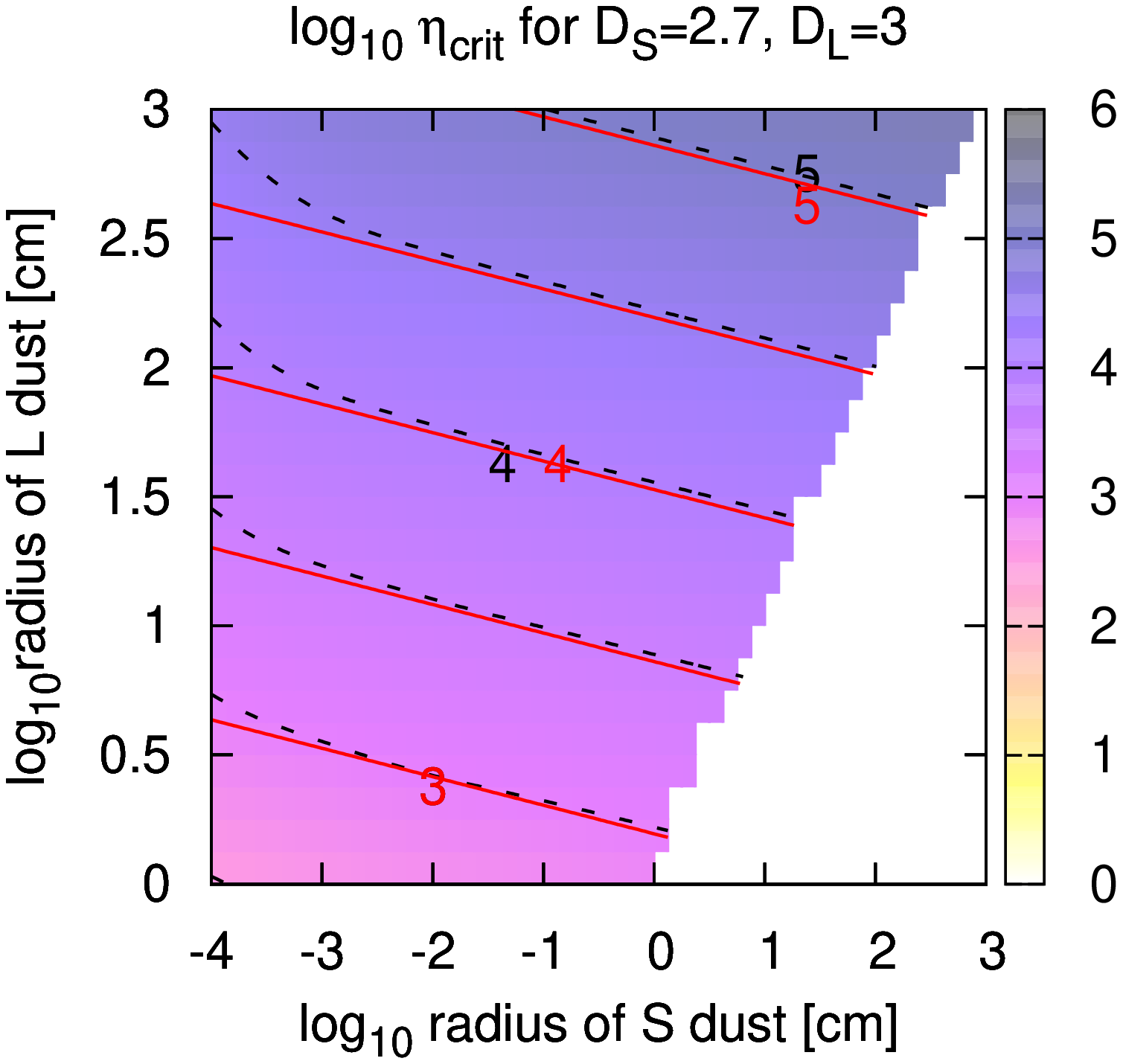} 
    \gridcut & \gridcut
    \includegraphics[scale=\gridscale,angle=270]{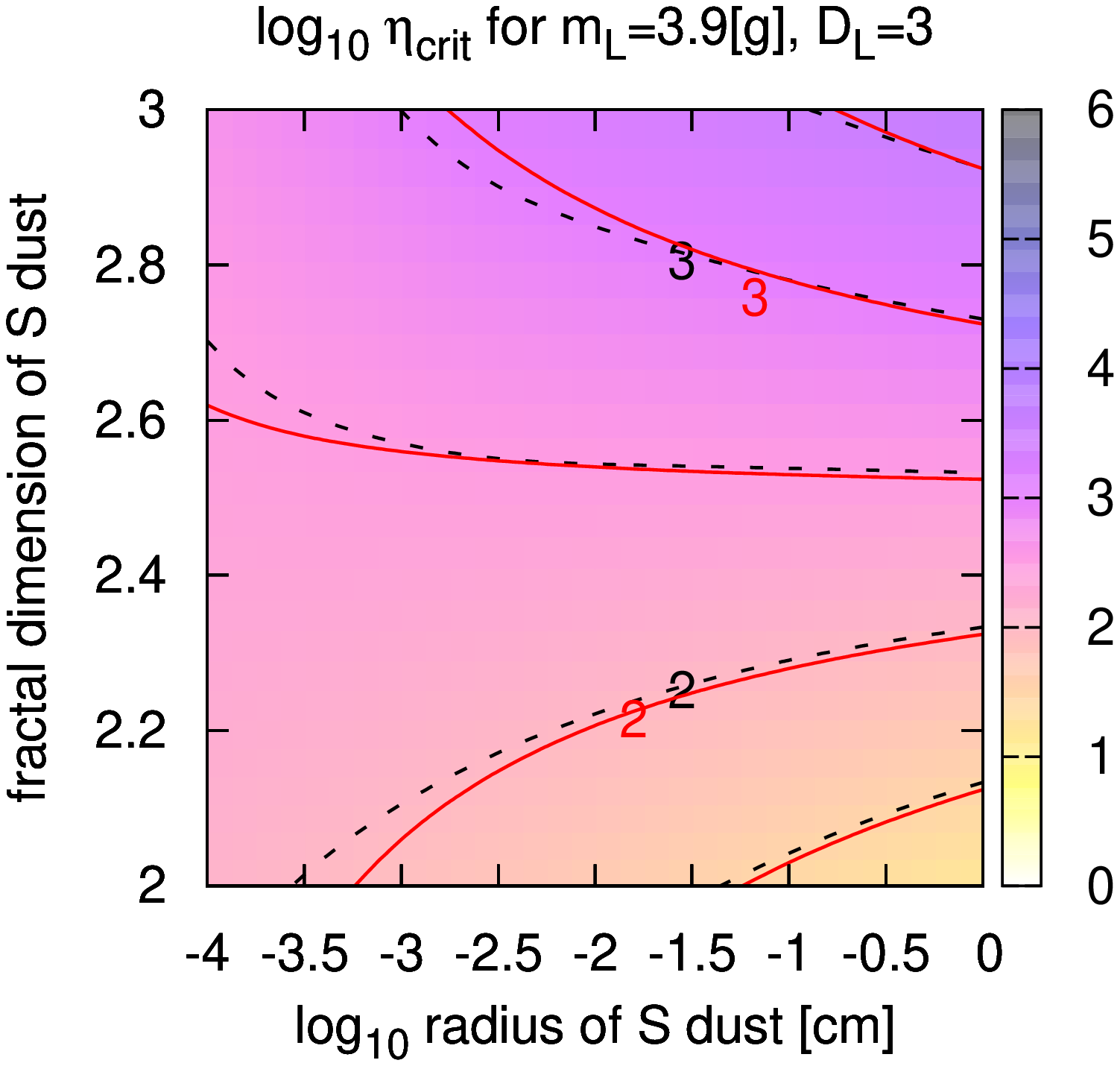} 
    \gridcut \\ \gridpush \gridpull
    \includegraphics[scale=\gridscale,angle=270]{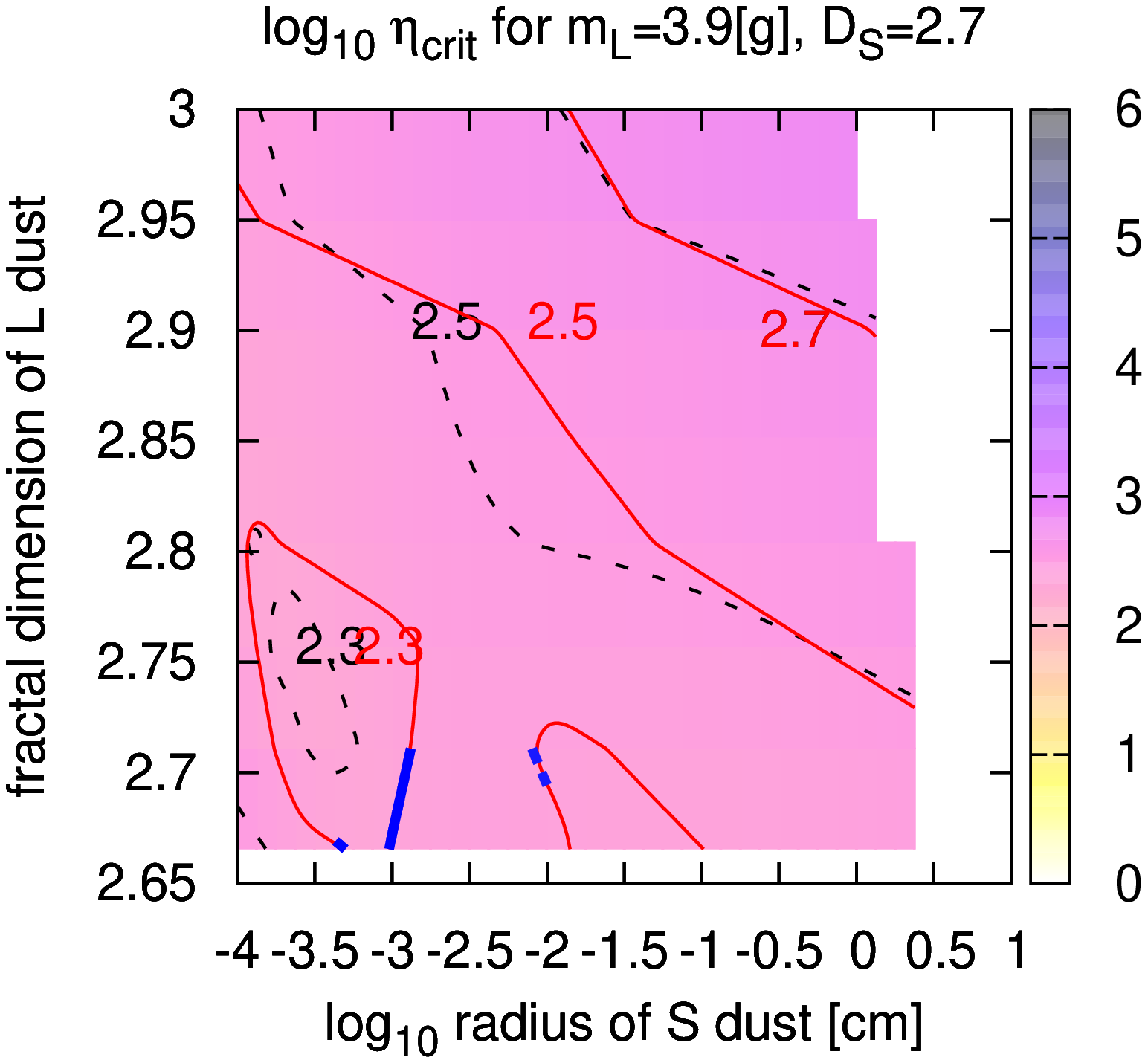} 
    \gridcut & \gridcut
    \includegraphics[scale=\gridscale,angle=270]{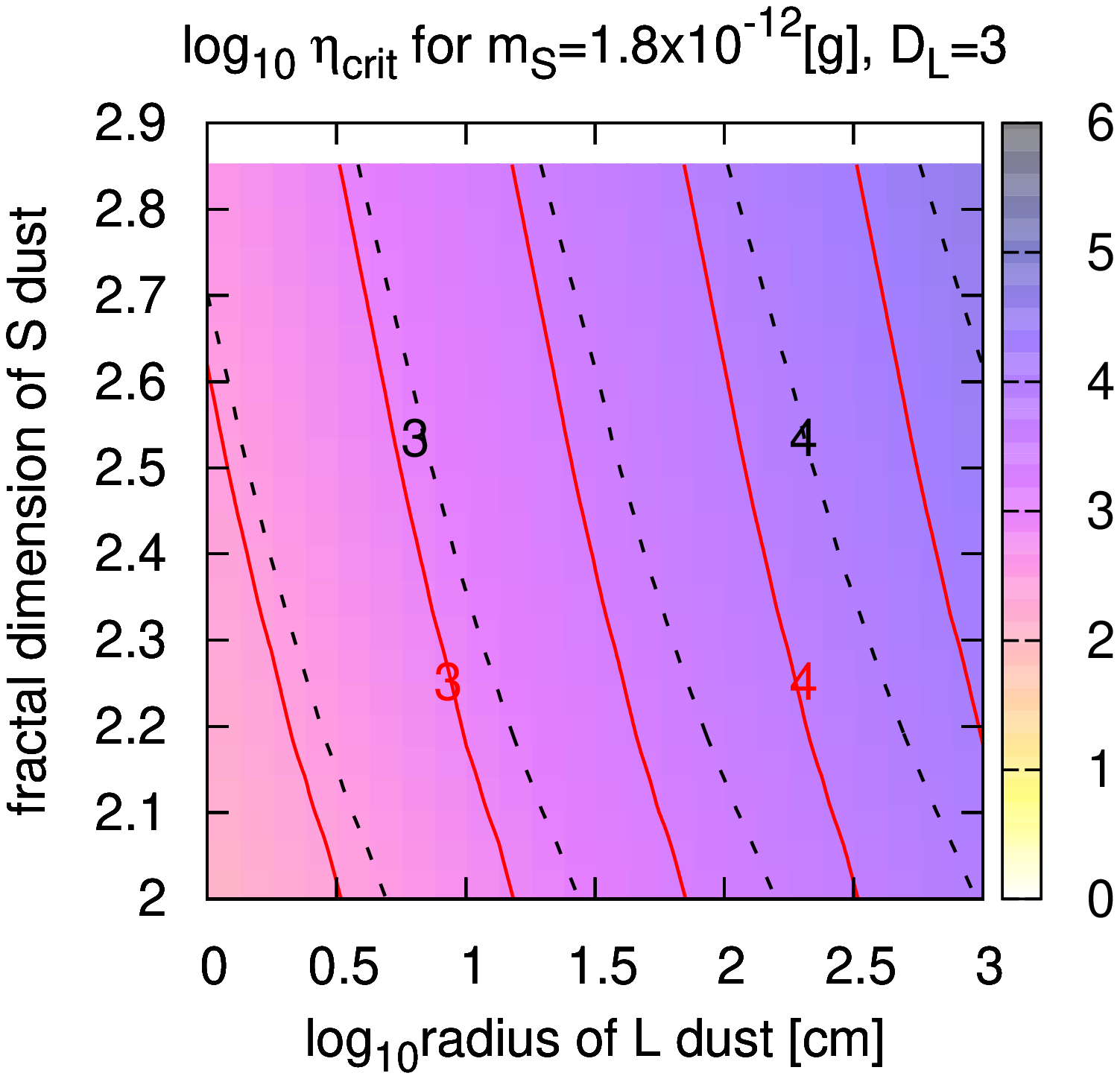} 
    \gridcut \\ \gridpush 
    \includegraphics[scale=\gridscale,angle=270]{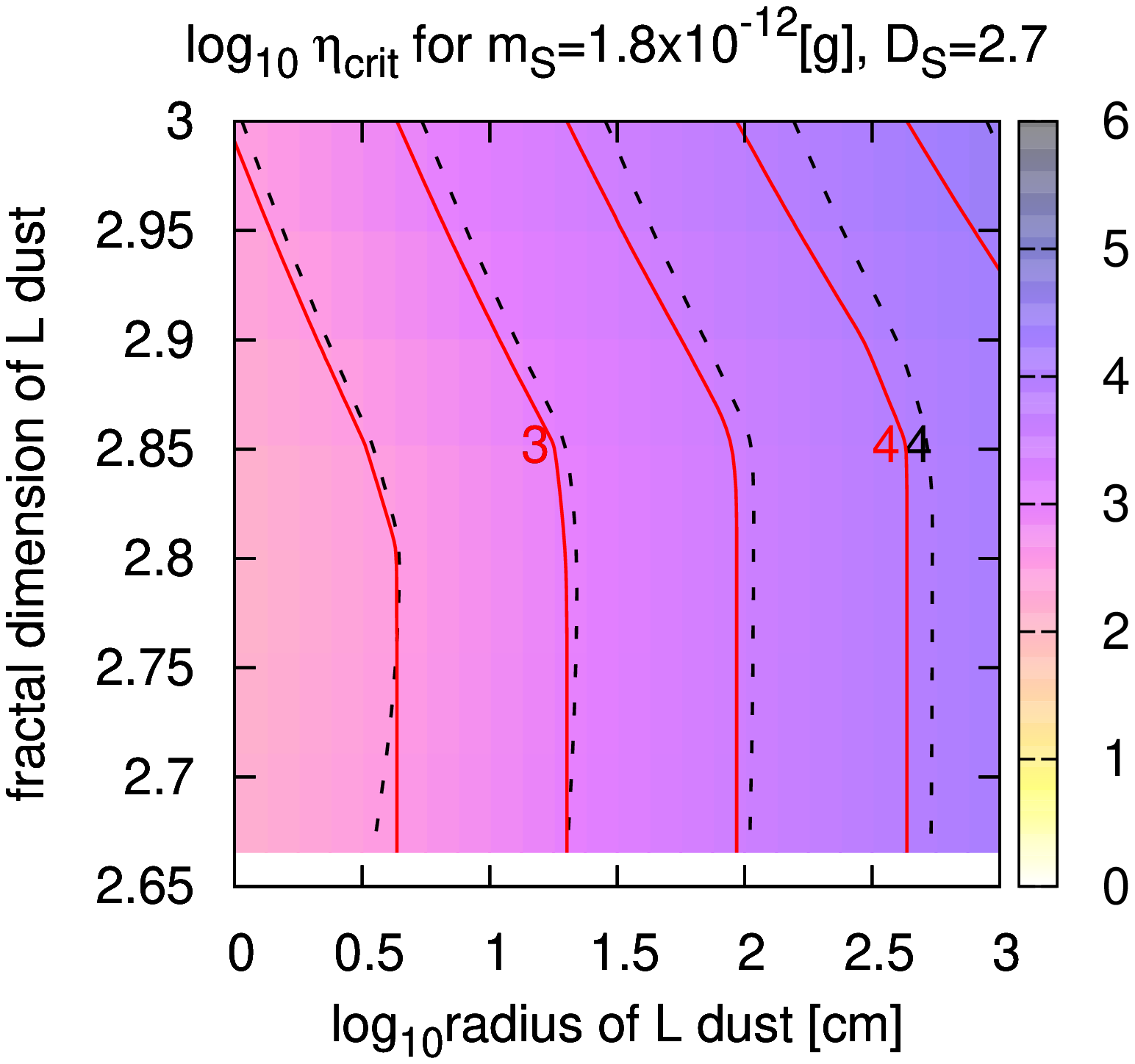} 
    \gridcut & \gridcut
    \includegraphics[scale=\gridscale,angle=270]{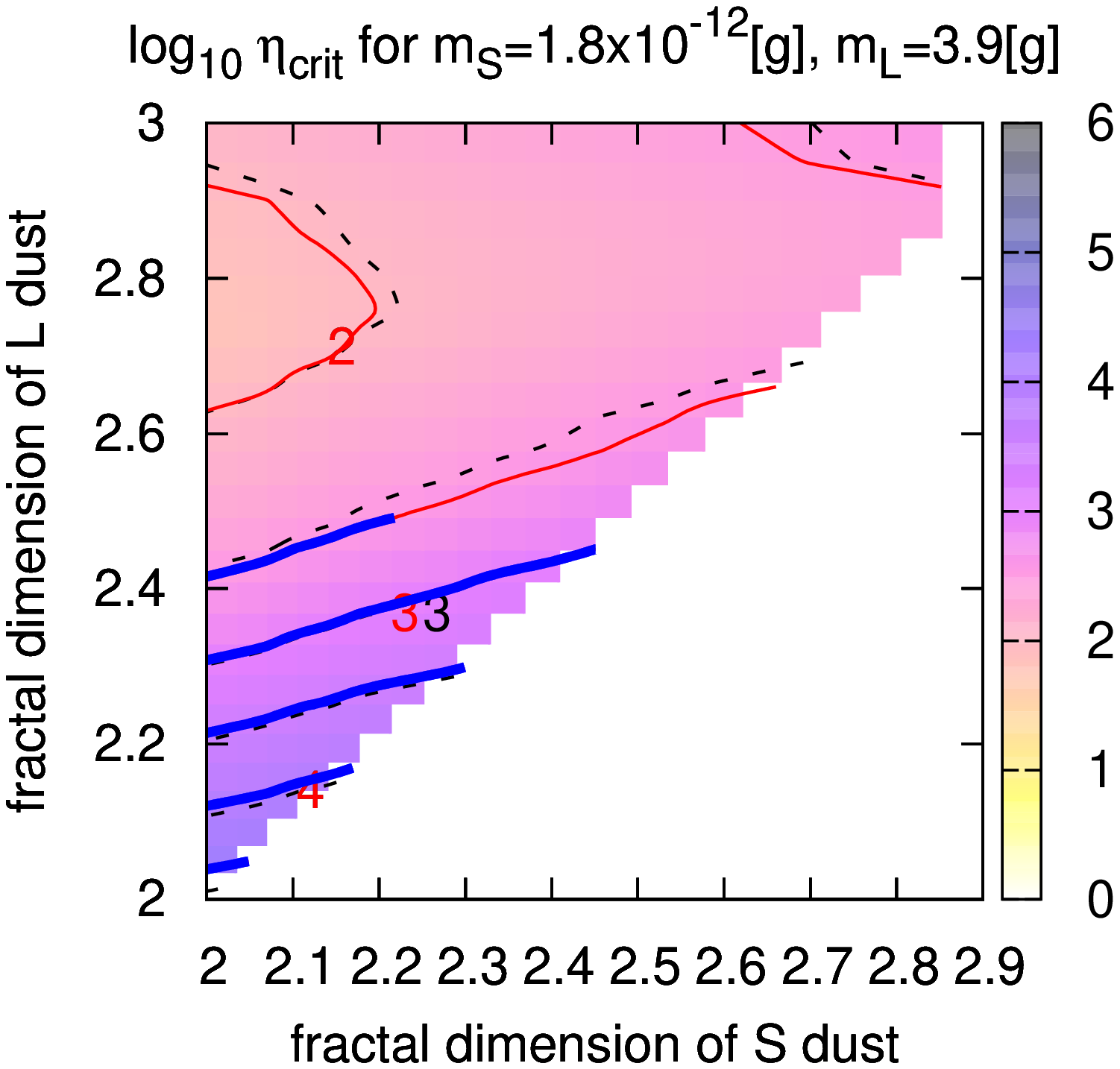} 
    \gridcut \\ 
  \end{tabular}
  \end{center}
  \caption {\small Value of $\eta_\crit$ as function of dust radius
    $r_\sm$, $r_\la$ and fractal dimension $D_\sm$, $D_\la$.  The base
    values are $r_\sm = 1.0 \times {10}^{-4}\unit{cm}$, $r_\la=1.0\unit{cm}$,
    $m_\sm = 1.9 \times {10}^{-12}\unit{g}$, $m_\la = 3.9\unit{g}$, $D_\sm =
    2.665$, and $D_\la = 3.0$.  We keep $m_\DPSub$
    constant when we vary $D_\DPSub$; we keep $D_\DPSub$ constant when
    we vary $r_\DPSub$.  Numerical results are in colour maps and black
    dashed contours; analytical values in coloured solid contours
    (c.f. \S \ref{section-analytic-combined} for the details of the
    plots.)  } \label{figure-result-hayashi}
\end{figure*}

\def\gridscale{0.4}
\def\gridpush{\hspace{-10pt}}
\def\gridpull{\vspace{-30pt}}
\def\gridcut{\hspace{-100pt}}
\begin{figure*}
  \begin{center}
  \begin{tabular}{cc} \vspace{40pt}
    \gridpush \gridpull
    \includegraphics[scale=\gridscale,angle=270]{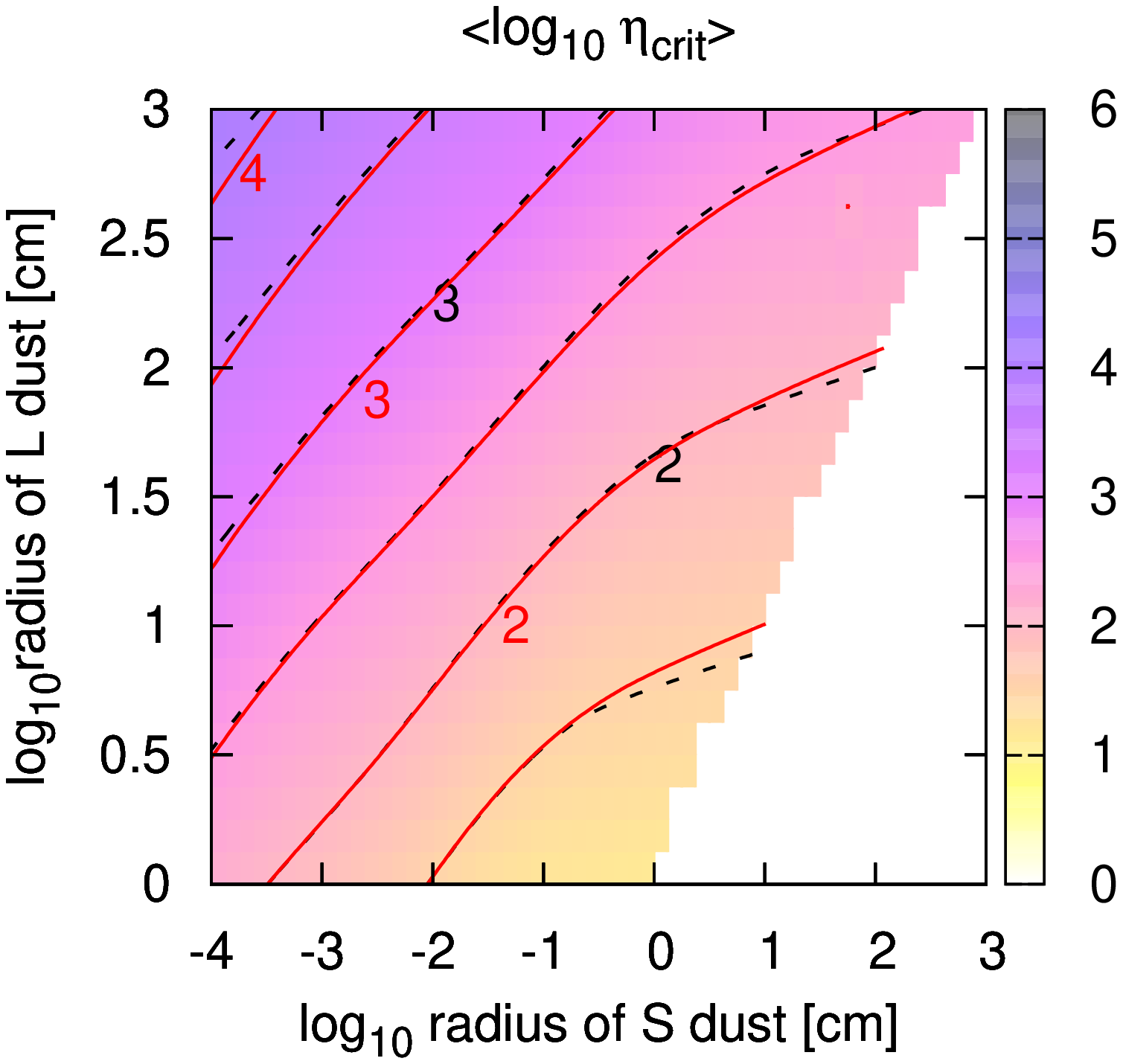} 
    \gridcut & \gridcut
    \includegraphics[scale=\gridscale,angle=270]{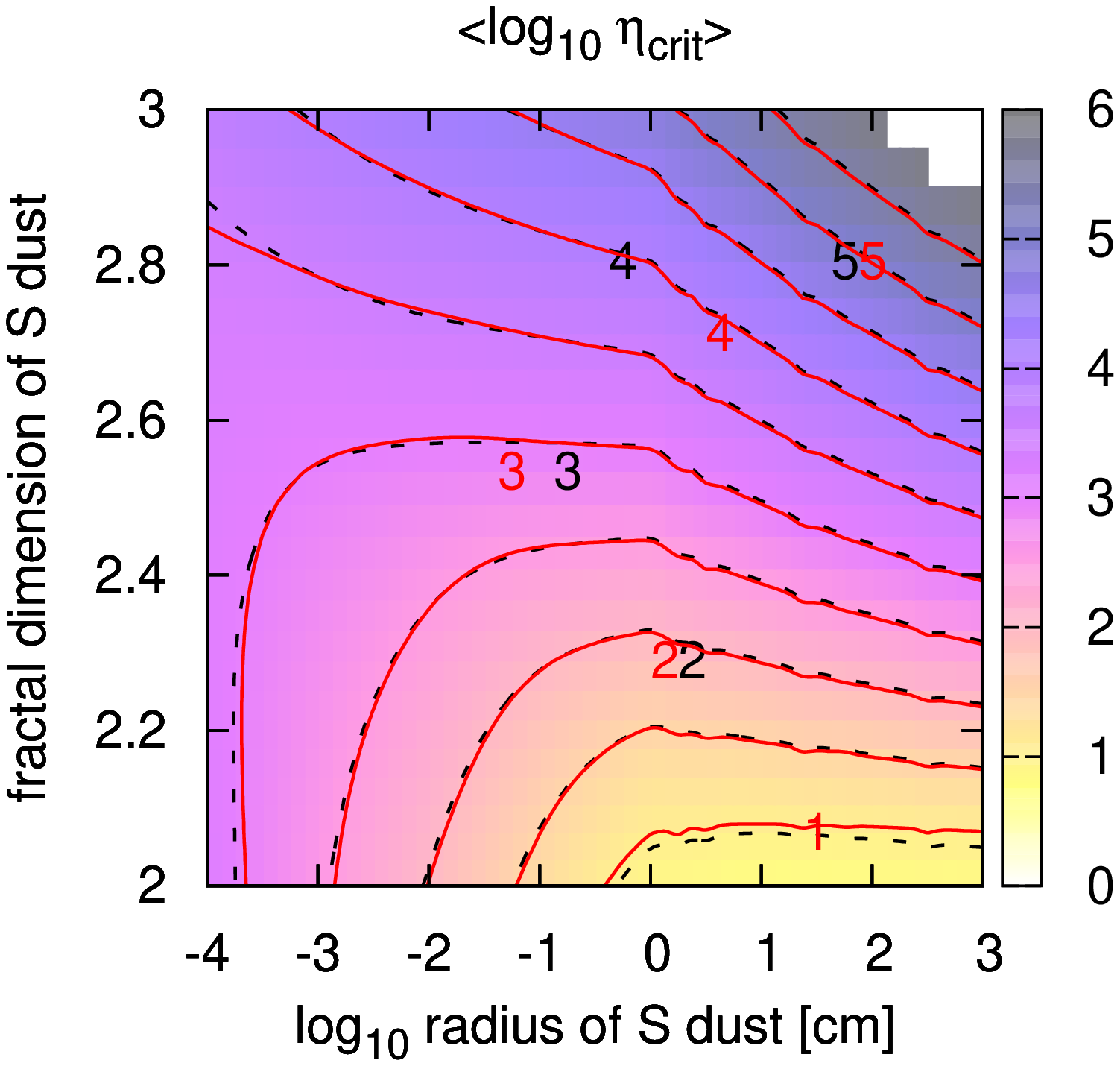} 
    \gridcut \\ \gridpush \gridpull
    \includegraphics[scale=\gridscale,angle=270]{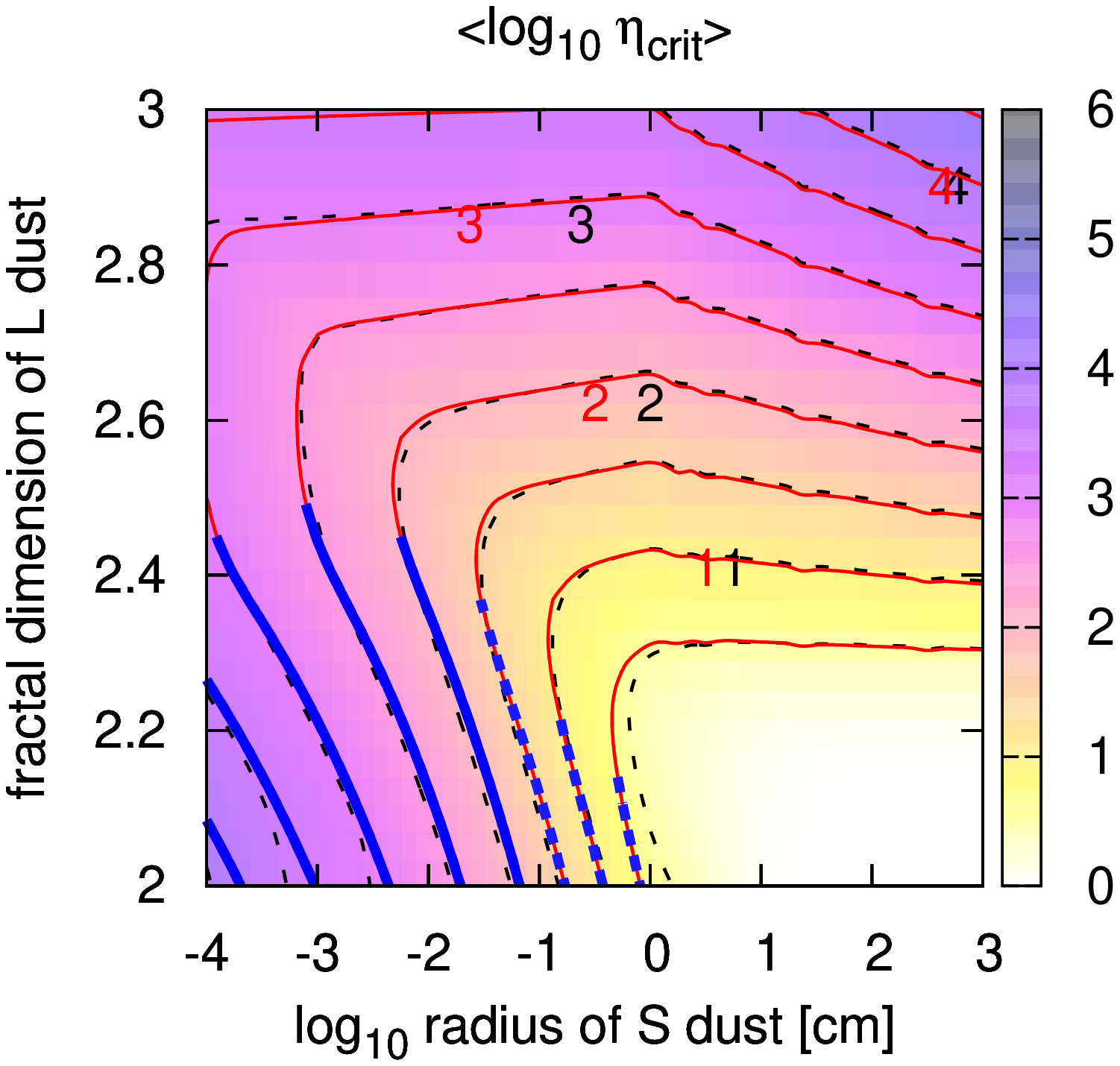} 
    \gridcut & \gridcut
    \includegraphics[scale=\gridscale,angle=270]{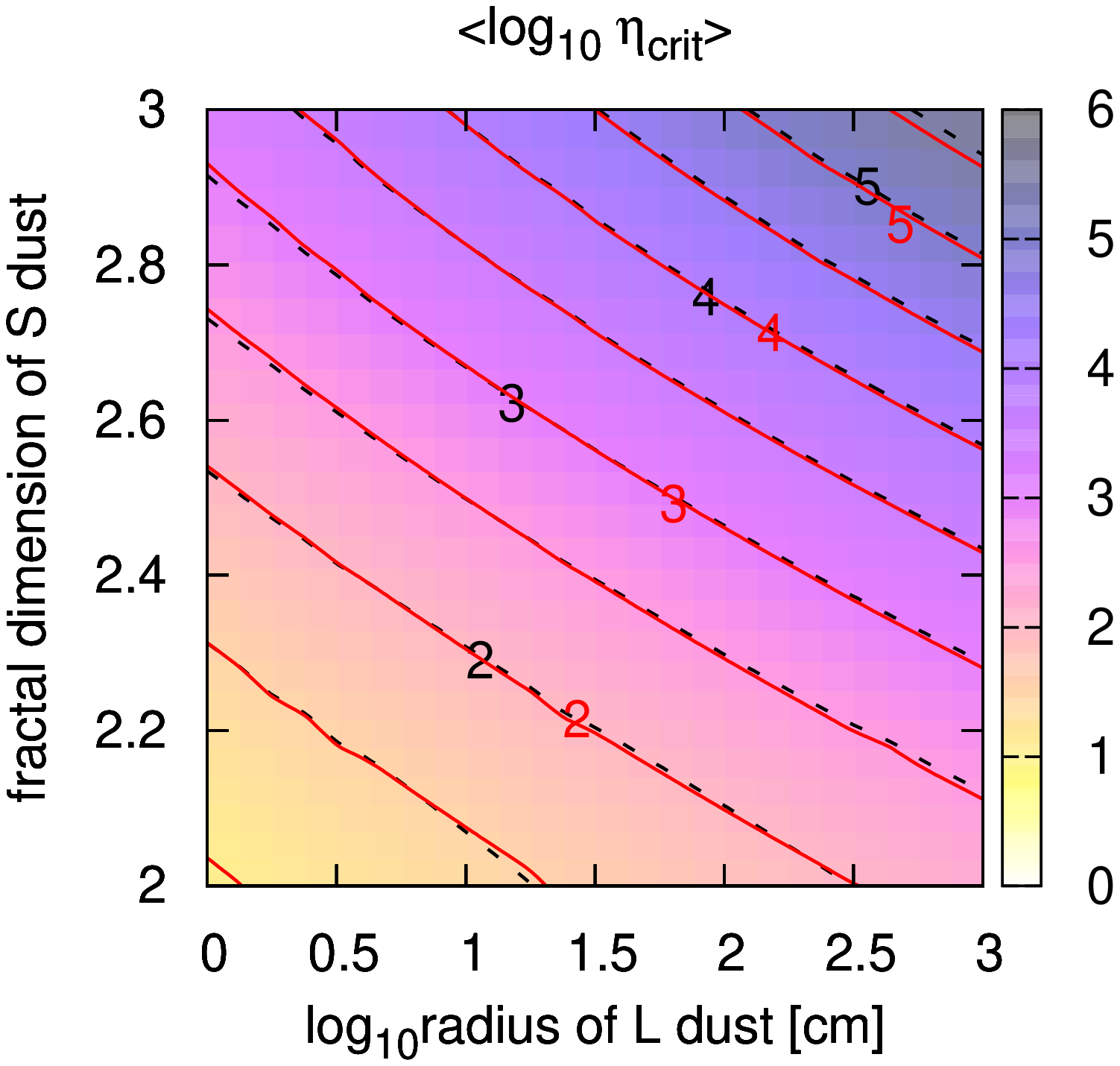} 
    \gridcut \\ \gridpush 
    \includegraphics[scale=\gridscale,angle=270]{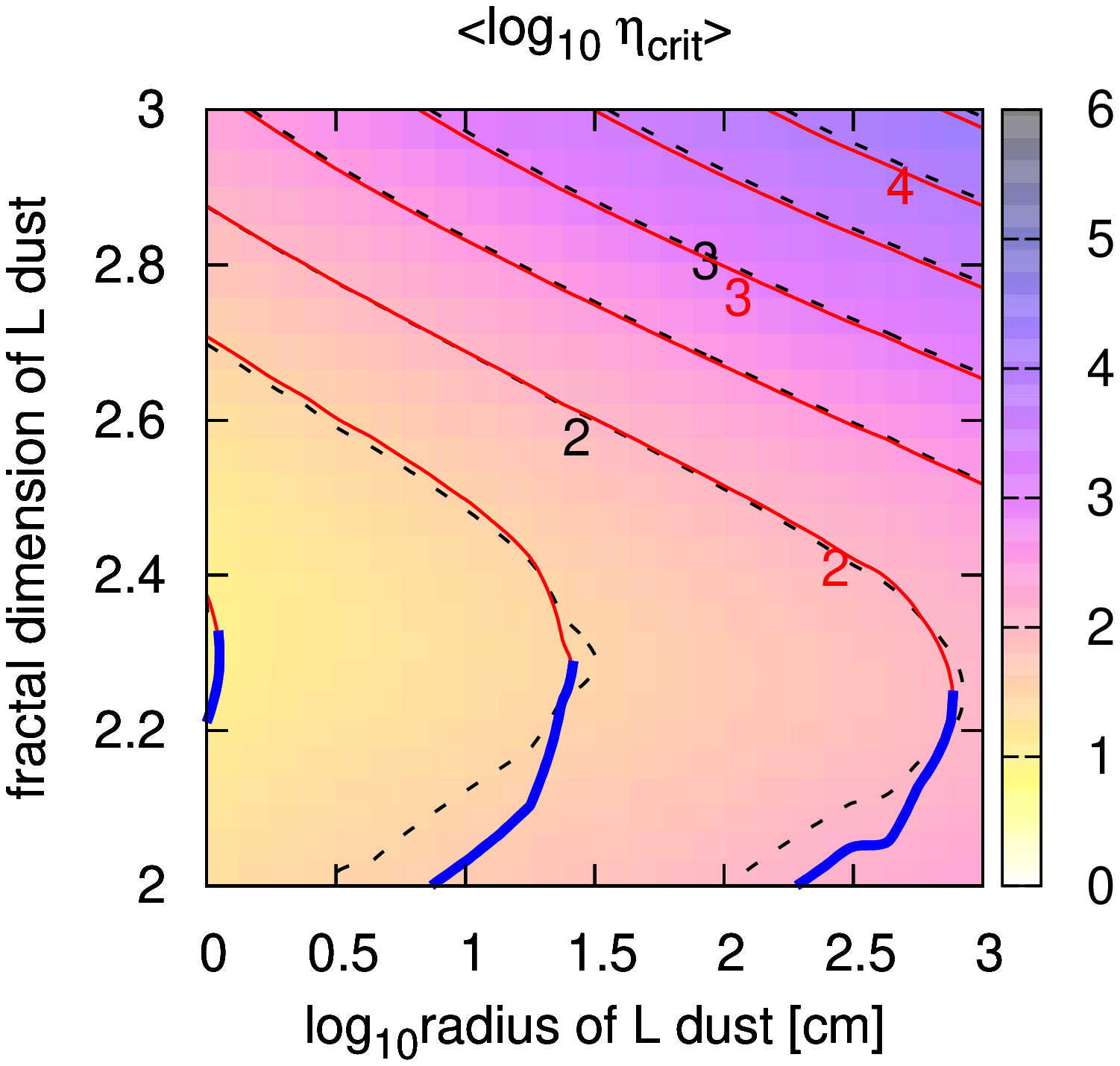} 
    \gridcut & \gridcut
    \includegraphics[scale=\gridscale,angle=270]{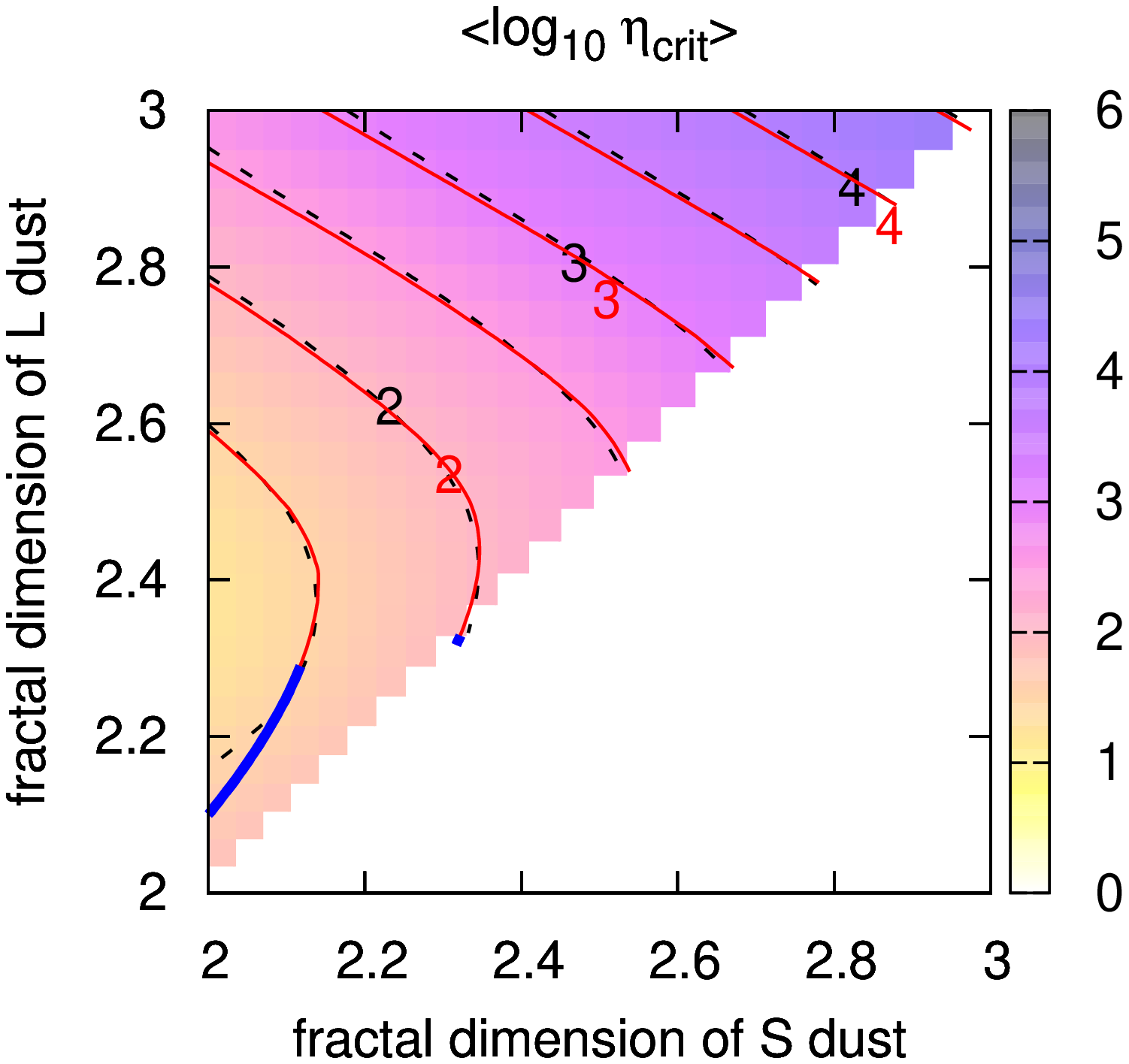} 
    \gridcut \\ 
  \end{tabular}
  \end{center}
  \caption {\small Value of $\eta_\crit$ as function of $r_\sm$,
    $r_\la$, $D_\sm$, and $D_\la$. Parameters do not appear in x-axis
    or y-axis are uniformly averaged over the parameter range we
    accept.  Numerical results are in colour maps and black dashed
    contours; analytical values in coloured solid contours (c.f. \S
    \ref{section-analytic-combined} for the details of the plots.)
  } \label{figure-result-averaged}
\end{figure*}

\def\gridscale{0.4}
\def\gridpush{\hspace{-10pt}}
\def\gridpull{\vspace{-30pt}}
\def\gridcut{\hspace{-100pt}}
\begin{figure*}
  \begin{center}
  \begin{tabular}{cc} \vspace{40pt}
    \gridpush \gridpull
    \includegraphics[scale=\gridscale,angle=270]{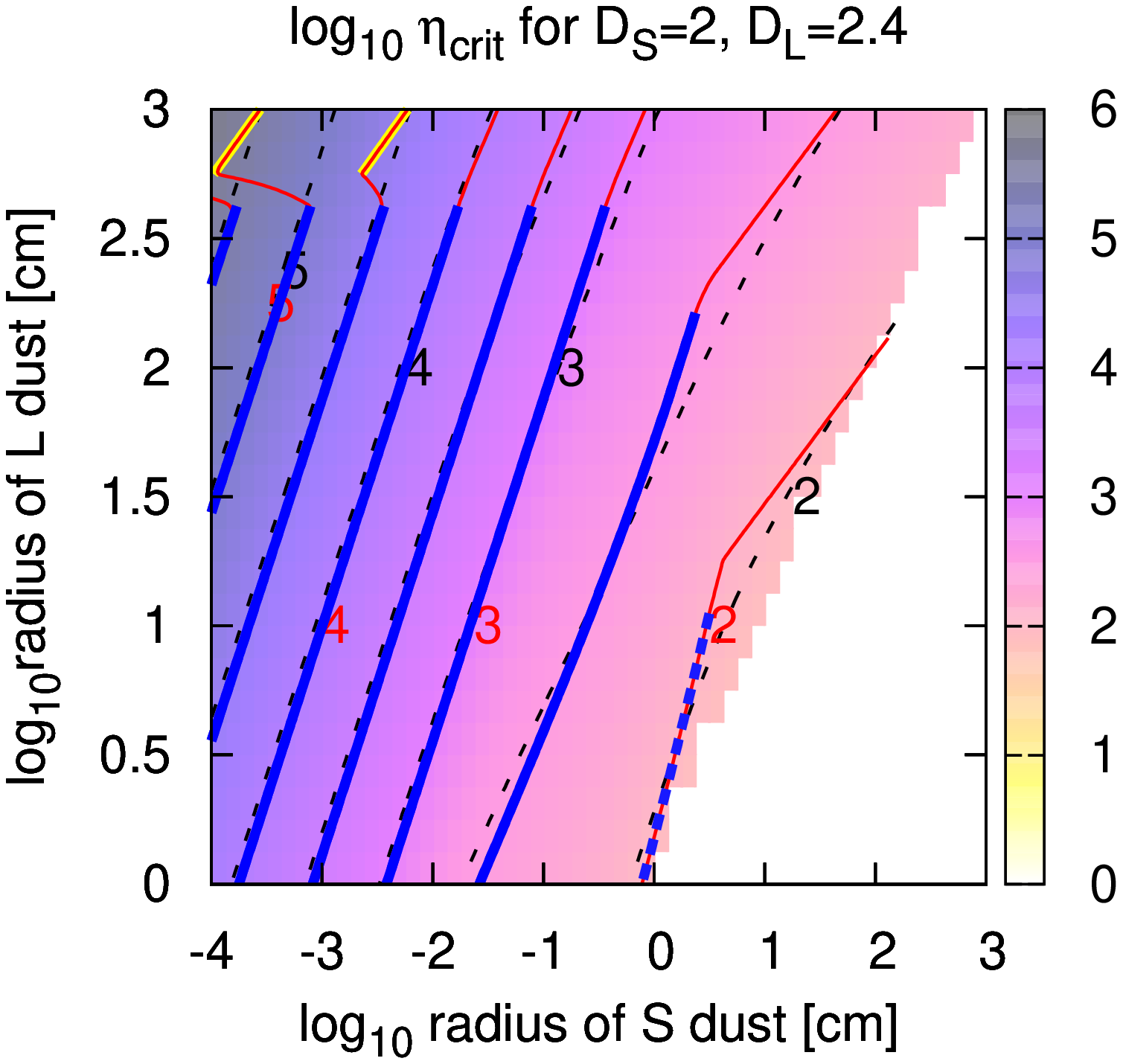} 
    \gridcut & \gridcut
    \includegraphics[scale=\gridscale,angle=270]{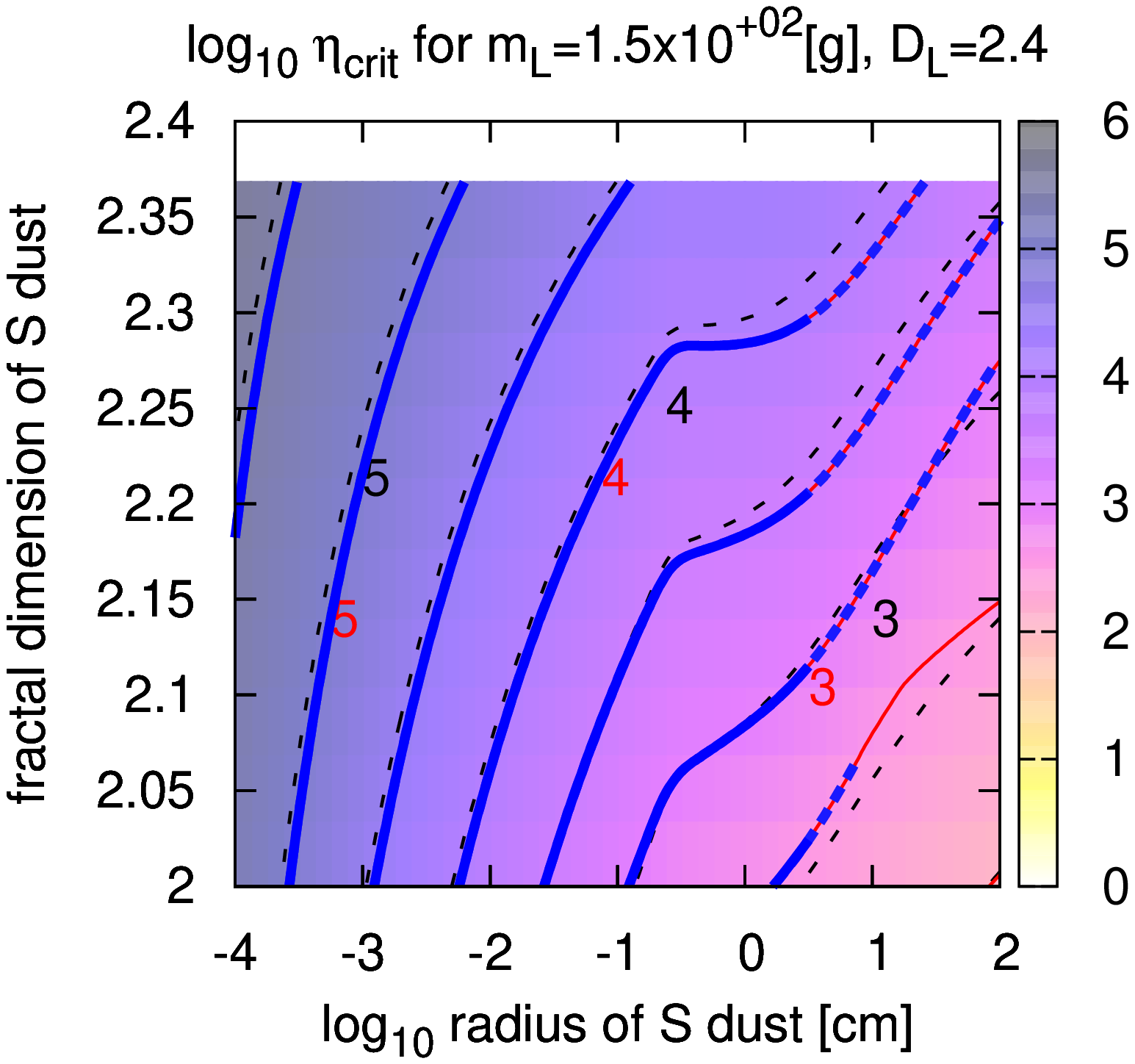} 
    \gridcut \\ \gridpush \gridpull
    \includegraphics[scale=\gridscale,angle=270]{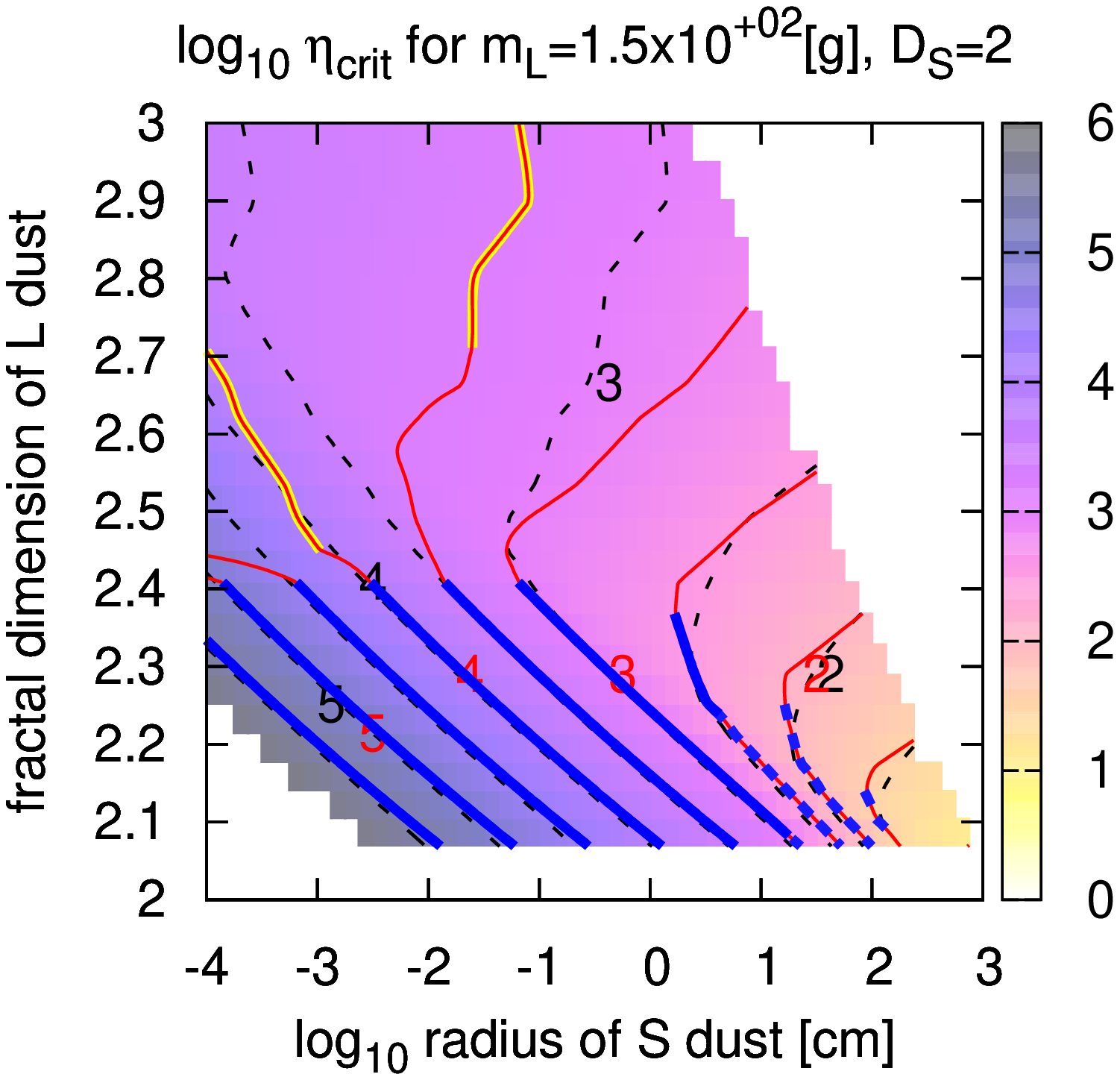} 
    \gridcut & \gridcut
    \includegraphics[scale=\gridscale,angle=270]{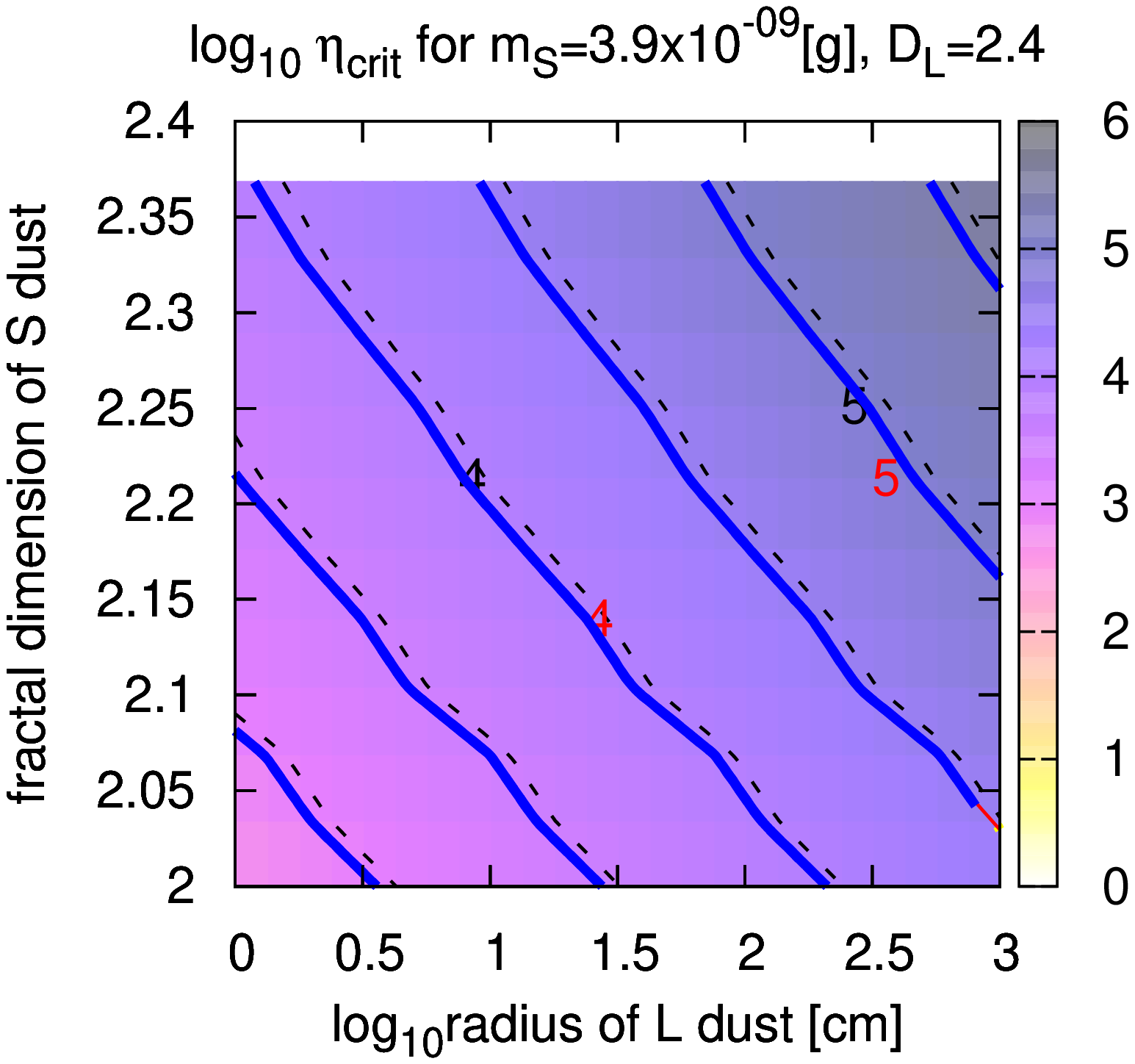} 
    \gridcut \\ \gridpush 
    \includegraphics[scale=\gridscale,angle=270]{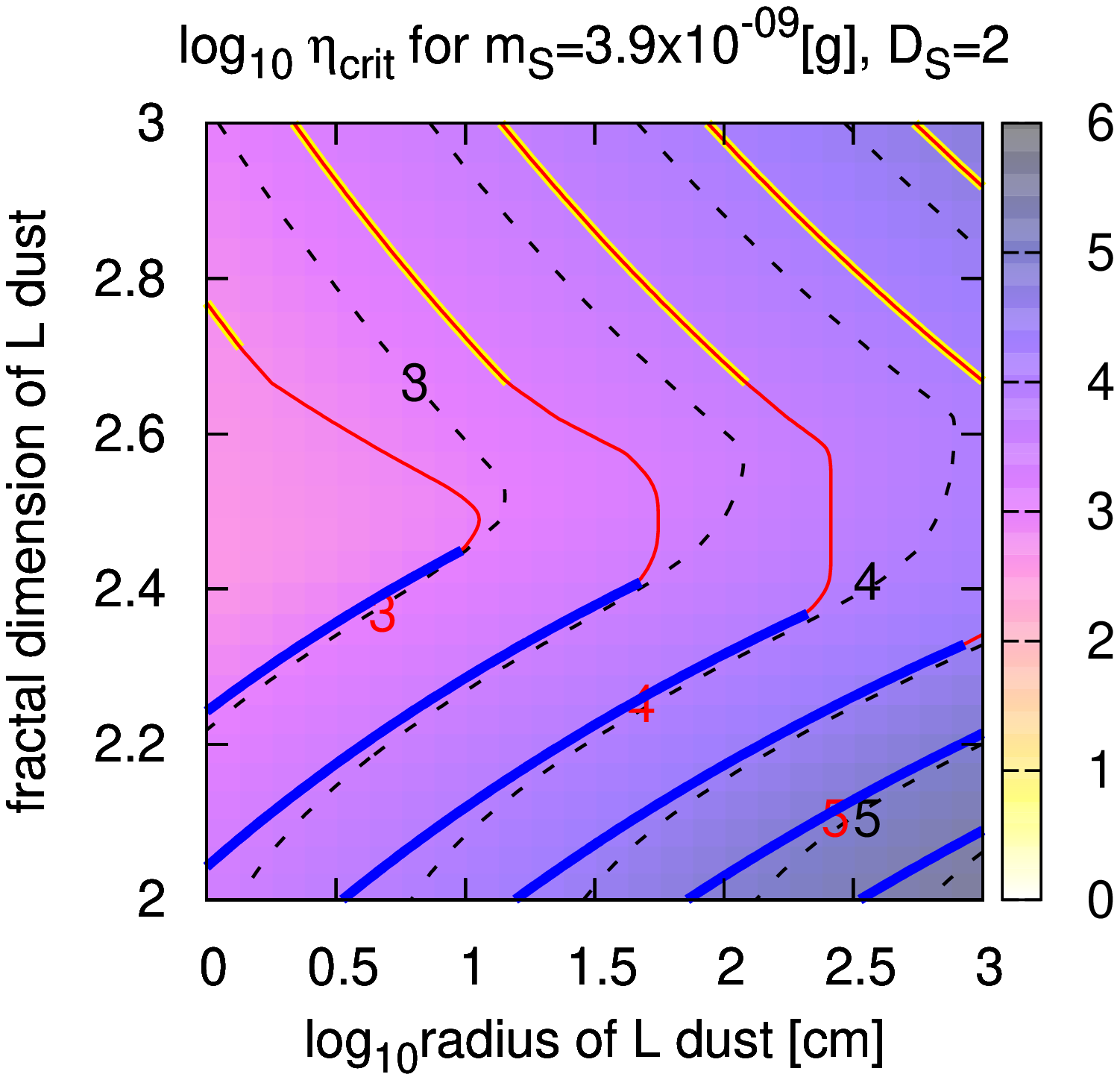} 
    \gridcut & \gridcut
    \includegraphics[scale=\gridscale,angle=270]{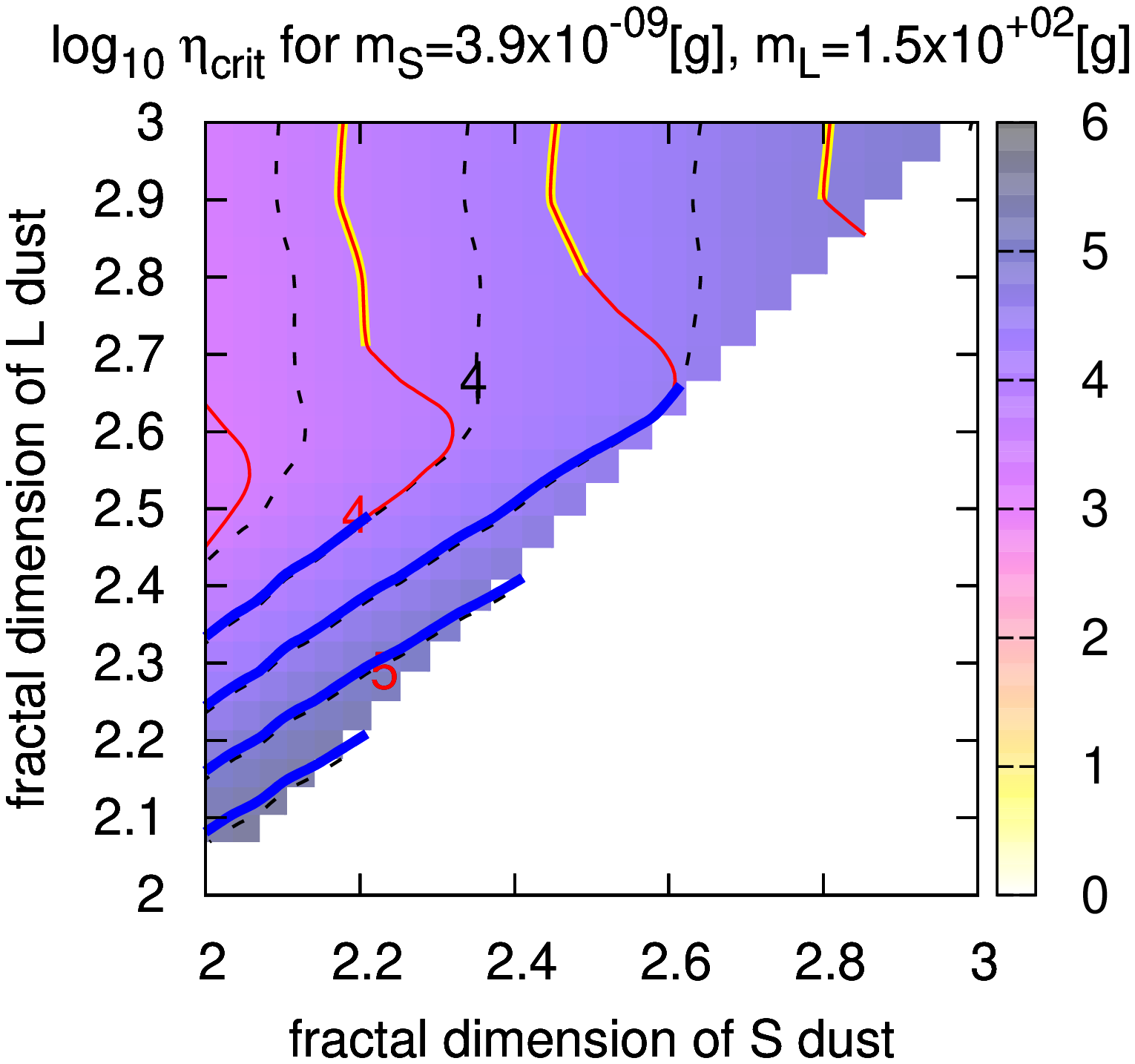} 
    \gridcut \\ 
  \end{tabular}
  \end{center}
  \caption {\small In the above plots, all the parameters but the
    charge separation efficiency is same as that of
    Fig. \ref{figure-result-tanaka}, while the the charge separation
    efficiency $\eta_\charge = 1.0 \times {10}^{-5}$ for this figure.  Numerical
    results are in colour maps and black dashed contours; analytical
    values in coloured solid contours (c.f. \S
    \ref{section-analytic-combined} for the details of the plots.)
  } \label{figure-result-hypertanaka}
\end{figure*}

\def\gridscale{0.4}
\def\gridpush{\hspace{-10pt}}
\def\gridpull{\vspace{-30pt}}
\def\gridcut{\hspace{-100pt}}
\begin{figure*}
  \begin{center}
  \begin{tabular}{cc} \vspace{40pt}
    \gridpush \gridpull
    \includegraphics[scale=\gridscale,angle=270]{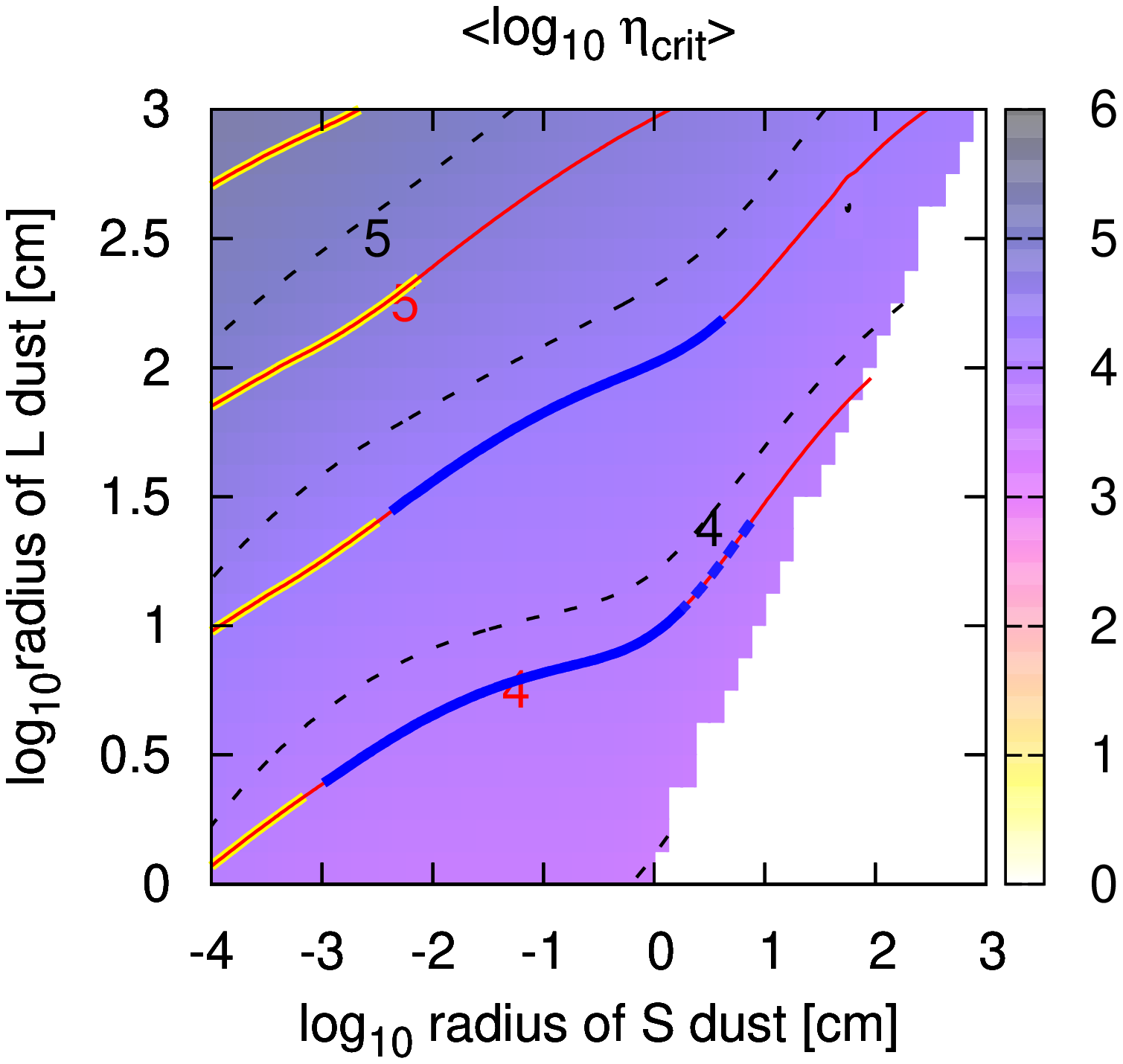} 
    \gridcut & \gridcut
    \includegraphics[scale=\gridscale,angle=270]{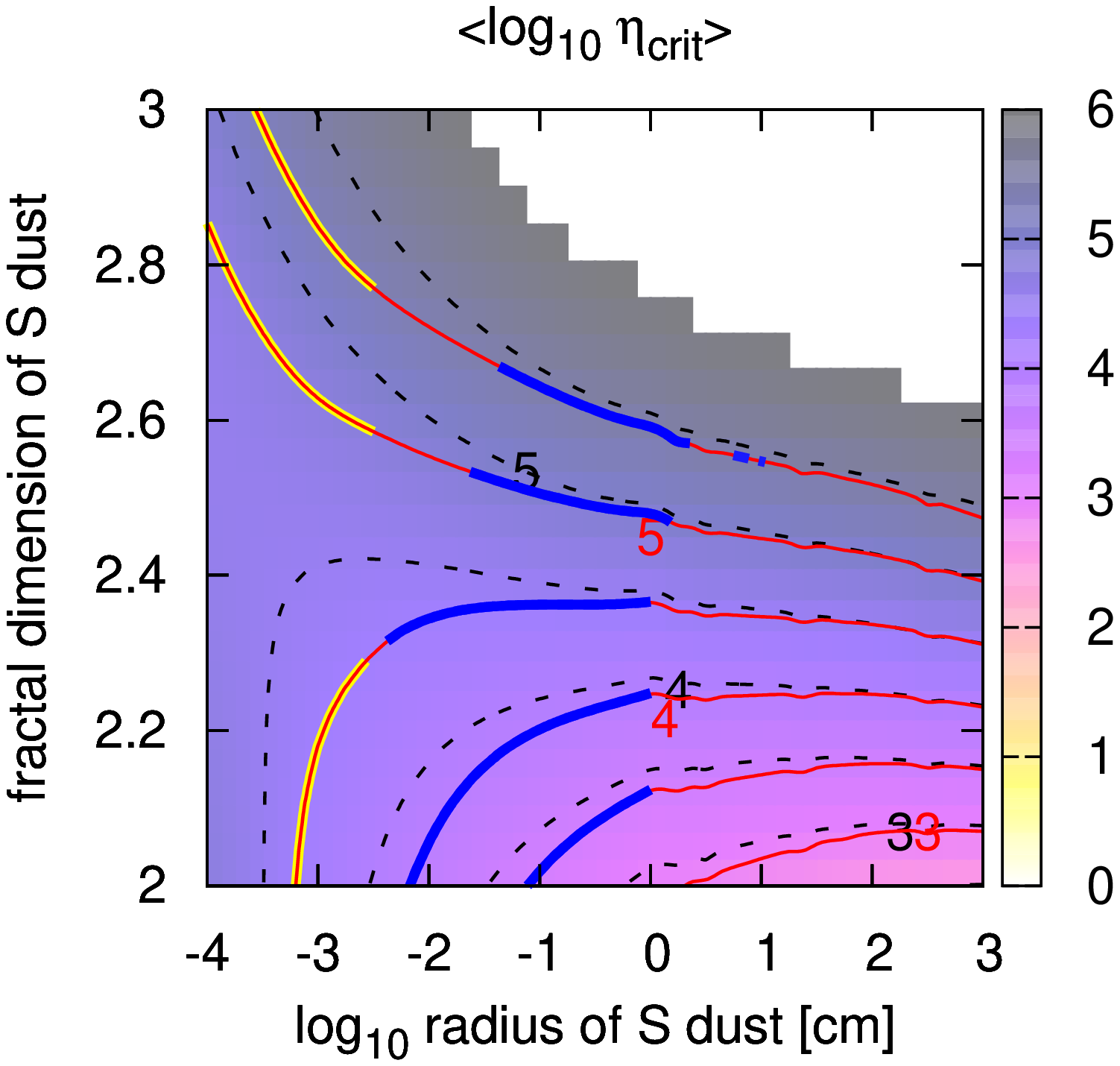} 
    \gridcut \\ \gridpush \gridpull
    \includegraphics[scale=\gridscale,angle=270]{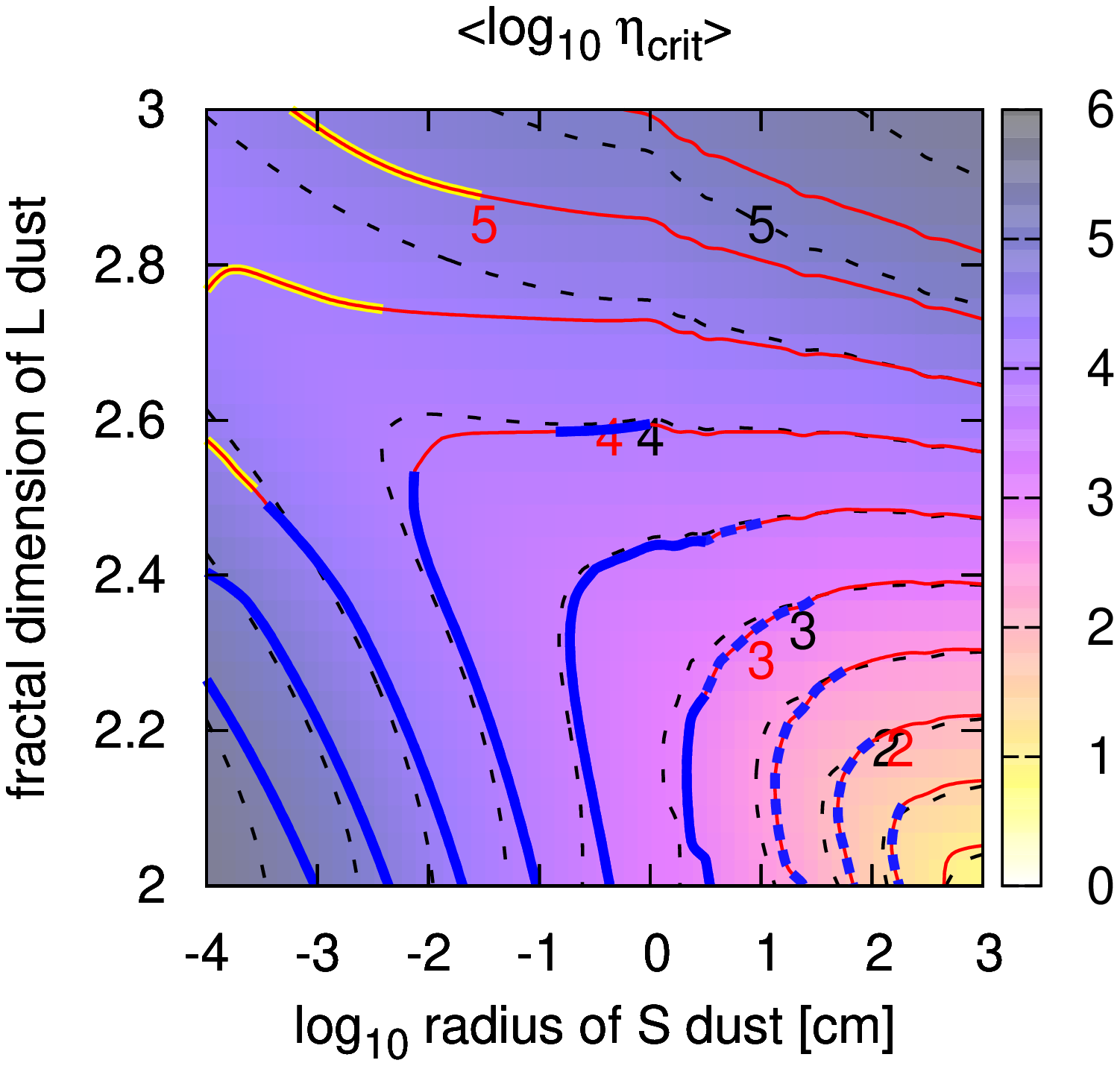} 
    \gridcut & \gridcut
    \includegraphics[scale=\gridscale,angle=270]{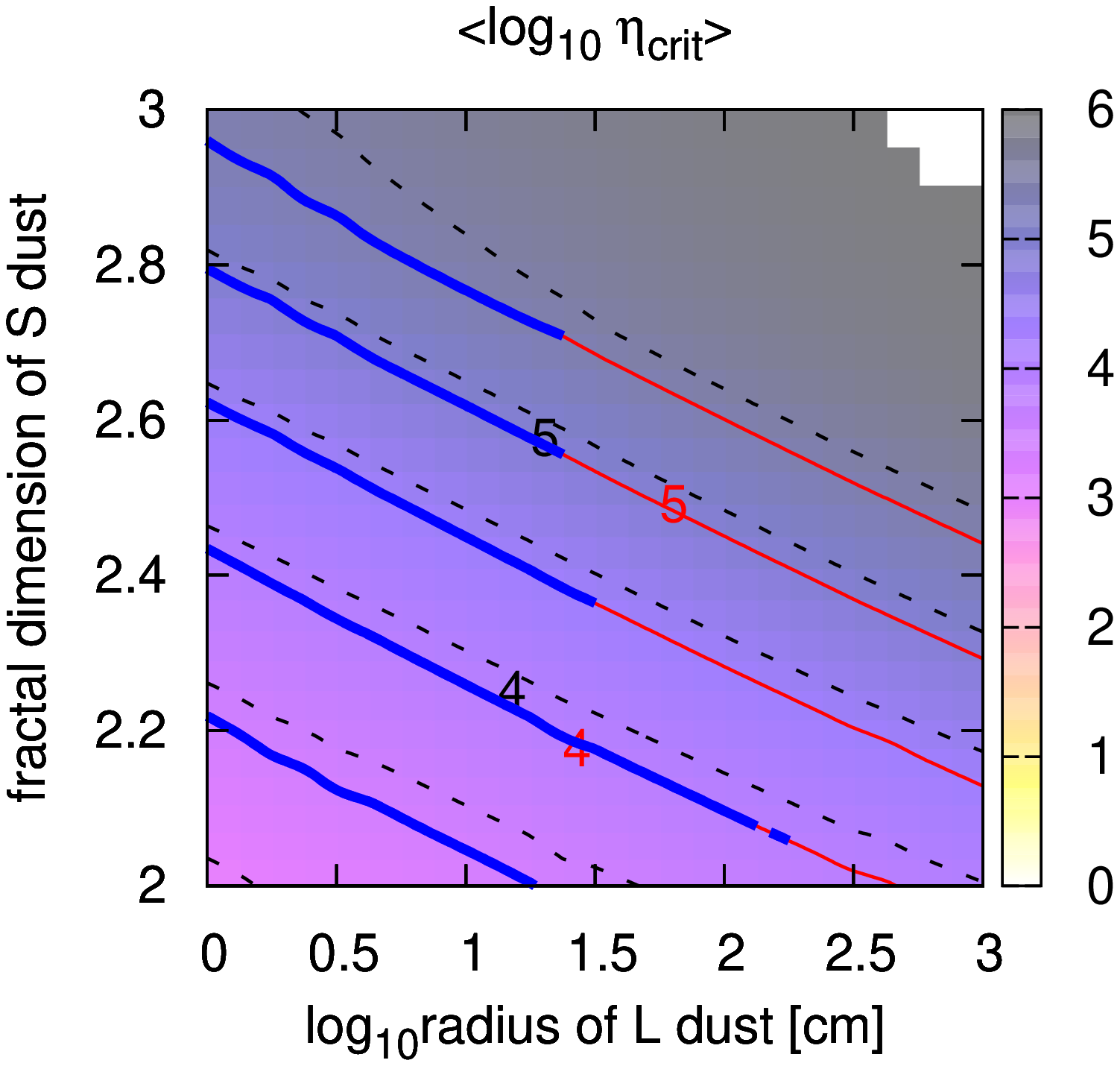} 
    \gridcut \\ \gridpush 
    \includegraphics[scale=\gridscale,angle=270]{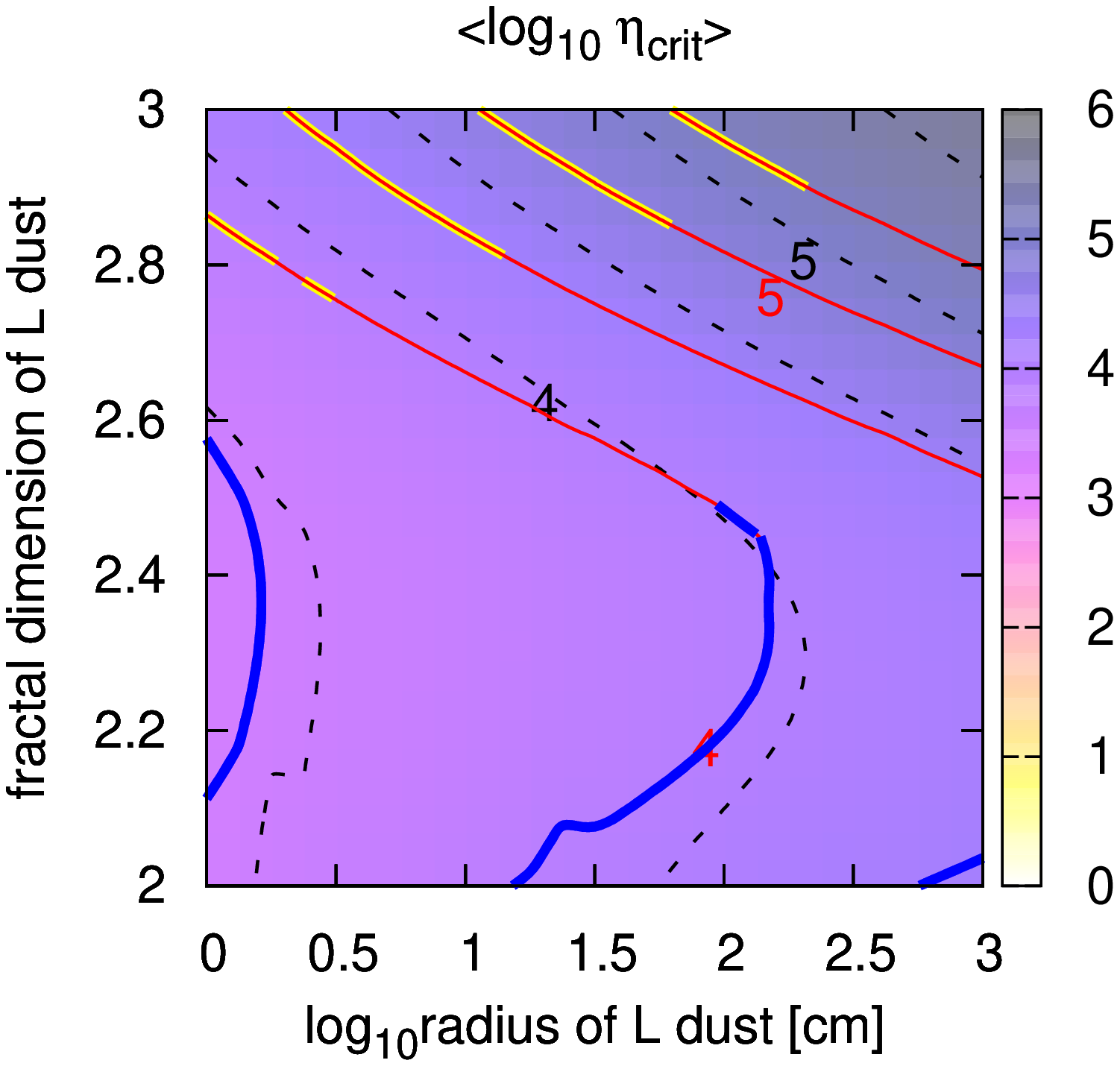} 
    \gridcut & \gridcut
    \includegraphics[scale=\gridscale,angle=270]{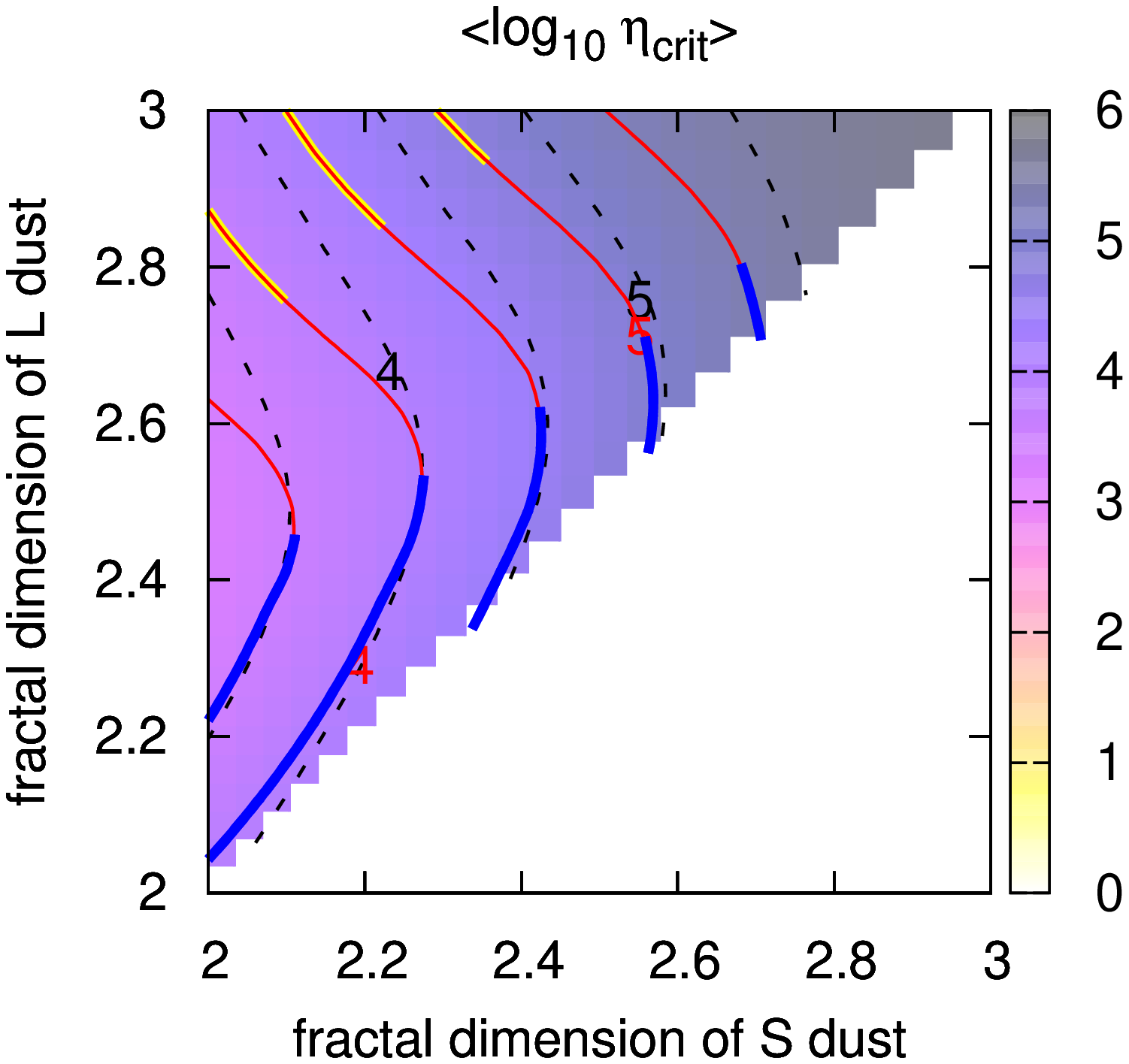} 
    \gridcut \\ 
  \end{tabular}
  \end{center}
  \caption {\small Value of $\eta_\crit$ as function of $r_\sm$,
    $r_\la$, $D_\sm$, and $D_\la$. Parameters do not appear in x-axis
    or y-axis are uniformly averaged over the parameter range as in Fig. \ref{figure-result-averaged},
    and we assume $\eta_\charge = 1.0 \times {10}^{-5}$ as in
    Fig. \ref{figure-result-hypertanaka}.
    Numerical results are in colour maps and black dashed
    contours; analytical values in coloured solid contours (c.f. \S
    \ref{section-analytic-combined} for the details of the plots.)
  } \label{figure-result-averaged-pessimistic}
\end{figure*}

\newpage

\appendix

\def\DP{{\mathbi{J}}}
\def\DPSub{{ \textbf{\tiny \em {J}}}}
\def\dDP{d\mathbi {J}}

\section[]{Cationic dust and anionic dust} \label{cationic-anionic-definition}

In this section we justify the two-dust picture introduced in
\S \ref{two-component-model}.
Protoplanetary discs consist of dust with various parameter $\DP$.
The parameter vector $\DP$ may include, but is
not limited to, dust radius, porosity, material and surface chemical potential.
We classify these dust into two groups according to their electric tendency; 
one is cationic dust
who receive positive charge through dust-dust collision and the other
is anionic dust 
who receive negative charge.
In this section we give the precise definition of cationic and anionic dust.

Let $n_\DPSub$ indicate the number density of the dust with parameter
$\DP$. Let $\sigma_{\DPSub', \DPSub}$, $\Delta v_{\DPSub',\DPSub}$, 
$\Delta q_{\DPSub', \DPSub}$
be collisional cross section, mean relative velocity, and mean amount of
charge that moves from dust $\DP'$ to dust $\DP$ in a collision, respectively.
Then we can calculate $\dot q_\DPSub$, charge received by dust $\DP$ per
unit time, by
\begin{eqnarray}
  \dot q_\DPSub=\sum_{\DP'} \Delta q_{\DPSub ', \DPSub} n_{\DPSub '} \sigma_{\DPSub', \DPSub}
  \Delta v_{\DPSub',\DPSub} \label{q-dot}.
\end{eqnarray}

We define cationic dust and anionic dust as $C \equiv \{\DP | \dot q_\DPSub >
0\}$ and $A \equiv \{\DP | \dot q_\DPSub < 0\}$. Cationic dust receive net
positive charge in dust-dust collision and tend to be cationic, while anionic
dust tend to be anionic. 

We assume the average dust distribution $n^{(0)}_\DPSub$ as that of
MMSN model. We also assume that at some local condensation regions,
dust number density is multiplied by carrying in dust from other
portions of the disk.  For simplicity we assume that the relative
number density $\eta$ is independent of dust parameter $\DPSub$ so
that $n_\DPSub=\eta n^{(0)}_\DPSub$ (e.g. this is the case when
collisional cascade equilibrium is faster than the migration).  We
define $\sigma_{\DPSub', \DPSub}$ and $\Delta q_{\DPSub', \DPSub}$ as
the value for neutral dust.  As the dust acquires charge, the amount
of charge exchanged in a single collision becomes smaller due to the
exchange of the charge they already have; we treat this deviation from
neutral dust as separate `neutralization' channel.  With these
assumptions, the sign of $\dot q_\DPSub$ (\ref{q-dot}) do not depend
on $\eta$, and the term `cationic dust' and `anionic dust' is well
defined independent of dust number density $\eta$.

Now we can simplify the problem by treat cationic and anionic dust as if they
are two discrete kinds of dust. 
Therefore we define the representative variables for cationic and anionic dust
as follows.
\begin{eqnarray}
  n_\C &\eq& \sum_{\DPSub \in C}  \ n_{\DPSub} 
  = \eta \sum_{\DPSub \in C}  \ n^{\left(0\right)}_{\DPSub}  \label{definition-nC} \\
  n_\A &\eq& \sum_{\DPSub \in A}  \ n_{\DPSub} 
  = \eta \sum_{\DPSub \in A}  \ n^{\left(0\right)}_{\DPSub}   \label{definition-nA}  \\
  v_{\A,\C} &\eq& \frac
  {\sum_{\DPSub \in A, \DPSub' \in C}  \  n_{\DPSub}  n_{\DPSub'} v_{\DPSub,\DPSub'}}
  {n_\C n_\A} \\
  \sigma_{\A,\C} &\eq& \frac{
    \sum_{\DPSub \in A, \DPSub' \in C}  \ 
    \sigma_{\DPSub,\DPSub'}  \, n_{\DPSub}  \, n_{\DPSub'} \, v_{\DPSub,\DPSub'}
  }{
    \sum_{\DPSub \in A, \DPSub' \in C}  \  n_{\DPSub}  n_{\DPSub'} v_{\DPSub,\DPSub'}
  } \\
    \Delta q_{\A,\C} &\eq&
    \frac{
    \sum_{\DPSub \in A, \DPSub' \in C}  \ \Delta q_{\DPSub,\DPSub'}
    \sigma_{\DPSub,\DPSub'}  \, n_{\DPSub}  \, n_{\DPSub'} \, v_{\DPSub,\DPSub'}    
  }{
    \sum_{\DPSub \in C,\DPSub' \in A}  \ 
    \sigma_{\DPSub,\DPSub'}  \, n_{\DPSub}  \, n_{\DPSub'} \, v_{\DPSub,\DPSub'}
  }.
\end{eqnarray}

\def\DP{{\mathbf{I}}}
\def\DPSub{{ \textbf{\tiny  {I}}}}
\def\dDP{d\mathbf {I}}

\section[]{Simulations} \label{section-simulations}

In this section, we briefly describe our numerical methods. 
We need to solve the equilibrium equations
(\ref{specific-equilibrium-begin}-\ref{specific-equilibrium-end}),
for various environmental parameters.
Especially we vary $\eta$ for each set of other parameters.
Then we know the minimum $\eta$ that satisfies the electric discharge condition
(\ref{equation-macroscopic-discharge}), for the each set of other parameters.

This kind of problem, a massive parameter parallelism,
is typically suitable for massively parallel computing hardware
\citep[e.g.][]{2009NewA...14..406F},
such as general purpose graphic processors (GPGPUs)
or GRAPE-DR \citep{2008IAUS..246..457M}.
We describe the CPU and GPGPU based programmes we used 
in this research to solve equations
(\ref{specific-eq-dynamic-begin}-\ref{specific-eq-dynamic-end}) 
in this section.

\subsection{Direct integral solver} \label{direct-integral-solver} The most
straightforward means of finding the equilibrium solutions
(\ref{specific-eq-dynamic-begin}-\ref{specific-eq-dynamic-end}) is to directly
integrate the dynamic equations
(\ref{specific-eq-dynamic-begin}-\ref{specific-eq-dynamic-end}).
Nevertheless, constant-time-step direct integral cannot solve
(\ref{specific-eq-dynamic-begin}-\ref{specific-eq-dynamic-end}) correctly for
some of the parameter range. This is because the current densities $Q_\DPSub$
differ many orders of magnitude for such parameters. We must choose the
integration time-step $dt_{i}$ carefully. This leads us to the use of a
adaptive time step.

The adaptive-time-step direct-integral solver follows the dynamic equations
(\ref{specific-eq-dynamic-begin}-\ref{specific-eq-dynamic-end}) in terms of 
discretized time $t_{i}$ where time $t_{i}$ is incremented by
 dynamic time-step $dt_{i}$:
\begin{eqnarray}
  Q_{\DPSub,i+1} &\eq& Q_{\DPSub, i} + \sum_{\DPSub'} J_{\DPSub',\DPSub} dt_{i} ,\\
  t_{i+1} &\eq& t_{i} + dt_{i}.
\end{eqnarray}
Our choice of the dynamical time-step $dt_{i}$ is as follows:
\begin{eqnarray}
  ratio\left(\DP\right) &\eq& \abs{\frac{Q_{\DPSub,i} - Q_{\DPSub,i-1}}{Q_{\DPSub,i}} }  \\
  dt_{i} &\eq& \frac{1.0 \times {10}^{-4} \cdot dt_{i-1}}{\max_{\DPSub} \left(ratio\left(\DP\right)\right)}
  \label{dynamical-timestep-equation}.
\end{eqnarray}

The direct integral solver is reliable, in sense that it is 
less prone to implementation mistakes
because it 
almost straightforwardly reflects the equations
(\ref{specific-eq-dynamic-begin}-\ref{specific-eq-dynamic-end}),
and that out of possible many equilibrium solutions
(\ref{specific-eq-dynamic-begin}-\ref{specific-eq-dynamic-end})
the solver will always find the desired equilibrium.

However, as $\eta$ become much larger or much smaller than unity,
we have found that charge distribution in the system get unbalanced.
As we try to update the species $\DP$ with little charge but large current,
the dynamic time-step $dt_{i}$ (\ref{dynamical-timestep-equation}) become 
the time-scale the equilibrium is reached,
and the simulations becomes time consuming.
Use of higher-order integral schemes are futile because we cannot take
time-step much larger than $dt_{i}$. 
In addition to that, floating point numbers mainly available on GPU
are single-precision floating point numbers. Computation of double-precision
floating point numbers are either not supported or order of magnitude slower on
common GPU.

\subsection{Binary search solver} \label{binary-search-solver}

So, we need to find an alternative method to solve the
equilibrium equations
(\ref{specific-equilibrium-begin}-\ref{specific-equilibrium-end})
without directly integrating the dynamic equation, 
avoiding the addition between numbers of different magnitude
as long as possible.

Binary search is a common method for finding zero point of a one-parameter function
$f$; solving equation $f(x) = 0$ for $x$.
To solve the system of equations
(\ref{specific-equilibrium-begin}-\ref{specific-equilibrium-end}),
we divide the problem into set of one-parameter problems,
and conquer by recursive use of binary search.
In doing so, we have to be careful in choosing which of equations
(\ref{specific-equilibrium-begin}-\ref{specific-equilibrium-end}) we use
to find zero point of which freedom $Q_\DPSub$. Wrong choice leads to
wrong result.

First, $Q_i$ and $Q_e$ can be analytically expressed in terms of $Q_\sm$ and
$Q_\la$ as follows:
\begin{eqnarray}
  Q_i &\eq& \frac{e \zeta n_g}
  {\left(n_\la \sigma_{\la,i} + n_\sm \sigma_{\sm, i} \right)v_i } \label{analytic-Q-i} , \\
  Q_e &\eq& \frac{-e \zeta n_g}
  {\left(n_\la \sigma_{\la, e} + n_\sm \sigma_{\sm, e} \right)v_e } \label{analytic-Q-e} . 
\end{eqnarray}
Where we made abbreviations 
\begin{eqnarray}
  \sigma_{\la,i} &\eq& \sigma_\cou\left(q_\la, e, r_\la, k_BT\right) ,\\
  \sigma_{\la,e} &\eq& \sigma_\cou\left(q_\la, -e, r_\la, k_BT\right)
\end{eqnarray}
and so on.  Further eliminations of freedoms is possible but complicated,
because of complicated and sign-sensitive form of the Coulomb cross sections
(\ref{sigma_coulomb_focusing+}-\ref{sigma_coulomb_focusing-}).  Instead we
resort to numerical methods to find out the equilibrium solution $Q_\sm$ and
$Q_\la$ for each $\eta$, and then find $\eta_\crit$, the minimum $\eta$ that
satisfies the electric discharge condition
(\ref{equation-macroscopic-discharge}).

We now describe how to solve the system of equations
(\ref{specific-kirch-L}),(\ref{specific-kirch-S}),(\ref{specific-charge-neutrality}),
and to find $\eta_\crit$,
provided that for any one-parameter $f$ 
we can solve $f(x) = 0$.

Let us name the left-hand-sides of equations 
(\ref{specific-kirch-L}), (\ref{specific-kirch-S}), 
(\ref{specific-charge-neutrality})
as $f_\la$, $f_\sm$, and $f_\Sigma$. 
We eliminate $Q_i$ and $Q_e$ from these using
(\ref{analytic-Q-i}) and (\ref{analytic-Q-e}),
and regard them as functions of $\eta, Q_\la, Q_\sm$ as follows:
\begin{eqnarray}
  f_\la\left(\eta, Q_\la, Q_\sm\right) &\equiv& -J_{\la,\sm} - J_{\la, i} - J_{\la, e} , \\
  f_\sm\left(\eta, Q_\la, Q_\sm\right) &\equiv& -J_{\la, \sm} - J_{\sm, i} - J_{\sm, e} , \\
  f_\Sigma\left(\eta, Q_\la, Q_\sm\right) &\equiv& Q_\la+Q_\sm+Q_i\left(\eta, Q_\la, Q_\sm\right)  \nonumber \\
  &+&Q_e\left(\eta, Q_\la, Q_\sm\right) , \\
  f_\crit\left(\eta, Q_\la, Q_\sm\right) &\equiv& 
 \frac{Q_\la  \ m_e v_e  u_\la}{Q_e\left(\eta, Q_\la, Q_\sm\right) \ \Wion} .
\end{eqnarray}
We have also defined $f_\crit$ according to
(\ref{equation-macroscopic-discharge}).

For each fixed set of $\eta$ and $Q_\la$, 
$f_\Sigma(\eta, Q_\la, Q_\sm)$ is a one-parameter function of $Q_\sm$.
According to the assumption we can solve $f_\Sigma(\eta, Q_\la, Q_\sm) = 0 $
for $Q_\sm$. 
We define $Q_\sm^0(\eta, Q_\la)$ to denote the solution, so that
\begin{eqnarray}
  f_\Sigma\left(\eta, Q_\la, Q_\sm^0\left(\eta, Q_\la\right)\right) = 0 
\end{eqnarray}
holds.

Then for each fixed $\eta$,
$f_\la(\eta, Q_\la, Q_\sm^0(\eta, Q_\la))$ is a one-parameter function of $Q_\la$.
According to the assumption we can solve 
$f_\la(\eta, Q_\la, Q_\sm^0(\eta, Q_\la)) = 0$
for $Q_\la$. 
We define $Q_\la^0(\eta)$ to denote the solution, so that
\begin{eqnarray}
  f_\la\left(\eta, Q_\la^0\left(\eta\right), Q_\sm^0\left(\eta, Q_\la^0\left(\eta\right)\right)\right) = 0 
\end{eqnarray}
holds.

Then 
$f_\crit(\eta, Q_\la^0(\eta), Q_\sm^0(\eta, Q_\la^0(\eta)))$ is a one-parameter function of $\eta$.
According to the assumption we can solve 
$f_\crit(\eta, Q_\la^0(\eta), Q_\sm^0(\eta, Q_\la^0(\eta))) = 0$
for $\eta$. 
We define $\eta^0$ to denote the solution, so that
\begin{eqnarray}
  f_\la\left(\eta^0, Q_\la^0\left(\eta^0\right), Q_\sm^0\left(\eta^0, Q_\la^0\left(\eta^0\right)\right)\right) = 0 
\end{eqnarray}
holds, which is the $\eta_\crit$ we are looking for.

With this method, whenever a solver approaches the zero point of one of $f$,
the $f$ will consist of two or more terms of same magnitude, and of other 
terms of smaller magnitude. Smaller terms are irrelevant to the equilibrium.
So we will always be comparing the terms of same magnitude. 
Because of this, the method gives sufficiently precise solutions 
even with single precision floating point numbers.

\subsection{Binary search solver on GPGPU}  \label{binary-search-solver-gpgpu}

Graphic processing units (GPUs) are processors specialized for computer graphics,
widely used in personal computers, workstations, and video game devices.
But as more and more realistic computer graphics have been demanded,
GPUs became capable of more and more types of calculations,
and finally evolved into general purpose graphic processing units (GPGPUs),
who are programmable for general computation, not limited to graphic processing.
Due to the nature of graphic processing tasks, 
GPUs' parallel computation capacities are
are one or two orders of magnitude larger compared to that of CPUs.
On the other hand marketplace competition and mass production keep the GPUs'
price low.
Although parallel programming has been a hard task for programmers,
the parallelism found in nature, together with GPGPU's power and price
makes it very alluring as next generation computational platform
for computational astrophysics, and computational natural science.
Use of GPGPU have already started in several fields of astronomy 
and astrophysics, such as
signal processing
\citep[e.g.][]{2008ExA....22..129H,2009arXiv0906.1887W},
N-body simulations of gravity
\citep[e.g.][]{2007astro.ph..3100H,2008NewA...13..103B,2008DPS....40.5809M},
gravitational lensing
\citep[e.g.][]{2009arXiv0905.2453T},
orbital dynamics
\citep[e.g.][]{2009NewA...14..406F},
radiation-transfer
\citep[e.g.][]{2009arXiv0907.3768J},
 and also in various other branches of
science
\citep[e.g.][]{2007arXiv0709.3225V,2008arXiv0809.1833A,2008arXiv0810.5365B,2009arXiv0903.3852J}.


With GPGPU we can challenge
problems that had been computationally formidable.
To begin this challenge, we have constructed {\em Tengu}, 
({\bf Ten}mon-{\bf G}PGP{\bf U} cluster; 
GPGPU cluster for astrophysical purposes. 
It consists of 10 computer nodes, each
equipped with NVIDIA's GPGPU.
We use the programming language {\sc {cuda}} to write codes for GPGPUs.
{\sc {cuda}} is compatible with {\sc {c++}}, so we benefit both 
from expressive power of {\sc {c++}} and computational power of GPGPUs.

Thanks to this, we organize our 
{\sc c++} and {\sc cuda} codes in the following way.
We made {\sc {c++}} classes representing the protoplanetary disc, 
dust plasma, problem initial conditions and solutions, 
and the numerical solvers.
Each solver inherit from an abstract solver class.
Thus the users of the solvers,
namely the programme parts that carries out  tests and numerical experiments
can use any solver they prefer, without noticing what algorithm
the solvers use nor on what hardware they run. 
We write most of the code in {\sc c++}
and compile them by {\sc gcc}.
The GPGPU related details are separated in several {\bf .cu} files
by means of pimpl idiom.
We compile {\bf .cu} files and link the object files using {\sc nvcc}, 
{\sc cuda} compiler provided by NVIDIA.

Another example of such benefit is {\sc thrust} ({\tt
  http://code.google.com/p/thrust/}), a {\sc cuda} counterpart of what
standard template library (STL) is in {\sc c++}.  With {\sc thrust}, for
example, device and host memory management is automated.  Memories are
allocated and freed automatically in the constructor and destructor of the
container classes.  Copying data between host memory and device memory are
simply expressed by substitution $=$ operators.


\subsection{Testing}

We choose the Test-Driven Development style for this study. 
We have tested
that the charge conservation and the current conservation conditions 
are held,
for each equilibrium solution that each solver give.
We have also tested that the value of the currents satisfy equations
by comparing them with the simplest implementation.

Why do we test our codes? We need tests in numerical physics 
because the codes must compile correctly, 
the codes must translate the algorithms correctly, and
algorithms represent the physical concepts correctly.

In order to assure these, we are accustomed to perform various tests during a
code development, 
by examining the internal states and outputs of the programme.
Furthermore, we want to make sure that criteria once tested
always meet thereafter, and that the tests cover all the important aspects of the code.
As the code grow, it becomes more and more effective to build up a system of
test rather than to perform tests manually. This technique is known as
Test-Driven Development \citep[e.g.][]{2005IEEE.31.3.2005}.

The programme is divided into many functional modules, and we test that these
modules give expected output for given input.  This is compared to
code-reading style of tests, where testers finds faults in the code by reading
them. Although code reading is capable of finding more mistakes
\citep{1987IEEE..SE13.12.1278} it is only effective when the code is short and
it is possible to trace the comprehensive behaviour of the code
line-by-line. Unit tests, on the other hand cares only on the input and
output. It is effective even if we are trying new languages or hardware, or we
cannot debug trace on them.  We use {\sc googletest}, Google's framework for
writing automated {\sc c++} tests ({\tt
  http://code.google.com/p/googletest/}).

We must also consider the time cost of the test. Because we do computationally 
heavy tasks, examining the entire behaviour of the programme is not practical.
With systematized tests, we can ensure the equivalence of the codes as 
we optimize them, or as we transplant it onto another language or
hardware. We further construct a `test ladder,' an analog of distance ladder
in cosmology. We develop a chain of successively faster algorithms, and feed
them with randomly generated inputs and check if their response is same up to 
required precision. At the one end of the ladder is a code that almost
directly trace the equations, correctness of whose implementation is 
self-evident. 
At the other end of the ladder is optimized,
massively parallel code running on GPGPU.

Finally, we evaluate the computational optimization achieved by
measuring the speed of each solvers in terms of 
how many problems they solve per wall clock time.
See Table \ref{table-optimisation} for optimization results.

\begin{table}
  \begin{center}
    \small
    \begin{tabular}{cccc}
       solver & \# of cases & time & optimization \\
      \hline
direct integral, CPU${}^1$ & 9885 & 636714 & $1.0$\\\\binary search, CPU${}^2$ & 19200 & 977.973 & $1.3 \times {10}^{3}$\\\\binary search, GPU${}^3$ & 19200 & 6.98705 & $1.8 \times {10}^{5}$\\\\final problem${}^4$ & 364325 & 226.469 & $1.0 \times {10}^{5}$
    \end{tabular}
  \end{center}
  \caption{\small
    The name of the computation runs, the size of initial conditions sets,
    and the wall clock time needed to solve $\eta_\crit$ 
    for all initial conditions
    in seconds.
    The speed of the codes are also listed, in terms of number of solved cases per time,
    relative to the first case.
    (1) We used the direct integral solver (\S \ref{direct-integral-solver}),
    running it in parallel on 40 CPU cores.
    (2) We used the binary search solver (\S \ref{binary-search-solver}),
    on a Core 2 Quad 9300 CPU in single thread. 
    (3) We used the {\sc cuda} version of binary search solver,
    (\S \ref{binary-search-solver-gpgpu}), 
    on single GTX280 GPGPU.
    (4) We used the same programme
    (\S \ref{binary-search-solver-gpgpu}) and the same GPU
    for actual numerical experiment.
    The run generates a set of data that corresponds to $\eta_\crit$
    as function of dust radius $r_\sm$, $r_\la$ and
    fractal dimension $D_\sm$, $D_\la$. 
    Or it corresponds to 
    one page of the result figure,
    e.g. Fig. \ref{figure-result-tanaka}.
   }   \label{table-optimisation}
\end{table}

\label{lastpage}


\begin{thebibliography}{}

\bibitem[\protect\citeauthoryear{Agmon}{Agmon}{1995}]{Agmon1995456}
Agmon N.,  1995, Chemical Physics Letters, 244, 456

\bibitem[\protect\citeauthoryear{{Andrecut}}{{Andrecut}}{2008}]{2008arXiv0809.%
1833A}
{Andrecut} M.,  2008, ArXiv e-prints

\bibitem[\protect\citeauthoryear{{Baker}, {Baker}, {Jayaratne}, {Latham} \&
  {Saunders}}{{Baker} et~al.}{1987}]{10.1002/qj.49711347807}
{Baker} B.,  {Baker} M.,  {Jayaratne} E.~R.,  {Latham} J.,    {Saunders} C.
  P.~R.,  1987, Quarterly Journal of the Royal Meteorological Society, 113,
  1193

\bibitem[\protect\citeauthoryear{{Balbus} \& {Hawley}}{{Balbus} \&
  {Hawley}}{1991}]{1991ApJ...376..214B}
{Balbus} S.~A.,  {Hawley} J.~F.,  1991, ApJ, 376, 214

\bibitem[\protect\citeauthoryear{{Barros}, {Babich}, {Brower}, {Clark} \&
  {Rebbi}}{{Barros} et~al.}{2008}]{2008arXiv0810.5365B}
{Barros} K.,  {Babich} R.,  {Brower} R.,  {Clark} M.~A.,    {Rebbi} C.,  2008,
  ArXiv e-prints

\bibitem[\protect\citeauthoryear{{Basili} \& {Selby}}{{Basili} \&
  {Selby}}{1987}]{1987IEEE..SE13.12.1278}
{Basili} V.~R.,  {Selby} R.~W.,  1987, IEEE Transactions on Software
  Engineering, pp 1278--1296

\bibitem[\protect\citeauthoryear{{Belleman}, {B{\'e}dorf} \& {Portegies
  Zwart}}{{Belleman} et~al.}{2008}]{2008NewA...13..103B}
{Belleman} R.~G.,  {B{\'e}dorf} J.,    {Portegies Zwart} S.~F.,  2008, New
  Astronomy, 13, 103

\bibitem[\protect\citeauthoryear{{Blum}}{{Blum}}{2004}]{2004ASPC..309..369B}
{Blum} J.,  2004, in {Witt} A.~N.,  {Clayton} G.~C.,   {Draine} B.~T.,  eds,
  Astrophysics of Dust Vol.~309 of Astronomical Society of the Pacific
  Conference Series.
pp 369--+

\bibitem[\protect\citeauthoryear{{Blum} \& {Wurm}}{{Blum} \&
  {Wurm}}{2000}]{2000Icar..143..138B}
{Blum} J.,  {Wurm} G.,  2000, Icarus, 143, 138

\bibitem[\protect\citeauthoryear{{Blum}, {Wurm}, {Kempf} \& {Henning}}{{Blum}
  et~al.}{1996}]{1996Icar..124..441B}
{Blum} J.,  {Wurm} G.,  {Kempf} S.,    {Henning} T.,  1996, Icarus, 124, 441

\bibitem[\protect\citeauthoryear{{Blum}, {Wurm}, {Poppe} \& {Heim}}{{Blum}
  et~al.}{1998}]{1998EM&P...80..285B}
{Blum} J.,  {Wurm} G.,  {Poppe} T.,    {Heim} L.-O.,  1998, Earth Moon and
  Planets, 80, 285

\bibitem[\protect\citeauthoryear{{Bondi}}{{Bondi}}{1964}]{JPhysChem.1964.68.44%
1}
{Bondi} A.,  1964, J. Phys. Chem., 68, 441

\bibitem[\protect\citeauthoryear{{Brauer}, {Dullemond} \& {Henning}}{{Brauer}
  et~al.}{2008}]{2008A&A...480..859B}
{Brauer} F.,  {Dullemond} C.~P.,    {Henning} T.,  2008, A\&A, 480, 859

\bibitem[\protect\citeauthoryear{{Chapillon}, {Guilloteau}, {Dutrey} \&
  {Pi{\'e}tu}}{{Chapillon} et~al.}{2008}]{2008A&A...488..565C}
{Chapillon} E.,  {Guilloteau} S.,  {Dutrey} A.,    {Pi{\'e}tu} V.,  2008, A\&A,
  488, 565

\bibitem[\protect\citeauthoryear{{Christian}, {Holmes}, {Bullock}, {Gaskell},
  {Illingworth} \& {Latham}}{{Christian} et~al.}{1980}]{QJRMetSoc1980.106.159}
{Christian} H.,  {Holmes} C.~R.,  {Bullock} J.~W.,  {Gaskell} W.,
  {Illingworth} A.~J.,    {Latham} J.,  1980, The Quarterly Journal of the
  Royal Meteorological Society, 106, 159

\bibitem[\protect\citeauthoryear{{Cuzzi} \& {Zahnle}}{{Cuzzi} \&
  {Zahnle}}{2004}]{2004ApJ...614..490C}
{Cuzzi} J.~N.,  {Zahnle} K.~J.,  2004, ApJ, 614, 490

\bibitem[\protect\citeauthoryear{{Dash}, {Mason} \& {Wettlaufer}}{{Dash}
  et~al.}{2001}]{JGR106(D17)2039520402}
{Dash} J.,  {Mason} B.,    {Wettlaufer} J.,  2001, Journal of Geophysical
  Research, 106, 20395

\bibitem[\protect\citeauthoryear{{Desch} \& {Cuzzi}}{{Desch} \&
  {Cuzzi}}{2000}]{2000Icar..143...87D}
{Desch} S.~J.,  {Cuzzi} J.~N.,  2000, Icarus, 143, 87

\bibitem[\protect\citeauthoryear{{Duley} \& {Williams}}{{Duley} \&
  {Williams}}{1984}]{1984inch.book.....D}
{Duley} W.~W.,  {Williams} D.~A.,  1984

\bibitem[\protect\citeauthoryear{{Dullemond} \& {Dominik}}{{Dullemond} \&
  {Dominik}}{2004}]{2004A&A...421.1075D}
{Dullemond} C.~P.,  {Dominik} C.,  2004, A\&A, 421, 1075

\bibitem[\protect\citeauthoryear{{Enoto}, {Tsuchiya}, {Yamada} \& {et
  al.}}{{Enoto} et~al.}{2008}]{2008ICRC....1..745E}
{Enoto} T.,  {Tsuchiya} H.,  {Yamada} S.,    {et al.} 2008, in International
  Cosmic Ray Conference Vol.~1 of International Cosmic Ray Conference.
pp 745--748

\bibitem[\protect\citeauthoryear{{Erdogmus}, {Morisio} \&
  {Torciano}}{{Erdogmus} et~al.}{2005}]{2005IEEE.31.3.2005}
{Erdogmus} H.,  {Morisio} M.,    {Torciano} M.,  2005, IEEE Transactions on
  Software Engineering, 31, 226

\bibitem[\protect\citeauthoryear{{Evans} II, {Rawlings}, {Shirley} \&
  {Mundy}}{{Evans} et~al.}{2001}]{2001ApJ...557..193E}
{Evans} II N.~J.,  {Rawlings} J.~M.~C.,  {Shirley} Y.~L.,    {Mundy} L.~G.,
  2001, ApJ, 557, 193

\bibitem[\protect\citeauthoryear{{Ford}}{{Ford}}{2009}]{2009NewA...14..406F}
{Ford} E.~B.,  2009, New Astronomy, 14, 406

\bibitem[\protect\citeauthoryear{{Gammie}}{{Gammie}}{1996}]{1996ApJ...457..355%
G}
{Gammie} C.~F.,  1996, ApJ, 457, 355

\bibitem[\protect\citeauthoryear{{Gaskell}, {Illingworth}, {Latham} \&
  {Moore}}{{Gaskell} et~al.}{1978}]{QJRMetSoc1978.104.447}
{Gaskell} W.,  {Illingworth} A.,  {Latham} J.,    {Moore} C.,  1978, The
  Quarterly Journal of the Royal Meteorological Society, 104, 447

\bibitem[\protect\citeauthoryear{{Gibbard}, {Levy} \& {Morfill}}{{Gibbard}
  et~al.}{1997}]{1997Icar..130..517G}
{Gibbard} S.~G.,  {Levy} E.~H.,    {Morfill} G.~E.,  1997, Icarus, 130, 517

\bibitem[\protect\citeauthoryear{Goncalves, Mathieu, Herlem \&
  Etcheberry}{Goncalves et~al.}{1999}]{Goncalves1999140}
Goncalves A.-M.,  Mathieu C.,  Herlem M.,    Etcheberry A.,  1999, Journal of
  Electroanalytical Chemistry, 477, 140

\bibitem[\protect\citeauthoryear{{G{\"u}ttler}, {Poppe}, {Wasson} \&
  {Blum}}{{G{\"u}ttler} et~al.}{2008}]{2008Icar..195..504G}
{G{\"u}ttler} C.,  {Poppe} T.,  {Wasson} J.~T.,    {Blum} J.,  2008, Icarus,
  195, 504

\bibitem[\protect\citeauthoryear{{Hamada} \& {Iitaka}}{{Hamada} \&
  {Iitaka}}{2007}]{2007astro.ph..3100H}
{Hamada} T.,  {Iitaka} T.,  2007, arXiv:astro-ph/0703100

\bibitem[\protect\citeauthoryear{{Harris}, {Haines} \&
  {Staveley-Smith}}{{Harris} et~al.}{2008}]{2008ExA....22..129H}
{Harris} C.,  {Haines} K.,    {Staveley-Smith} L.,  2008, Experimental
  Astronomy, 22, 129

\bibitem[\protect\citeauthoryear{{Hayashi}}{{Hayashi}}{1981}]{1981PThPS..70...%
35H}
{Hayashi} C.,  1981, Progress of Theoretical Physics Supplement, 70, 35

\bibitem[\protect\citeauthoryear{{Herczeg} \& {Hillenbrand}}{{Herczeg} \&
  {Hillenbrand}}{2008}]{2008ApJ...681..594H}
{Herczeg} G.~J.,  {Hillenbrand} L.~A.,  2008, ApJ, 681, 594

\bibitem[\protect\citeauthoryear{{Inutsuka} \& {Sano}}{{Inutsuka} \&
  {Sano}}{2005}]{2005ApJ...628L.155I}
{Inutsuka} S.,  {Sano} T.,  2005, ApJL, 628, L155

\bibitem[\protect\citeauthoryear{{Januszewski} \& {Kostur}}{{Januszewski} \&
  {Kostur}}{2009}]{2009arXiv0903.3852J}
{Januszewski} M.,  {Kostur} M.,  2009, ArXiv e-prints

\bibitem[\protect\citeauthoryear{{Jonsson} \& {Primack}}{{Jonsson} \&
  {Primack}}{2009}]{2009arXiv0907.3768J}
{Jonsson} P.,  {Primack} J.,  2009, ArXiv e-prints

\bibitem[\protect\citeauthoryear{{Kempf}, {Pfalzner} \& {Henning}}{{Kempf}
  et~al.}{1999}]{1999Icar..141..388K}
{Kempf} S.,  {Pfalzner} S.,    {Henning} T.~K.,  1999, Icarus, 141, 388

\bibitem[\protect\citeauthoryear{{Koshak} \& {Krider}}{{Koshak} \&
  {Krider}}{1989}]{JGR1989.94.1165}
{Koshak} W.,  {Krider} E.,  1989, Journal of Geophisical Research, 94, 1165

\bibitem[\protect\citeauthoryear{{Kretke} \& {Lin}}{{Kretke} \&
  {Lin}}{2007}]{2007ApJ...664L..55K}
{Kretke} K.~A.,  {Lin} D.~N.~C.,  2007, ApJL, 664, L55

\bibitem[\protect\citeauthoryear{Kudin \& Car}{Kudin \&
  Car}{2008}]{10.1021/ja077205t}
Kudin K.~N.,  Car R.,  2008, Journal of the American Chemical Society, 130,
  3915

\bibitem[\protect\citeauthoryear{{Levasseur-Regourd}, {Mukai}, {Lasue} \&
  {Okada}}{{Levasseur-Regourd} et~al.}{2007}]{2007P&SS...55.1010L}
{Levasseur-Regourd} A.~C.,  {Mukai} T.,  {Lasue} J.,    {Okada} Y.,  2007,
  Planet. Space Sci., 55, 1010

\bibitem[\protect\citeauthoryear{{Lin}, {Uman}, {Tiller}, {Brantley},
  {Beasley}, {Krider} \& {Weidman}}{{Lin} et~al.}{1979}]{JGR1979.84.6307}
{Lin} Y.~T.,  {Uman} M.~A.,  {Tiller} J.~A.,  {Brantley} R.~D.,  {Beasley}
  W.~H.,  {Krider} E.~P.,    {Weidman} C.~D.,  1979, Journal of Geophisical
  Research, 84, 6307–6314

\bibitem[\protect\citeauthoryear{{Makino}}{{Makino}}{2008}]{2008IAUS..246..457%
M}
{Makino} J.,  2008, in {Vesperini} E.,  {Giersz} M.,   {Sills} A.,  eds, IAU
  Symposium Vol.~246 of IAU Symposium.
pp 457--466

\bibitem[\protect\citeauthoryear{{Mason} \& {Dash}}{{Mason} \&
  {Dash}}{2000}]{JGR105(D16)10185}
{Mason} B.,  {Dash} J.,  2000, Journal of Geophysical Research, 105, 10185

\bibitem[\protect\citeauthoryear{{Miura}, {Nakamoto} \& {Doi}}{{Miura}
  et~al.}{2008}]{2008Icar..197..269M}
{Miura} H.,  {Nakamoto} T.,    {Doi} M.,  2008, Icarus, 197, 269

\bibitem[\protect\citeauthoryear{{Moore} \& {Quillen}}{{Moore} \&
  {Quillen}}{2008}]{2008DPS....40.5809M}
{Moore} A.~J.,  {Quillen} A.,  2008, in Bulletin of the American Astronomical
  Society Vol.~40 of Bulletin of the American Astronomical Society.
pp 504--+

\bibitem[\protect\citeauthoryear{{Nomura} \& {Millar}}{{Nomura} \&
  {Millar}}{2005}]{2005A&A...438..923N}
{Nomura} H.,  {Millar} T.~J.,  2005, A\&A, 438, 923

\bibitem[\protect\citeauthoryear{{Okuzumi}}{{Okuzumi}}{2009}]{2009ApJ...698.11%
22O}
{Okuzumi} S.,  2009, ApJ, 698, 1122

\bibitem[\protect\citeauthoryear{{Ormel} \& {Cuzzi}}{{Ormel} \&
  {Cuzzi}}{2007}]{2007A&A...466..413O}
{Ormel} C.~W.,  {Cuzzi} J.~N.,  2007, A\&A, 466, 413

\bibitem[\protect\citeauthoryear{{Ormel}, {Spaans} \& {Tielens}}{{Ormel}
  et~al.}{2007}]{2007A&A...461..215O}
{Ormel} C.~W.,  {Spaans} M.,    {Tielens} A.~G.~G.~M.,  2007, A\&A, 461, 215

\bibitem[\protect\citeauthoryear{{Ossenkopf}}{{Ossenkopf}}{1993}]{1993A&A...28%
0..617O}
{Ossenkopf} V.,  1993, A\&A, 280, 617

\bibitem[\protect\citeauthoryear{{P{\'e}rez}, {McCollum}, {van den Ancker} \&
  {Joner}}{{P{\'e}rez} et~al.}{2008}]{2008A&A...486..533P}
{P{\'e}rez} M.~R.,  {McCollum} B.,  {van den Ancker} M.~E.,    {Joner} M.~D.,
  2008, A\&A, 486, 533

\bibitem[\protect\citeauthoryear{{Pilipp}, {Hartquist} \& {Morfill}}{{Pilipp}
  et~al.}{1992}]{1992ApJ...387..364P}
{Pilipp} W.,  {Hartquist} T.~W.,    {Morfill} G.~E.,  1992, ApJ, 387, 364

\bibitem[\protect\citeauthoryear{{Rakov} \& {Uman}}{{Rakov} \&
  {Uman}}{1998}]{1998IEEE.40.403}
{Rakov} V.~A.,  {Uman} M.~A.,  1998, IEEE Transactions on Electromagnetic
  Compatibility, 40, 403

\bibitem[\protect\citeauthoryear{{Roussel-Dupr{\'e}} \&
  {Gurevich}}{{Roussel-Dupr{\'e}} \& {Gurevich}}{1996}]{1996JGR...101.2297R}
{Roussel-Dupr{\'e}} R.,  {Gurevich} A.~V.,  1996, JGR, 101, 2297

\bibitem[\protect\citeauthoryear{{Sano}, {Inutsuka} \& {Miyama}}{{Sano}
  et~al.}{1998}]{1998ApJ...506L..57S}
{Sano} T.,  {Inutsuka} S.,    {Miyama} S.~M.,  1998, ApJL, 506, L57

\bibitem[\protect\citeauthoryear{{Sano}, {Inutsuka}, {Turner} \&
  {Stone}}{{Sano} et~al.}{2004}]{2004ApJ...605..321S}
{Sano} T.,  {Inutsuka} S.,  {Turner} N.~J.,    {Stone} J.~M.,  2004, ApJ, 605,
  321

\bibitem[\protect\citeauthoryear{{Shakura} \& {Sunyaev}}{{Shakura} \&
  {Sunyaev}}{1973}]{1973AnA....24..337S}
{Shakura} N.~I.,  {Sunyaev} R.~A.,  1973, A\&A, 24, 337

\bibitem[\protect\citeauthoryear{{Sickafoose}, {Colwell}, {Hor{\'a}nyi} \&
  {Robertson}}{{Sickafoose} et~al.}{2001}]{2001JGR...106.8343S}
{Sickafoose} A.,  {Colwell} J.,  {Hor{\'a}nyi} M.,    {Robertson} S.,  2001,
  JGR, 106, 8343

\bibitem[\protect\citeauthoryear{{Sirono}}{{Sirono}}{1999}]{1999A&A...347..720%
S}
{Sirono} S.,  1999, A\&A, 347, 720

\bibitem[\protect\citeauthoryear{{Somorjai}}{{Somorjai}}{1994}]{1994.somorjai.%
book}
{Somorjai} G.~A.,  1994

\bibitem[\protect\citeauthoryear{{Spergel}, {Kasdin} \& {Belikov}}{{Spergel}
  et~al.}{2009}]{2009AAS...21345804S}
{Spergel} D.~N.,  {Kasdin} J.,    {Belikov} R. e.~a.,  2009, in Bulletin of the
  American Astronomical Society Vol.~41 of Bulletin of the American
  Astronomical Society.
pp 362--+

\bibitem[\protect\citeauthoryear{{Spitzer}}{{Spitzer}}{1941}]{1941ApJ....93..3%
69S}
{Spitzer} L.~J.,  1941, ApJ, 93, 369

\bibitem[\protect\citeauthoryear{{Suyama}, {Wada} \& {Tanaka}}{{Suyama}
  et~al.}{2008}]{2008ApJ...684.1310S}
{Suyama} T.,  {Wada} K.,    {Tanaka} H.,  2008, ApJ, 684, 1310

\bibitem[\protect\citeauthoryear{{Takahashi}}{{Takahashi}}{2005}]{TakahashiBub%
ble}
{Takahashi} M.,  2005, The journal of physical chemistry, B, 109, 21858

\bibitem[\protect\citeauthoryear{{Takahashi}}{{Takahashi}}{1978}]{TakahashiThu%
nder}
{Takahashi} T.,  1978, Journal of the Atmospheric Sciences, 35, 1536

\bibitem[\protect\citeauthoryear{{Thompson}, {Fluke}, {Barnes} \&
  {Barsdell}}{{Thompson} et~al.}{2009}]{2009arXiv0905.2453T}
{Thompson} A.~C.,  {Fluke} C.~J.,  {Barnes} D.~G.,    {Barsdell} B.~R.,  2009,
  ArXiv e-prints

\bibitem[\protect\citeauthoryear{{Tsuchiya}, {Enoto}, {Yamada}, {Yuasa},
  {Kawaharada}, {Kitaguchi}, {Kokubun}, {Kato}, {Okano}, {Nakamura} \&
  {Makishima}}{{Tsuchiya} et~al.}{2007}]{2007PhRvL..99p5002T}
{Tsuchiya} H.,  {Enoto} T.,  {Yamada} S.,  {Yuasa} T.,  {Kawaharada} M.,
  {Kitaguchi} T.,  {Kokubun} M.,  {Kato} H.,  {Okano} M.,  {Nakamura} S.,
  {Makishima} K.,  2007, Physical Review Letters, 99, 165002

\bibitem[\protect\citeauthoryear{{Turner}, {Sano} \& {Dziourkevitch}}{{Turner}
  et~al.}{2007}]{2007ApJ...659..729T}
{Turner} N.~J.,  {Sano} T.,    {Dziourkevitch} N.,  2007, ApJ, 659, 729

\bibitem[\protect\citeauthoryear{{Umebayashi} \& {Nakano}}{{Umebayashi} \&
  {Nakano}}{2009}]{2009ApJ...690...69U}
{Umebayashi} T.,  {Nakano} T.,  2009, ApJ, 690, 69

\bibitem[\protect\citeauthoryear{{van Meel}, {Arnold}, {Frenkel}, {Portegies
  Zwart} \& {Belleman}}{{van Meel} et~al.}{2007}]{2007arXiv0709.3225V}
{van Meel} J.~A.,  {Arnold} A.,  {Frenkel} D.,  {Portegies Zwart} S.~F.,
  {Belleman} R.~G.,  2007, ArXiv e-prints

\bibitem[\protect\citeauthoryear{{Wada}, {Tanaka}, {Suyama}, {Kimura} \&
  {Yamamoto}}{{Wada} et~al.}{2008a}]{2008ApJ...677.1296W}
{Wada} K.,  {Tanaka} H.,  {Suyama} T.,  {Kimura} H.,    {Yamamoto} T.,  2008a,
  ApJ, 677, 1296

\bibitem[\protect\citeauthoryear{{Wada}, {Tanaka}, {Suyama}, {Kimura} \&
  {Yamamoto}}{{Wada} et~al.}{2008b}]{2008LPI....39.1545W}
{Wada} K.,  {Tanaka} H.,  {Suyama} T.,  {Kimura} H.,    {Yamamoto} T.,  2008b,
  in Lunar and Planetary Institute Science Conference Abstracts Vol.~39 of
  Lunar and Planetary Institute Science Conference Abstracts.
pp 1545--+

\bibitem[\protect\citeauthoryear{{Wayth}, {Greenhill} \& {Briggs}}{{Wayth}
  et~al.}{2009}]{2009arXiv0906.1887W}
{Wayth} R.~B.,  {Greenhill} L.~J.,    {Briggs} F.~H.,  2009, ArXiv e-prints

\bibitem[\protect\citeauthoryear{{Weidenschilling}}{{Weidenschilling}}{1977}]{%
1977MNRAS.180...57W}
{Weidenschilling} S.~J.,  1977, MNRAS, 180, 57

\bibitem[\protect\citeauthoryear{{Weidling}, {G{\"u}ttler}, {Blum} \&
  {Brauer}}{{Weidling} et~al.}{2009}]{2009ApJ...696.2036W}
{Weidling} R.,  {G{\"u}ttler} C.,  {Blum} J.,    {Brauer} F.,  2009, ApJ, 696,
  2036

\bibitem[\protect\citeauthoryear{{Weidnschilling}}{{Weidnschilling}}{1997}]{19%
97LPI....28.1515W}
{Weidnschilling} S.~J.,  1997, in Lunar and Planetary Institute Science
  Conference Abstracts Vol.~28 of Lunar and Planetary Inst. Technical Report.
pp 1515--+

\bibitem[\protect\citeauthoryear{{Williams}}{{Williams}}{2001}]{PhysicsToday00%
319228}
{Williams} E.,  2001, Physics Today, 54, 41

\bibitem[\protect\citeauthoryear{{Wurm} \& {Blum}}{{Wurm} \&
  {Blum}}{1998}]{1998Icar..132..125W}
{Wurm} G.,  {Blum} J.,  1998, Icarus, 132, 125

\bibitem[\protect\citeauthoryear{{Zsom} \& {Dullemond}}{{Zsom} \&
  {Dullemond}}{2008}]{2008A&A...489..931Z}
{Zsom} A.,  {Dullemond} C.~P.,  2008, A\&A, 489, 931

\end{thebibliography}
\end{document}